\documentclass[preprint,showpacs,preprintnumbers,amsmath,amssymb]{revtex4}
\newcommand{\newwidth}{0.675\textwidth}
\newcommand{\newheight}{0.45\textwidth}
\newcommand{\newwidthprime}{0.225\textwidth}
\newcommand{\newheightprime}{0.30\textwidth}

\usepackage{graphicx,verbatim}
\usepackage{dcolumn}
\usepackage{bm,amssymb}
\usepackage{sidecap}
\graphicspath{{/user/rhaertle/epsfiles/Propplots/},{/user/rhaertle/Berlin08/},
{/user/rhaertle/epsfiles/Rate-Report/},{/user/rhaertle/epsfiles/RateCoul/},
{/user/rhaertle/epsfiles/Benesch/},{/user/rhaertle/PaperI/},{/user/rhaertle/PaperII/},
{/user/rhaertle/epsfiles/Cizek/}}

\begin{document}

\title{
Vibrationally Induced Decoherence in Single-Molecule Junctions
}

\author{R.\ H\"artle}
\altaffiliation{Present address: Department of Physics, Columbia University, New York, New York 10027, USA}
\author{M.\ Butzin}
\altaffiliation{Present address: Robert Bosch GmbH, Corporate Sector Research and Advanced Engineering Applied Research 1, 70049 Stuttgart, Germany}
\author{M.\ Thoss}
\affiliation{
Institut f\"ur Theoretische Physik und Interdisziplin\"ares Zentrum f\"ur Molekulare 
Materialien, \\
Friedrich-Alexander-Universit\"at Erlangen-N\"urnberg,\\ 
Staudtstr.\ 7/B2, D-91058 Erlangen, Germany
}

\date{\today}

\begin{abstract}
We investigate the interplay of quantum interference effects and electronic-vibrational coupling 
in electron transport through single-molecule junctions, employing a 
nonequilibrium Green's function approach. 
Our findings show that inelastic processes lead, in general, to a quenching of 
quantum interference effects. This quenching is more pronounced 
for increasing bias voltages and levels of vibrational excitation. 
As a result of this vibrationally induced decoherence, vibrational signatures 
in the transport characteristics of a molecular contact may strongly deviate 
from a simple Franck-Condon picture. This includes signatures in both the resonant 
and the non-resonant transport regime.
Moreover, it is shown that local cooling by electron-hole pair creation processes can 
influence the transport characteristics profoundly, giving rise to a significant 
temperature dependence of the electrical current. 
\end{abstract}


\maketitle

\section{Introduction}

In the past decades, electronic components have become continuously 
smaller and reached the nanoscale. At these scales, electron 
transport can no longer be understood  on purely classical grounds but quantum 
mechanical effects, such as, for example, the quantization of energy levels, quantum interference 
effects and tunneling processes need to be taken into account. 
Single-molecule junctions, where a molecule is contacted by two macroscopic 
(in most cases, metallic) electrodes, represent prime examples of nanoelectronic devices. 
These junctions allow to investigate molecules under controllable nonequilibrium 
conditions and may facilitate an ultimate step in the miniaturization 
of nanoelectronic devices \cite{Nitzan01,Galperin07,cuevasscheer2010,Zimbovskaya2011}. 
It is thus highly desirable to understand the electron transport properties 
of these junctions.

Experimental studies employing a variety of techniques, such as mechanically controlled break junctions 
\cite{Reed97,Reichert02,Smit02,Boehler04,Elbing05,Martin2010,Secker2010,Ballmann2010,Kim2011,Ballmann2012}, 
scanning tunneling microscopy \cite{Ho98,Pascual03,Ogawa07,Repp2009,Brumme2010,Tao2010,Frederiksen2010,Venkataraman2011,Toher2011,Limot2012,Franke2012} 
and electromigration \cite{Sapmaz05,Leon2008,Huettel2009,Osorio2010},
have shown that molecular junctions exhibit a large variety of interesting transport phenomena, 
including, for example, switching \cite{Blum05,Riel2006,Choi,Quek2009,Berndt2009,Evers2009,Molen2010,Mohn2010}, 
negative differential resistance \cite{Pop2005,Sapmaz05,Tal2008,Osorio2010,Limot2012}, diode- and/or transistor-like 
behavior \cite{Elbing05,Yeganeh2007,Song2009}.
Many of these transport properties are not fully understood yet. 
On one hand, this is due to ambiguities in the contact geometry, 
which may lead to large variations in the recorded current-voltage 
characteristics \cite{Boehler04,Osorio2007,Tao2010,Secker2010}. 
While strategies to control the contact geometry exist, for example, by the use of 
special linker groups \cite{Hybertsen2008,Quek2009}, 
direct binding schemes \cite{Kiguchi2008,Venkataraman2011}, 
or click-chemistry approaches \cite{Mayor2009,Braunschweig2009}, 
the electrodes themselves may show irregular behavior. It is therefore expedient to identify 
transport properties of molecules that are robust with respect to the 
contact geometry of the contact, such as, for example, 
inelastic electron tunneling spectra \cite{Ho98,Lorente2000,Troisi2006,GalpScience08,Godby2012}). 
On the other hand, electron transport through a molecular conductor represents a 
complex many-body problem, 
where, aside from multiple electronic states, the coupling to 
numerous vibrational degrees of freedom has to be 
taken into account \cite{Galperin07,Hartle09,Secker2010,Hartle2011b,Ballmann2012}. 
In contrast to other nanoelectronic devices (\emph{e.g.}\ quantum dots \cite{Alhassid2000}), 
electronic-vibrational coupling is typically rather strong in molecular systems. This is due to their 
small size and mass and can lead to pronounced 
vibrational effects in the respective transport characteristics 
\cite{Pascual03,Leon2008,Natelson2008,Huettel2009,Repp2009,Tao2010,Ballmann2010,
Jewell2010,Osorio2010,Frederiksen2010,Secker2010,Kim2011,Natelson2011,Ballmann2012}. 
Thereby, the interplay of electronic and the vibrational degrees of freedom 
of a molecular conductor often involves strong nonequilibrium 
\cite{Hartle09,Hartle2010b,Thoss2012} as well as quantum mechanical effects \cite{Hartle2011b,Ballmann2012}. 

A fundamental quantum mechanical effect is quantum interference. 
The occurrence and observability of quantum interference effects 
in electron transport through molecular junctions has received great attention recently 
\cite{Hartle2011b,Guedon2011,Ballmann2012,Viewpoint2012}. 
Quantum interference effects in molecular junctions arise because, 
very similar to a double-slit experiment, an electron may take different pathways through a molecule. 
Thereby, pathways can be spatially different paths in molecules with an appropriate topology 
and/or originate from different paths in energy space, \emph{e.g.}, for molecules 
that exhibit quasidegenerate electronic states 
\cite{Hod2006,Begemann2008,Solomon2008,Hod2008,Markussen2010,Hartle2011b,Ballmann2012}. 
Recently, strong experimental evidence for quantum interference effects in molecular junctions
has been reported \cite{Guedon2011,Ballmann2012}.
Moreover, quantum interference effects have been observed in the closely related field of 
electron transport through quantum dots that are 
set up as Aharonov-Bohm interferometers \cite{Umansky1998,Holleitner2001,Wegscheider2007}. 
Depending on the magnetic flux that is threading the quantum dot, the transmission 
amplitudes of the electrons interfere constructively or destructively with each other, leading to 
sizable oscillations in the conductance of such a device \cite{Holleitner2001,Kubala2002,Wegscheider2007}. 
A great deal of theoretical work \cite{Kalyanaraman2002,Collepardo2004,Ernzerhof2005,Hod2006,Goyer2007,
Solomon2008b,Begemann2008,Darau2009,Stadler2009,Brisker2010,
Schultz2010,Markussen2010,Stafford2011,Markussen2011,Lambert2011,Brisker2012} has been 
devoted to study quantum interference effects in molecular junctions, 
which is, on one hand, due to their fundamental importance but, on the other hand, also 
due to possible device applications, such as, for example, in transistors \cite{Stafford2007,QuIETPatent}, 
thermoelectric devices \cite{Bergfield2009,Bergfield2010b} or spin filters \cite{Herrmann2010,Herrmann2011}. 
Thereby, in most of the studies, the effect of electron-electron interactions and electronic-vibrational coupling 
has been neglected. 

In this article, we investigate the influence of electronic-vibrational coupling on  
quantum interference effects in single-molecule junctions. 
Thereby, we extend our earlier studies \cite{Hartle2011b,Ballmann2012} and analyze in detail
the mechanism of vibrationally induced decoherence.
Moreover, we consider the effect of local cooling by electron-hole pair creation processes 
and show how these processes facilitate a mechanism that allows to control 
quantum interference effects by an external parameter, that is, the temperature of the electrodes. 
Only recently, this mechanism has been experimentally verified by Ballmann \emph{et al.}\ \cite{Ballmann2012} 
for a number of different junctions, including different molecular species. 
Note that, in contrast to Refs.\ \onlinecite{Segal2000,Lehmann04,Brisker2012}, we do not 
consider the vibrational degrees of freedom as being part of a thermal bath 
but, on the contrary, as active degrees of freedom. 

For our studies, we employ a nonequilibrium Green's function approach developed by 
Galperin \emph{et al.}\ \cite{Galperin06}, which was recently extended by some 
of us to account for multiple vibrational \cite{Hartle} as well as multiple electronic degrees of freedom 
\cite{Hartle09,Hartle2010,Volkovich2011b}. The approach is based on a separation of 
electronic and vibrational time scales and, therefore, allows a nonperturbative description of 
electronic-vibrational coupling as it is required in this nonequilibrium transport problem. 
Moreover, it suits to describe quasidegenerate levels, which exhibit pronounced 
quantum interference effects. Besides this approach, a variety of other methods has been 
developed to describe vibrationally coupled electron transport through nanoelectronic devices. 
This includes other nonequilibrium Green's function approaches, which are based on either perturbation theory 
\cite{Mitra04,Galperin04,Frederiksen04,Haupt2009,Avriller2009,Schmidt2009,Urban2010,Wohlman2010} or 
nonperturbative schemes \cite{Flensberg03,Galperin0908,Fehske2010}, 
scattering theory \cite{Cizek04,Toroker07,Zimbovskaya09,Seidemann10}, 
master equation methodologies  \cite{May02,Mitra04,Lehmann04,Pedersen05,Reckermann2008,Leijnse09,Esposito2010,Timm2010,Kosov2012}  
and a number of numerically exact schemes \cite{Han2006,Eckel2010,Thoss2011,Thoss2012}.

The article is organized as follows: In Sec.\ \ref{hamiltonian}, we introduce the 
model Hamiltonian that we use to describe vibrationally coupled electron transport through 
a single-molecule junction. The nonequilibrium Green's function approach, which we employ to calculate 
the corresponding transport characteristics, is outlined in Sec.\ \ref{NEGFapproach}. 
In Sec.\ \ref{BasIntSec}, we introduce and analyze in detail the basic quantum interference effects 
that occur in single-molecule junctions. This facilitates the discussion of the effect 
of vibrationally induced decoherence in the following sections, 
Secs.\ \ref{SequTunReg} -- \ref{anomal}. Thereby, we study first simplified electronic-vibrational 
coupling scenarios that allow to investigate basic decoherence phenomena due to vibrations in the resonant 
(Sec.\ \ref{SequTunReg}) and the non-resonant transport regime (Sec.\ \ref{CoTunEffects}). 
In addition, we also discuss results for more realistics coupling scenarios. 
In Secs.\ \ref{localcool} and \ref{anomal}, we  show the importance of cooling by 
electron-hole pair creation processes 
in the presence of strong destructive quantum interference effects and how this cooling mechanism 
can be used to control interference effects by varying the temperature of the electrodes.

\section{Theory}

\subsection{Model Hamiltonian}
\label{hamiltonian}

We consider quantum interference effects and vibrationally induced decoherence 
in electron transport through a single molecule that is bound to two metal leads 
\cite{Galperin07,cuevasscheer2010,Zimbovskaya2011}. 
This transport problem can be described by a set of discrete electronic states, which 
are localized on the molecular bridge (M), and a continuum of electronic states, which 
are localized in the left (L) and the right (R) lead, respectively.  
Thereby, the states on the molecular bridge are coupled to the states in the leads. 
The corresponding model Hamiltonian can be written as \cite{Hartle2010b}
\begin{eqnarray}
\label{hel}
H_{\text{el}} &=& \sum_{m\in\text{M}} \epsilon_{m} c_{m}^{\dagger}c_{m} + \sum_{m<n\in\text{M}} U_{mn} c_{m}^{\dagger}c_{m} 
 c_{n}^{\dagger}c_{n} 
+ \sum_{k\in\text{L,R}} \epsilon_{k} c_{k}^{\dagger}c_{k}  + \sum_{k\in\text{L,R};m\in\text{M}} ( V_{mk} c_{k}^{\dagger}c_{m} + \text{h.c.} ), \hspace{1cm}
\end{eqnarray}
where the $\epsilon_{k}$ denote the energies of the
lead states and $c_{k}^{\dagger}$ and $c_{k}$ the 
corresponding creation and annihilation operators. 
Similarly, the $m$th electronic 
state on the molecular bridge is addressed by the creation and annihilation 
operators $c_{m}^{\dagger}$ and $c_{m}$, while its energy is given by $\epsilon_{m}$. 
Coulomb interactions may give rise to additional charging energies. These energies are 
accounted for in ${\text{Hamiltonian (\ref{hel})}}$ by Hubbard-like 
electron-electron interaction terms, 
$U_{mn} c_{m}^{\dagger}c_{m} c_{n}^{\dagger}c_{n}$ \footnote{Note at this point 
that we focus on electronic states of the molecular bridge 
that are located above the Fermi level of the junction. If there are states that are located 
below the Fermi level, the structure of the Hamiltonian is, in general, slightly more 
complex, since the parameters $\epsilon_{m}$, $U_{mn}$ etc.\ need to be determined with respect to 
a reference state (see Refs.\ \cite{Cederbaum74,Benesch06,Benesch08,Hartle2010b}). 
For the effects and mechanisms that are described below, it is, however, 
not decisive whether the states are located above or below the Fermi level of the junction.}. 
In principle, the index $m$ distinguishes different molecular orbitals, including the spin of the electrons.  
However, as interference effects occur only for transport through orbitals with the same spin, 
we suppress the spin degree of freedom in the following.  
The coupling matrix elements $V_{mk}$ in the fourth term of $H_{\text{el}}$ characterize the 
strength of the interaction between the electronic states of  
the molecular bridge and the leads and determine the so-called level-width 
functions $\Gamma_{K,mn}(\epsilon)=2\pi\sum_{k\in K} V_{mk}^{*} V_{nk} 
\delta(\epsilon-\epsilon_{k})$ ($K$=L,R). 
Modeling the leads as semi-infinite tight-binding chains with an internal 
hopping parameter $\gamma=2$\,eV, these functions are given by \cite{Cizek04}
\begin{eqnarray}
 \Gamma_{K,mn}(\epsilon) &=& \frac{ \nu_{K,m} \nu_{K,n}}{\gamma^{2}} \sqrt{4\gamma^{2}-(\epsilon-\mu_{K})^{2}},  
\end{eqnarray}
where, similar to $V_{mk}$, the parameters $\nu_{K,m}$ denote the coupling strength 
of state $m$ to lead $K$. Thereby, following Refs.\ \cite{Datta1997,Mujica2002,Koenemann2006}, we 
assume a symmetric drop of the bias voltage $\Phi$ at the contacts, 
\emph{i.e.}\ the chemical potentials in the left and the right lead 
are given by $\mu_{\text{L}}=e\Phi/2$ and $\mu_{\text{R}}=-e\Phi/2$, respectively. 
Note that throughout the paper, 
the Fermi energy of the leads is set to $\epsilon_{\text{F}}=0$\,eV.

Applying a bias voltage an electrical current is flowing through the junction. 
The molecular bridge may respond to this current, in particular, to the 
fluctuations of its charge state, which are induced by the tunneling electrons, by 
adapting its geometrical structure. If the 
equilibrium positions of the nuclei are different in the different charge states 
that are probed by the transferred electrons (which is usually the case), 
transitions between the different vibrational levels become available, 
and  the vibrational degrees of freedom of the junction may be excited according to the standard Franck-Condon picture
\cite{Semmelhack,Hartle,Hartle09,Romano10,Hartle2010,Hartle2010b,Hartle2011,Volkovich2011b}. 
The vibrational degrees of freedom are described in our model as harmonic oscillators that are 
linearly coupled to the electron densities $c_{m}^{\dagger}c_{m}$ 
on the molecular bridge \cite{Cederbaum74,Benesch08,Han2010},   
\begin{eqnarray}
\label{Hvib}
H_{\text{vib}} &=& \sum_{\alpha} \Omega_{\alpha} a_{\alpha}^{\dagger}a_{\alpha} 
+ \sum_{m\alpha} \lambda_{m\alpha} Q_{\alpha} c_{m}^{\dagger}c_{m},
\end{eqnarray}
where the operator $a^{\dagger}_{\alpha}$ denotes the creation operator of the 
$\alpha$th oscillator with frequency $\Omega_{\alpha}$ and  
$Q_{\alpha}=a_{\alpha}+a_{\alpha}^{\dagger}$ the corresponding 
vibrational displacement operators. The respective 
coupling strengths are denoted by $\lambda_{m\alpha}$. 
Finally, the Hamiltonian of the overall system is given by the sum 
\begin{eqnarray}
H &=& H_{\text{el}} + H_{\text{vib}}. 
\end{eqnarray}

For the nonequilibrium Green's function approach that we use in this article 
(cf.\ Sec.\ \ref{NEGFapproach}),  
it is expedient to remove the direct electronic-vibrational coupling term in the 
Hamiltonian $H$. To this end, we employ 
the small polaron transformation \cite{Mahan81,Mitra04,Galperin06,Hartle}  
\begin{eqnarray}
\label{transformedHamiltonian}
\overline{H} &=& \text{e}^{S} H \text{e}^{-S} \\ 
&=& \sum_{m} \overline{\epsilon}_{m} c_{m}^{\dagger}c_{m} + \sum_{m<n} \overline{U}_{mn} c_{m}^{\dagger}c_{m}
c^{\dagger}_{n}c_{n}  
 + \sum_{k} \epsilon_{k} c_{k}^{\dagger}c_{k} + \sum_{km} ( V_{mk} X_{m} 
c_{k}^{\dagger}c_{m} + \text{h.c.} ) + \sum_{\alpha} \Omega_{\alpha} a^{\dagger}_{\alpha}a_{\alpha} , \nonumber
\end{eqnarray}
with 
\begin{eqnarray}
S &=& - i \sum_{m\alpha} \frac{\lambda_{m\alpha}}{\Omega_{\alpha}}  c^{\dagger}_{m}c_{m}  
 P_{\alpha}, \\
X_{m} &=& \text{exp}[i\sum_{\alpha}\frac{\lambda_{m\alpha}}{\Omega_{\alpha}}
P_{\alpha}], 
\end{eqnarray}
and $P_{\alpha}=-i(a_{\alpha}-a_{\alpha}^{\dagger})$ the momentum operator associated with vibrational mode $\alpha$. 
Note that, although there is no explicit electronic-vibrational coupling in $\overline{H}$, 
it appears in the transformed Hamiltonian $\overline{H}$ 
at three different places: $i$) in the polaron-shifted energies 
${\overline{\epsilon}_{m}=\epsilon_{m}-\sum_{\alpha}
(\lambda_{m\alpha}^{2}/\Omega_{\alpha})}$, $ii$) in additional electron-electron interactions, 
which add to the original electron-electron interaction terms 
 ${\overline{U}_{mn}=U_{mn}-2\sum_{\alpha}(\lambda_{m\alpha}
\lambda_{n\alpha}/\Omega_{\alpha})}$ and $iii$) in the molecule-lead coupling 
term $\sum_{km}( V_{mk} X_{m} c_{k}^{\dagger}c_{m} + \text{h.c.} )$, which 
is renormalized by the shift operators $X_{m}$. 

Because we focus in this article on the effect of 
electronic-vibrational coupling on the transport characteristics of a single-molecule 
junction, we suppress the effect of electron-electron interactions in the following, 
that is we set the renormalized electron-electron interaction strengths $\overline{U}_{mn}$ to zero. 
A discussion of the effect of electron-electron interactions 
on the transport characteristics of a molecular junction can be found, for example, 
in our earlier work, Refs.\ \onlinecite{Hartle09,Hartle2010b,Volkovich2011b}, or in 
Refs.\ \onlinecite{Fransson2005,Paaske2005,Sela,Galperin2007,Han2010,Segal2011}. 
In this context, it is interesting to note that for high bias voltage and
within the wide-band approximation, electron-electron interactions have no
substantial influence on the electrical current. This was shown by Gurvitz and
Prager \cite{Gurvitz96,Gurvitz98} based on their exact microscopic rate equation methodology. 
At small (finite) bias voltages, however, it has already 
been demonstrated that in the presence of electron-electron interactions quantum interference effects 
become quenched by spin-flip processes \cite{Gefen2001,Iye2001,Gefen2002,Wegscheider2007}. 
We like to stress that these effects are interesting on their own and very important, 
but that they are beyond the scope of this work.

\subsection{Nonequilibrium Green's function approach}
\label{NEGFapproach}

The central quantities  
in (nonequilibrium) Green's function theory are the single-particle 
Green's functions, which are given for the electronic degrees of freedom by
 \begin{eqnarray}
G_{mn}(\tau,\tau') &=& -i \langle \text{T}_{c}
c_{m}(\tau)c_{n}^\dagger(\tau') \rangle_{H}.
\end{eqnarray}
They allow to calculate 
all single-particle observables, such as, \emph{e.g.}, the population of levels 
or the current flowing through the junction. 
We employ the following ansatz to calculate 
these Green's functions \cite{Galperin06}: 
\begin{eqnarray}
\label{decoupling}
G_{mn}(\tau,\tau') &=& -i \langle \text{T}_{c}
c_{m}(\tau)c_{n}^\dagger(\tau') \rangle_{H} \,=\, -i \langle \text{T}_{c} c_{m}(\tau)X_m(\tau)c_{n}^\dagger(\tau') X^\dagger_{n}(\tau')\rangle_{\overline{H}} \\
&\approx& \bar{G}_{mn}(\tau,\tau') \langle \text{T}_{c} X_m(\tau)X_{n}^\dagger(\tau') \rangle_{\overline{H}}, \qquad \nonumber
\end{eqnarray}
with the electronic Green's functions  
$\bar{G}_{mn}(\tau,\tau')= -i \langle \text{T}_{c} c_{m}(\tau)c_{n}^\dagger(\tau') \rangle_{\overline{H}}$ 
and $\text{T}_{c}$ the time-ordering operator on the Keldysh contour.  
Thereby, the indices $H/\overline{H}$ indicate the Hamiltonian that 
is used to evaluate the respective expectation values. 
The effective factorization of the single-particle 
Green's functions $G_{mn}$ into 
a product of the electronic Green's functions, $\bar{G}_{mn}$, 
and a correlation function of shift operators, 
$\langle \text{T}_{c} X_m(\tau)X_{n}^\dagger(\tau') \rangle_{\overline{H}}$, 
is justified, if the dynamics of the electronic and 
the vibrational degrees of freedom take place on different time scales. 
This is conceptually similar to the 
Born-Oppenheimer approximation \cite{Born1927,Domcke04}. In the present context, however,
we do not address the adiabatic regime, where the Born-Oppenheimer approximation applies, 
but rather the opposite regime, that is the anti-adiabatic regime, 
where the time scales for electron tunneling events are much longer than 
the ones for vibrational motion such that the nuclei of 
the molecule can follow the corresponding charge fluctuations.

Employing the equations of motion for the electronic Green's functions $\bar{G}_{mn}$, 
\begin{eqnarray}
 (i\partial_{\tau}-\bar{\epsilon}_{m}) \bar{G}_{mn}(\tau,\tau') 
(-i\partial_{\tau'}-\bar{\epsilon}_{n}) &=& \delta(\tau,\tau') 
(-i\partial_{\tau'}-\bar{\epsilon}_{n}) + \Sigma_{\text{L},mn}(\tau,\tau')+ \Sigma_{\text{R},mn}(\tau,\tau'), \nonumber\\
\end{eqnarray}
the respective self-energies due to the coupling 
of the molecule to the left and the right leads, $\Sigma_{\text{L},mn}(\tau,\tau')$ 
and $\Sigma_{\text{R},mn}(\tau,\tau')$, can be obtained. 
These self-energies are given to second order in the molecule-lead coupling by   
\begin{eqnarray}
\label{elselfen}
\Sigma_{\text{L/R},mn}(\tau,\tau')=\sum_{k\in \text{L/R}} V_{mk}^{*} V_{nk} g_{k}(\tau,\tau')\langle \text{T}_{c} X_{n}(\tau')X_{m}^\dagger(\tau) \rangle_{\overline{H}},
\end{eqnarray}
where $g_{k}(\tau,\tau')$ denotes the free Green's function associated with lead state $k$. 
The real-time projections of these self-energies determine the electronic part of 
the single-particle Green's functions $G_{mn}$. 
In the energy domain the corresponding Dyson-Keldysh equations read 
\begin{eqnarray}
\label{elDy}
 \bar{G}^{\text{r/a}}_{mn}(\epsilon) &=& \bar{G}^{0,\text{r/a}}_{mn}(\epsilon) + \sum_{op} \bar{G}^{0,\text{r/a}}_{mo}(\epsilon) \left(\Sigma^{\text{r/a}}_{\text{L},op}(\epsilon) + \Sigma^{\text{r/a}}_{\text{R},op}(\epsilon)\right) \bar{G}^{\text{r/a}}_{pn}(\epsilon),\\
\label{elKe}
 \bar{G}^{</>}_{mn}(\epsilon) &=& \sum_{op} \bar{G}^{\text{r}}_{mo}(\epsilon) \left(\Sigma^{</>}_{\text{L},op}(\epsilon) + \Sigma^{</>}_{\text{R},op}(\epsilon)\right) \bar{G}^{\text{a}}_{pn}(\epsilon),
\end{eqnarray}
with
\begin{eqnarray}
 \bar{G}^{0,\text{r/a}}_{mn}(\epsilon) &=& \delta_{mn} \frac{1}{\epsilon-\overline{\epsilon}_{m}+i \eta},
\end{eqnarray}
where $\eta$ denotes a positive infinitesimal number. 
Note that Eqs.\ (\ref{elDy}) and (\ref{elKe}) give the exact 
result in the non-interacting limit, that is for $\lambda_{m\alpha}\rightarrow0$, as well as for 
an isolated molecular bridge, where $V_{mk}\rightarrow0$.

Besides the electronic Green's functions $\bar{G}_{mn}$, 
we also need to evaluate the correlation functions of the shift operators, 
$\langle \text{T}_{c} X_m(\tau)X_{n}^\dagger(\tau')
\rangle_{\overline{H}}$, in order to obtain the full 
single-particle Green's functions $G_{mn}$. To this end, we use 
a second-order cumulant expansion in the dimensionless coupling parameters 
$\frac{\lambda_{m\alpha}}{\Omega_{\alpha}}$ \cite{Galperin06,Hartle,Hartle2010}
\begin{eqnarray}
\langle \text{T}_{c} X_m(\tau)X_{n}^\dagger(\tau')
\rangle_{\overline{H}}=\text{exp}\left(\sum_{\alpha\beta}i\frac{\lambda_{m\alpha}\lambda_{n\beta}}{\Omega_{\alpha}
\Omega_{\beta}}D_{\alpha\beta}(\tau,\tau')-i\frac{\lambda_{m\alpha}\lambda_{m\beta}+\lambda_{n\alpha}\lambda_{n\beta}}{2\Omega_{\alpha}
\Omega_{\beta}}D_{\alpha\beta}(\tau,\tau)\right),
\end{eqnarray}
with the momentum correlation functions 
\begin{eqnarray}
D_{\alpha\beta}=-i\langle \text{T}_{c} P_{\alpha}(\tau)P_{\beta}(\tau')\rangle_{\overline{H}}.
\end{eqnarray}
Employing the equation of motion for $D_{\alpha\beta}$
\begin{eqnarray}
\frac{1}{4\Omega_{\alpha}\Omega_{\beta}} (-\partial^{2}_{\tau}-\Omega_{\alpha}^{2}) D_{\alpha\beta}(\tau,\tau') (-\partial^{2}_{\tau'}-\Omega_{\beta}^{2}) &=& \delta(\tau,\tau') \delta_{\alpha\beta} (-\partial^{2}_{\tau'}-\Omega_{\beta}^{2}) \frac{1}{2\Omega_{\beta}} + \Pi_{\text{el},\alpha\beta}(\tau,\tau'), \nonumber\\
\end{eqnarray}
we determine the corresponding self-energy matrix $\Pi_{\text{el},\alpha\beta}$. The self-energy matrix 
$\Pi_{\text{el},\alpha\beta}$, which describes  
the interactions between the vibrational modes and 
the electronic degrees of freedom of the molecular bridge, 
is evaluated up to second order 
in the molecule-lead coupling \cite{Galperin06,Hartle,Hartle2010,Volkovich2011b} 
\begin{eqnarray}
\label{Piel}
\Pi_{\text{el},\alpha\beta}(\tau,\tau')=-i\sum_{mn}\frac{\lambda_{m\alpha}\lambda_{n\beta}}{\Omega_{\alpha} \Omega_{\beta}}(\Sigma_{mn}(\tau,\tau')\bar{G}_{nm}(\tau',\tau)+\Sigma_{nm}(\tau',\tau)\bar{G}_{mn}(\tau,\tau')). 
\end{eqnarray}
Since $\Pi_{\text{el},\alpha\beta}$ depends on the electronic self-energies 
$\Sigma_{mn}=\Sigma_{\text{L},mn}+\Sigma_{\text{R},mn}$ and Green's functions $\bar{G}_{mn}$, 
Eqs.\ (\ref{elselfen}) -- (\ref{Piel}) constitute a closed set of coupled nonlinear equations 
that needs to be solved iteratively in terms of a self-consistent scheme \cite{Galperin06,Hartle}.

With the Green's functions $D_{\alpha\beta}$ and $G_{mn}$, 
the average vibrational excitation of each vibrational mode is obtained according to \cite{Hartle,Hartle2010,Volkovich2011b} 
\begin{eqnarray}
\label{formulavibex}
\langle a^\dagger_{\alpha} a_{\alpha} \rangle_{H}&\approx&-\frac{1}{2}\text{Im}\left[D^{<}_{\alpha\alpha}(t=0)\right] -\frac{1}{2} +\sum_{m}\frac{\lambda_{m\alpha}^{2}}{\Omega_{\alpha}^2}\text{Im}[\bar{G}^<_{mm}(t=0)] \\
&&+2\sum_{m<n}\frac{\lambda_{m\alpha}\lambda_{n\alpha}}{\Omega_{\alpha}^2}\text{Im}[\bar{G}^<_{mm}(t=0)]\text{Im}[\bar{G}^<_{nn}(t=0)] \nonumber\\
&&-2\sum_{m<n}\frac{\lambda_{m\alpha}\lambda_{n\alpha}}{\Omega_{\alpha}^2}\text{Im}[\bar{G}^<_{mn}(t=0)]\text{Im}[\bar{G}^<_{nm}(t=0)], \nonumber 
\end{eqnarray}
where we use the Hartree-Fock factorization
\begin{eqnarray} 
 \langle c_{m}^{\dagger}c_{m} c_{n}^{\dagger}c_{n} \rangle_{\overline{H}} \approx 
\langle c_{m}^{\dagger}c_{m} \rangle\langle c_{n}^{\dagger}c_{n} \rangle-\langle c_{m}^{\dagger}c_{n} \rangle\langle c_{n}^{\dagger}c_{m} \rangle  
\end{eqnarray}
for $m\neq n$. 
The current is calculated employing the Meir-Wingreen-like formula \cite{Meir92,Nitzan01,Hartle,Hartle09}
\begin{eqnarray}
\label{currentformula}
I &=& 2e\int\frac{\text{d}\epsilon}{2\pi}\, \sum_{mn} \left( \Sigma_{\text{L},mn}^{<}(\epsilon)\bar{G}^{>}_{nm}(\epsilon)-\Sigma_{\text{L},mn}^{>}(\epsilon)\bar{G}^{<}_{nm}(\epsilon) \right) .
\end{eqnarray}
Note that this expression includes a factor of two to account for spin degeneracy.

\cleardoublepage

\section{Results}
\label{secResults}

In this section we discuss quantum interference effects and vibrationally induced decoherence in 
single-molecule junctions based on the methodology introduced above. Thereby,
we consider models of increasing complexity. 
First, in Sec.\ \ref{BasIntSec}, we discuss the basic quantum interference effects 
that occur in electron transport through single molecules. 
Thereby, we introduce the framework and the methodology that we use to analyze quantum 
interference effects in general. 
Next, in Secs.\ \ref{SequTunReg} and \ref{CoTunEffects}, we study 
the effect of electronic-vibrational coupling on the transport
characteristics of a single-molecule contact. 
Thereby, we use both simplified coupling scenarios, which facilitate the 
interpretation of the results and provide a first understanding 
of the effect of vibrationally induced decoherence, and 
more complex (yet more realistic) scenarios, 
which, nevertheless, can be understood on the same grounds. 
Finally, in Secs.\ \ref{localcool} and \ref{anomal}, we 
discuss the effect of local cooling due to electron-hole pair creation processes, 
which, in the presence of strong destructive quantum interference effects, can become 
dominant (Sec.\ \ref{localcool}). 
As will be shown in Sec.\ \ref{anomal}, this results in a strong temperature 
dependence of the electrical current that cannot be explained in terms of the 
temperature dependence of the Fermi distribution functions in the 
leads \cite{Nitzan01,Selzer2004,Poot2006,Choi2008,Sedghi2011}.

The parameters for all the models employed in this article are summarized in Tab.\ \ref{parameters}. 
They have been chosen to represent typical values for molecular
junctions and are similar to those employed in a number of 
first-principles studies of molecular junctions 
\cite{Benesch06,Troisi2006,Benesch08,Monturet2010,Frederiksen2010,Buerkle2011}. 
It should be noted that we consider only models, where the coupling between 
the molecule and the electrodes is rather weak, i.e., where the effective factorization 
of the single-particle Green's functions, Eq.\ (\ref{decoupling}), is valid.
To this end, the coupling parameter $\nu$, which determines the molecule-lead
interaction strengths (see Tab.\ \ref{parameters}), has been chosen 
as $\nu = 0.1$\,eV, meeting the condition for the 
antiadiabatic approximation (\ref{decoupling}): 
$\vert\Gamma_{\text{L/R},mn}(\epsilon)\vert\leq2\nu^{2}/\gamma\lesssim\Omega_{1}/10\ll\Omega_{1}$.

\begin{table}
\begin{center}
\begin{tabular}{|*{13}{ccc|}}
\hline \hline 
&model &&& $\epsilon_{1}$&&&$\epsilon_{2}$&&&$U_{12}$&&& $\nu_{\text{L},1/2}$&&&$\nu_{\text{R},1}$&&&$\nu_{\text{R},2}$&&&
$\Omega_{1}$&&&$\lambda_{11}$&&&$\lambda_{21}$ &&&$\gamma$&\\ \hline 
& DES &&&0.5&&&0.505&&&0&&&$\nu$&&&$\nu$&&&-$\nu$&&&--&&&--&&&-- &&&2&\\
& CON &&&0.5&&&0.505&&&0&&&$\nu$&&&$\nu$&&&$\nu$&&&--&&&--&&&-- &&&2&\\
& DESVIB &&&0.5&&&0.53&&&0&&&$\nu$&&&$\nu$&&&-$\nu$&&&0.1&&&0&&&0.05 &&&2&\\
& CONVIB &&&0.5&&&0.53&&&0&&&$\nu$&&&$\nu$&&&$\nu$&&&0.1&&&0&&&0.05&&&2& \\
& DESVIB2 &&&0.52401&&&0.53&&&0.049&&&$\nu$&&&$\nu$&&&-$\nu$&&&0.1&&&0.049&&&0.05 &&&2&\\
& DESCOOL &&&0.05123&&&0.05126&&&0.00248&&&$\frac{\nu}{10}$&&&$\frac{\nu}{10}$&&&-$\frac{\nu}{10}$&&&0.005&&&0.00248&&&0.0025&&&0.25& \\
\hline \hline
\end{tabular}
\end{center}
\caption{\label{parameters}
Parameters for the models of single-molecule junctions that are 
investigated in this article. All energy values are given in $\mathrm{eV}$. 
In all models the temperature of the electrodes $T_{\text{L/R}}$ is set to $10$\,K. 
The value of the molecule-lead coupling parameter is $\nu = 0.1$\,eV.  
}
\end{table}

\subsection{Basic Interference Effects}
\label{BasIntSec}

Quantum interference effects can be constructive or destructive, depending on
the model, the physical parameters and/or the specific observable. In this section, we 
study basic interference effects for two different models of single-molecule junctions. 
The first model exhibits pronounced destructive interference effects  in its conduction properties 
and is referred to as model DES. The second model, CON, shows   
constructive interference effects. 
Note that both systems have been extensively studied 
before \cite{Kalyanaraman2002,Collepardo2004,Ernzerhof2005,Hod2006,Goyer2007,
Solomon2008b,Begemann2008,Darau2009,Stadler2009,Brisker2010,
Schultz2010,Markussen2010,Stafford2011,Markussen2011,Lambert2011,Brisker2012}.  
They are thus very useful to set the stage for the discussion of vibrationally induced 
decoherence, which is presented in the subsequent sections, Secs.\ \ref{SequTunReg} -- \ref{anomal}. 
It is also noted that similar models have been used to study interference effects 
in quantum dot arrays, in particular in the context of Aharonov-Bohm 
interferometers \cite{Kubala2002,Ueda2007,Goldstein2007,Segal2012}.

\begin{figure}
\begin{center}
\begin{tabular}{llll}
(a) & (b) & (c) & (d)\\
\includegraphics[width=0.235\textwidth]{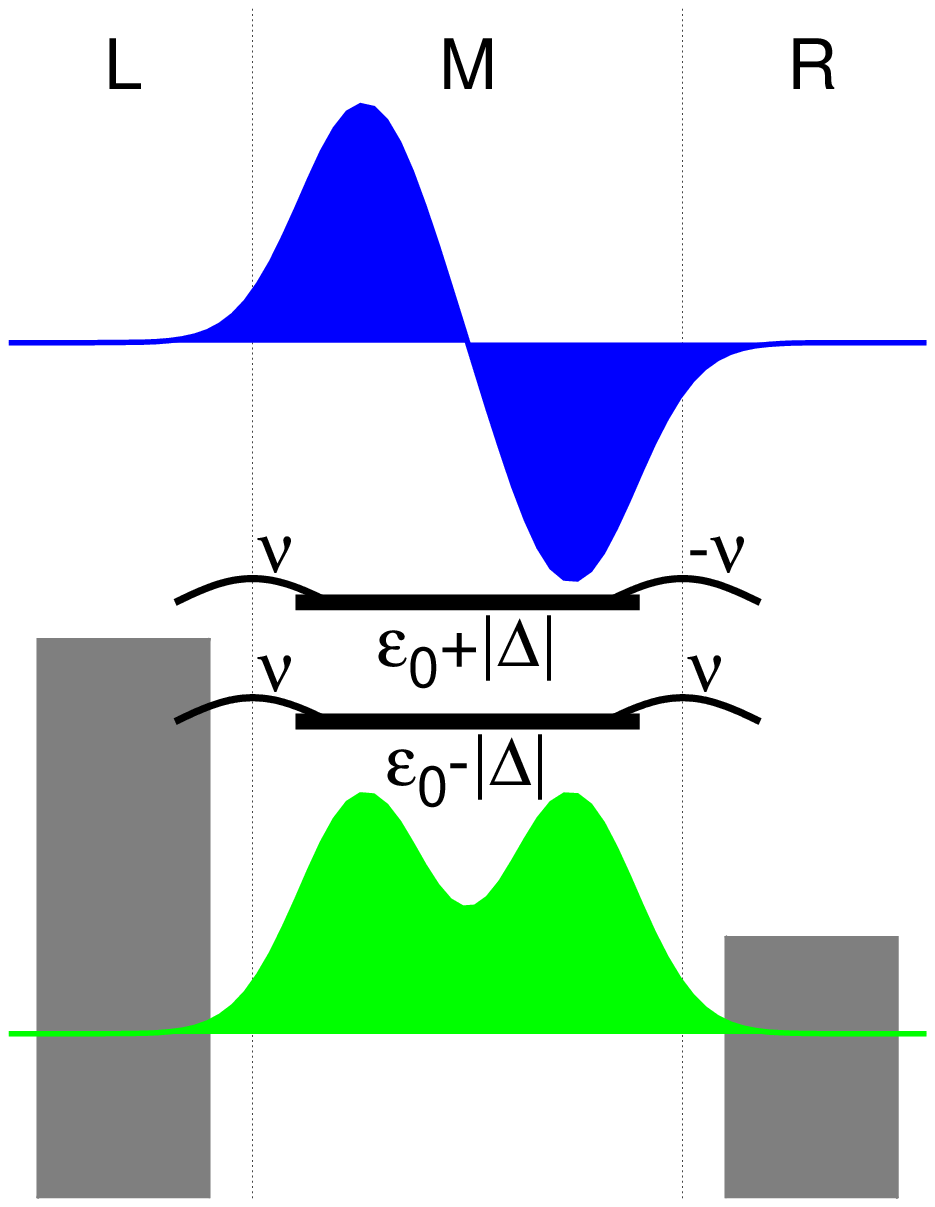}
 & 
\includegraphics[width=0.235\textwidth]{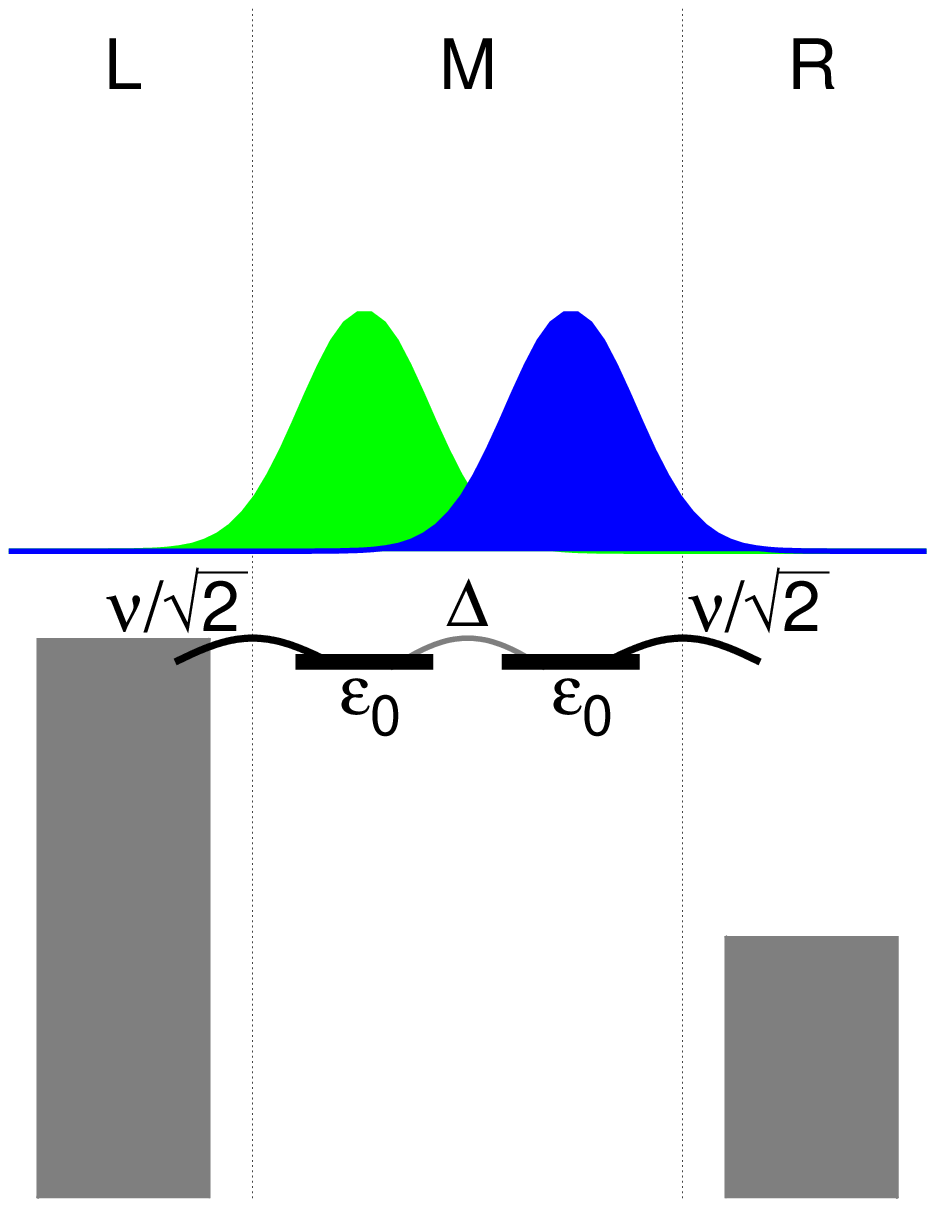} 
& 
\includegraphics[width=0.235\textwidth]{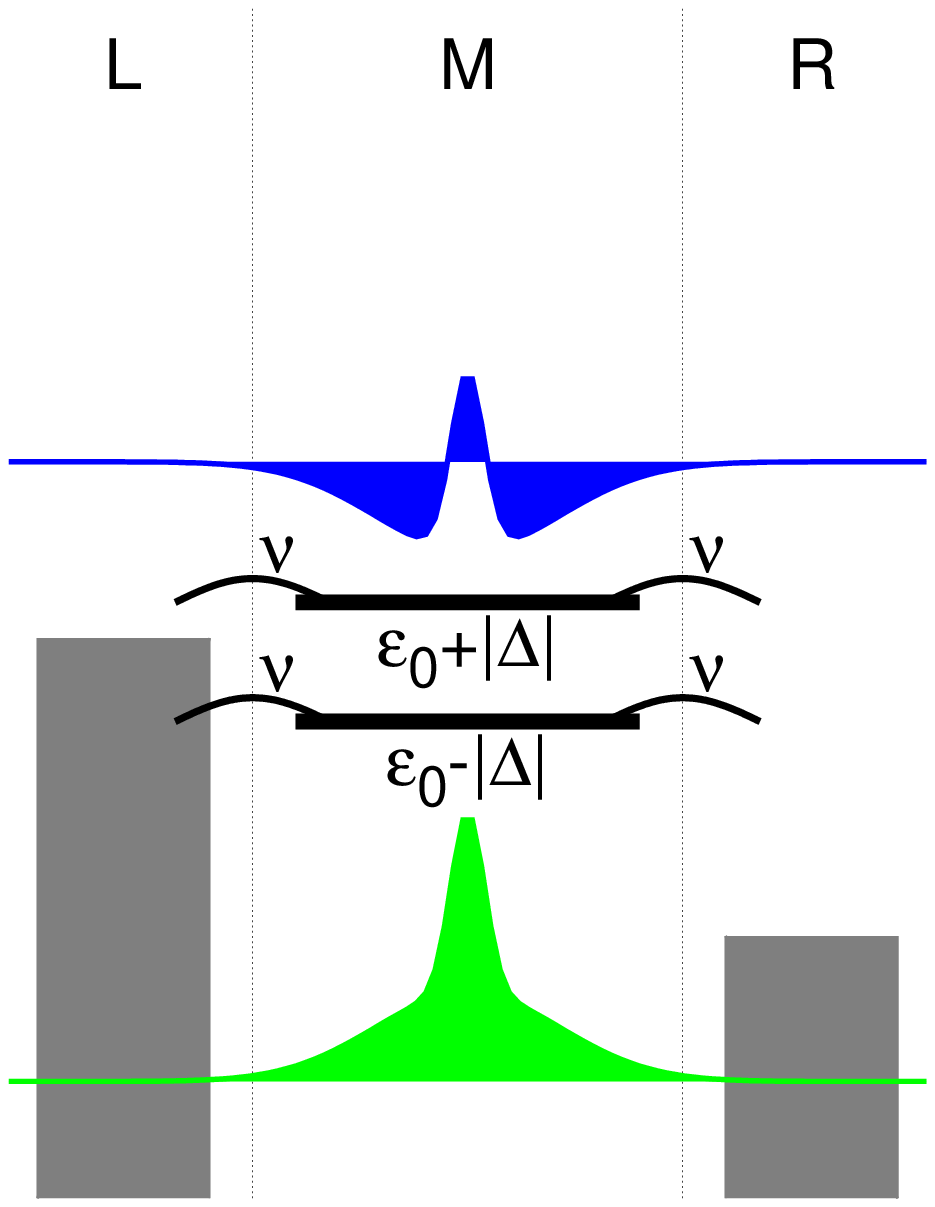}
 & 
\includegraphics[width=0.235\textwidth]{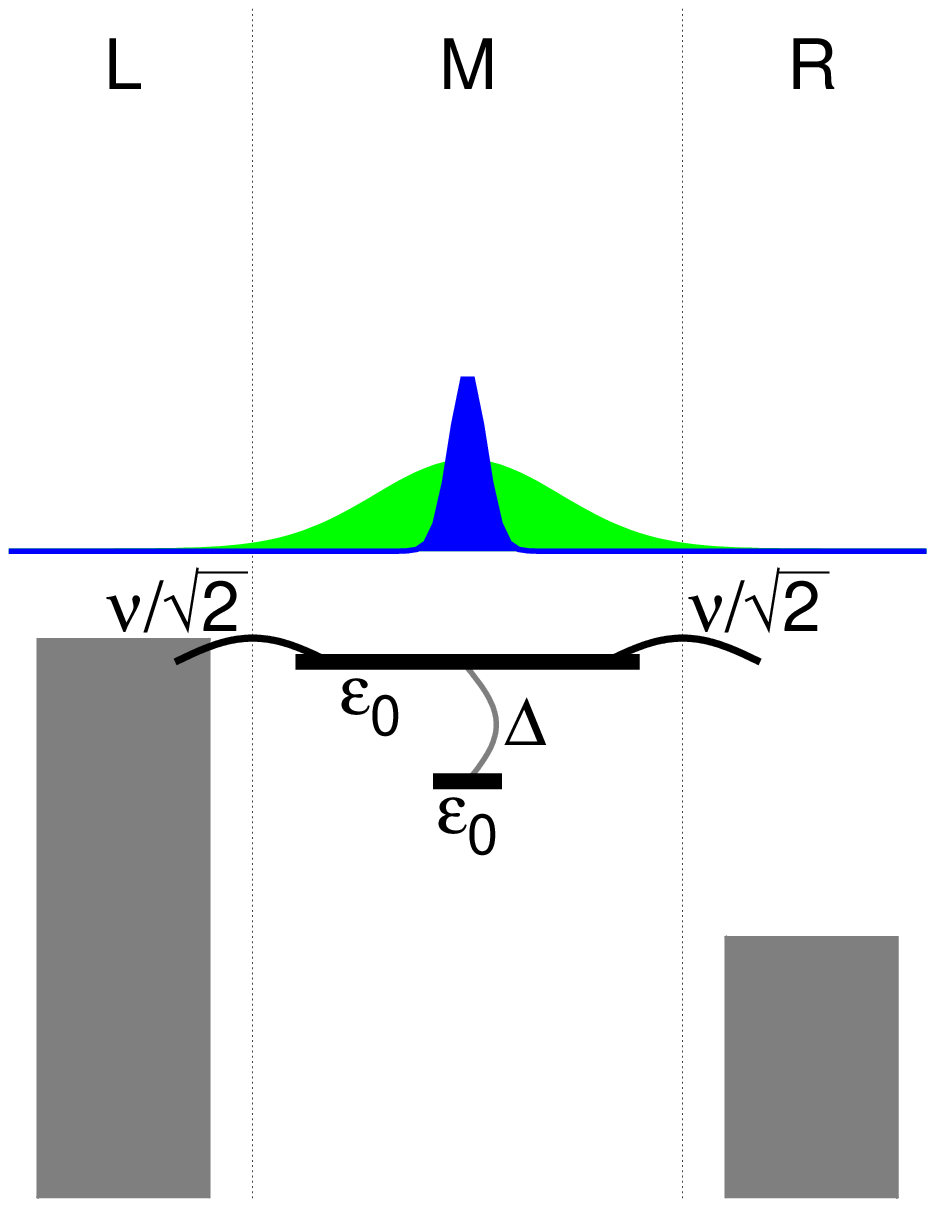} \\ 
\end{tabular}
\end{center}
\caption{\label{LinConduct} 
Panel (a): Schematic representation of example wavefunctions of the electronic eigenstates of model DES. 
Panel (b): Localized molecular orbitals corresponding to the eigenstates of model DES. 
Panel (c): Schematic representation of example wavefunctions of the electronic eigenstates of model CON. 
Panel (d): Localized molecular orbitals corresponding to the eigenstates of model CON. 
}
\end{figure}

Interference effects arise in electron transport through molecular junctions  
whenever electrons can traverse the junction along different but equivalent pathways. 
This is the case, for example, if the electrical current of a junction 
is carried by two quasidegenerate electronic states. 
The first model system that we study, model DES (see Tab.\ \ref{parameters} 
for a summary of the respective parameters), 
comprises two electronic states, which are located 
at $\epsilon_{1}=0.5$\,eV and $\epsilon_{2}=0.505$\,eV 
above the Fermi level of the junction. 
Both of these states are coupled to the left lead 
with coupling strengths $\nu_{\text{L},1/2}=0.1$\,eV and to the right lead 
with coupling strengths $\nu_{\text{R},1}=0.1$\,eV and $\nu_{\text{R},2}=-0.1$\,eV, respectively. 
These coupling strengths, in particular the different sign in the coupling to the right lead, 
reflect the different spatial symmetry of the two states, which represent  
symmetric and antisymmetric combinations of localized molecular orbitals 
(see Figs.\ \ref{LinConduct}a and \ref{LinConduct}b 
for a graphical representation of the two electronic states 
and the corresponding localized molecular orbitals). 
The coupling to the electrodes induces a broadening of the 
two states of $\approx 20$\,meV that exceeds their level spacing 
$\epsilon_{2}-\epsilon_{1}=5$\,meV, \emph{i.e.}\ they are quasidegenerate.

\begin{figure}
\begin{center}
\begin{tabular}{l}
\resizebox{\newwidth}{\newheight}{
\includegraphics{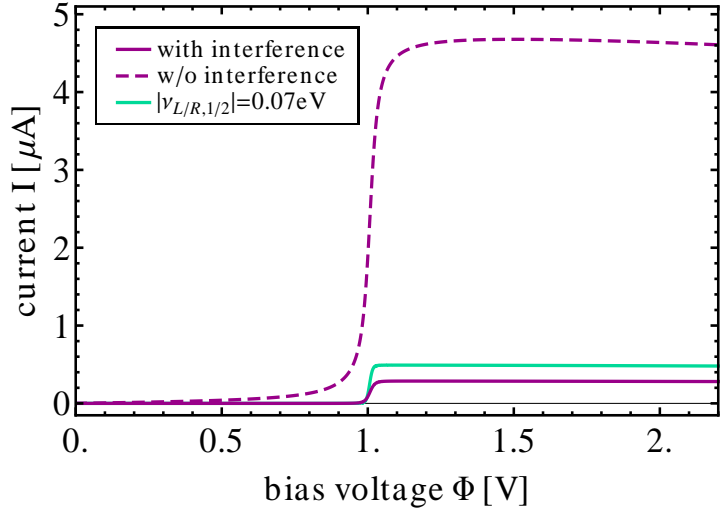}
} \\
\end{tabular}
\end{center}
\caption{(Color online)\label{BasicInterferenceLinearFig} 
Current-voltage characteristics of the linear molecular conductor described by model DES 
(see Figs.\ \ref{LinConduct}a and \ref{LinConduct}b). 
The solid purple line depicts the current-voltage characteristic of this junction, including 
quantum interference effects. The dashed purple line is obtained by discarding them. If 
the absolute value of the coupling strengths $\nu_{\text{L/R},1/2}$ of model DES is reduced 
to $0.07$\,meV, the current-voltage characteristic depicted by the
solid turquoise line is obtained. 
}
\end{figure}

The current-voltage characteristic of junction DES 
is represented in Fig.\ \ref{BasicInterferenceLinearFig} by the solid purple line. 
Due to destructive quantum interference effects, it shows rather low levels of 
electrical current. This is demonstrated by comparison  
with the dashed purple line, which depicts the current-voltage characteristic 
of this junction disregarding quantum interference effects, that is, the incoherent sum of the 
electrical currents that are flowing through either of the two electronic states. 
As can be seen, quantum interference effects suppress the current level in this junction 
by more than an order of magnitude. 
This applies throughout the whole range of bias voltages considered, 
including the non-resonant as well as the resonant transport regime, 
which are separated by the step that appears at $e\Phi\approx2\epsilon_{1/2}$ in both characteristics.

\begin{figure}
\begin{center}
\begin{tabular}{llll}
(a) & (b)& (c) & (d) \\
\includegraphics[width=0.235\textwidth]{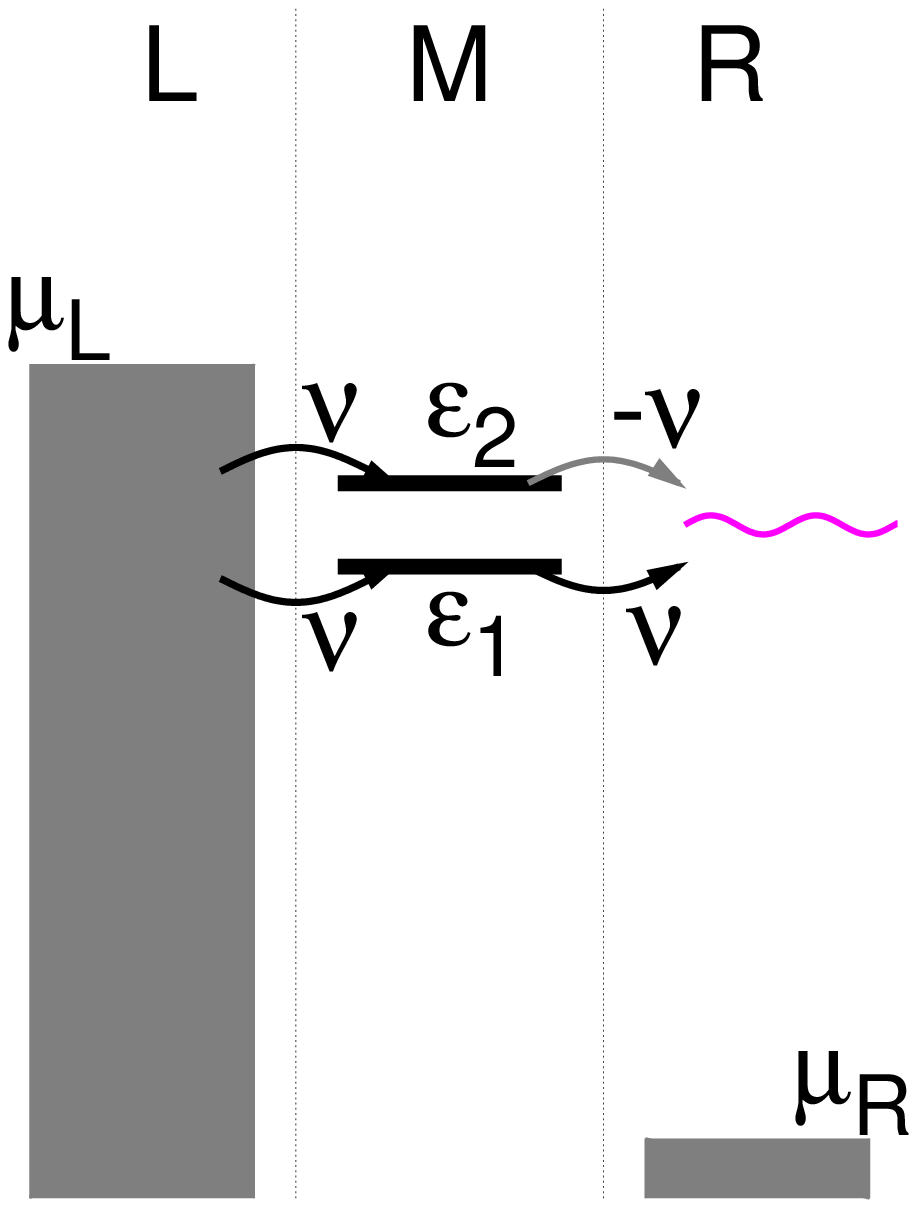}
 & 
\includegraphics[width=0.235\textwidth]{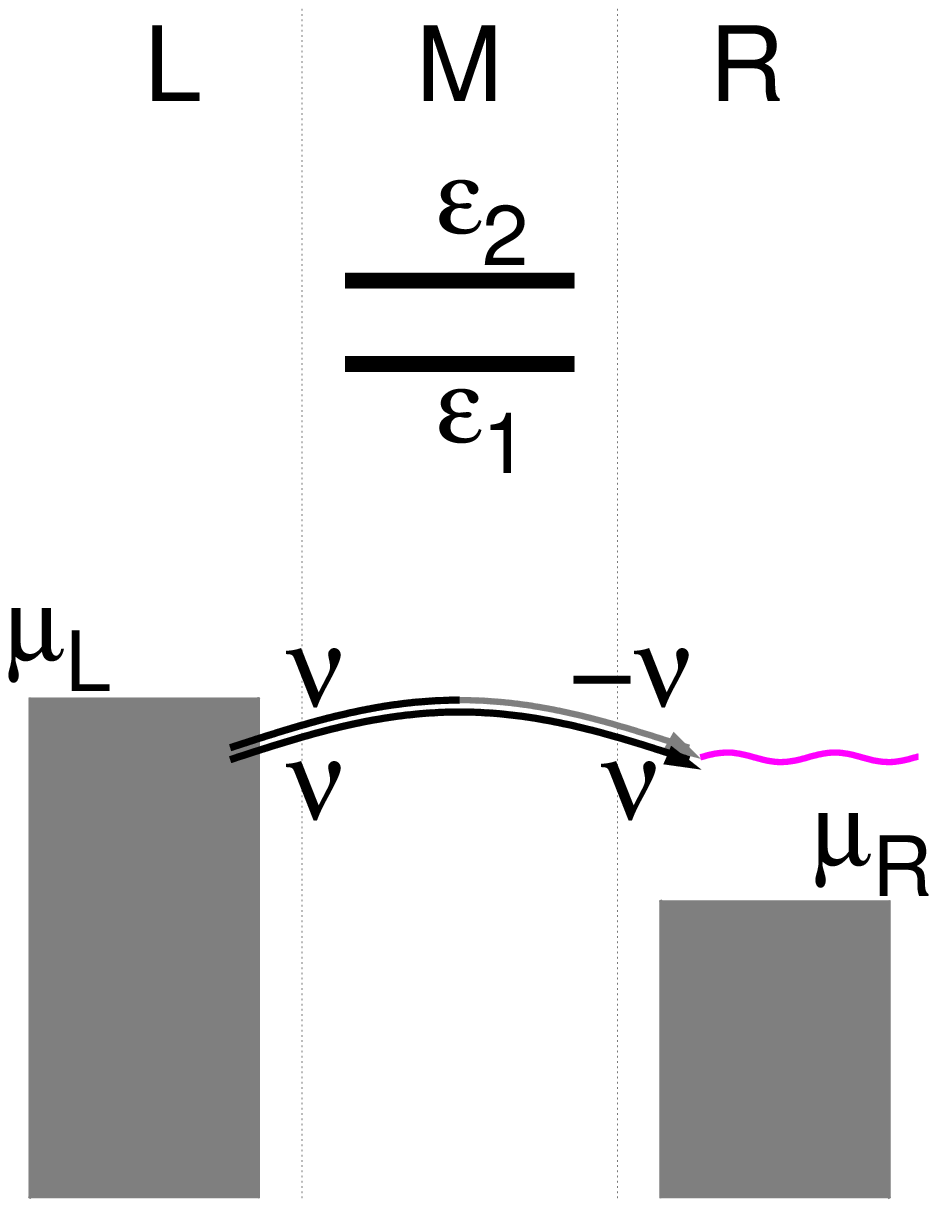}
 & 
\includegraphics[width=0.235\textwidth]{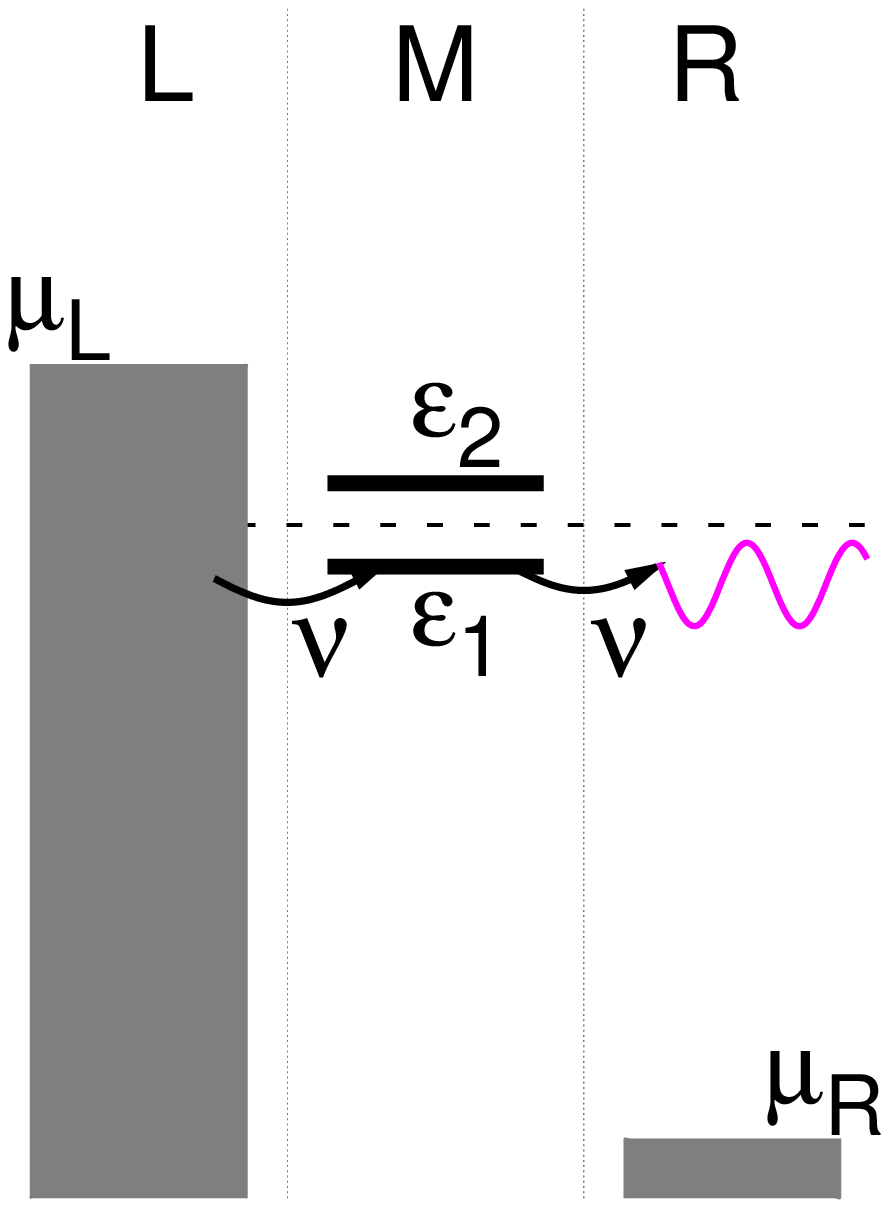}
 & 
\includegraphics[width=0.235\textwidth]{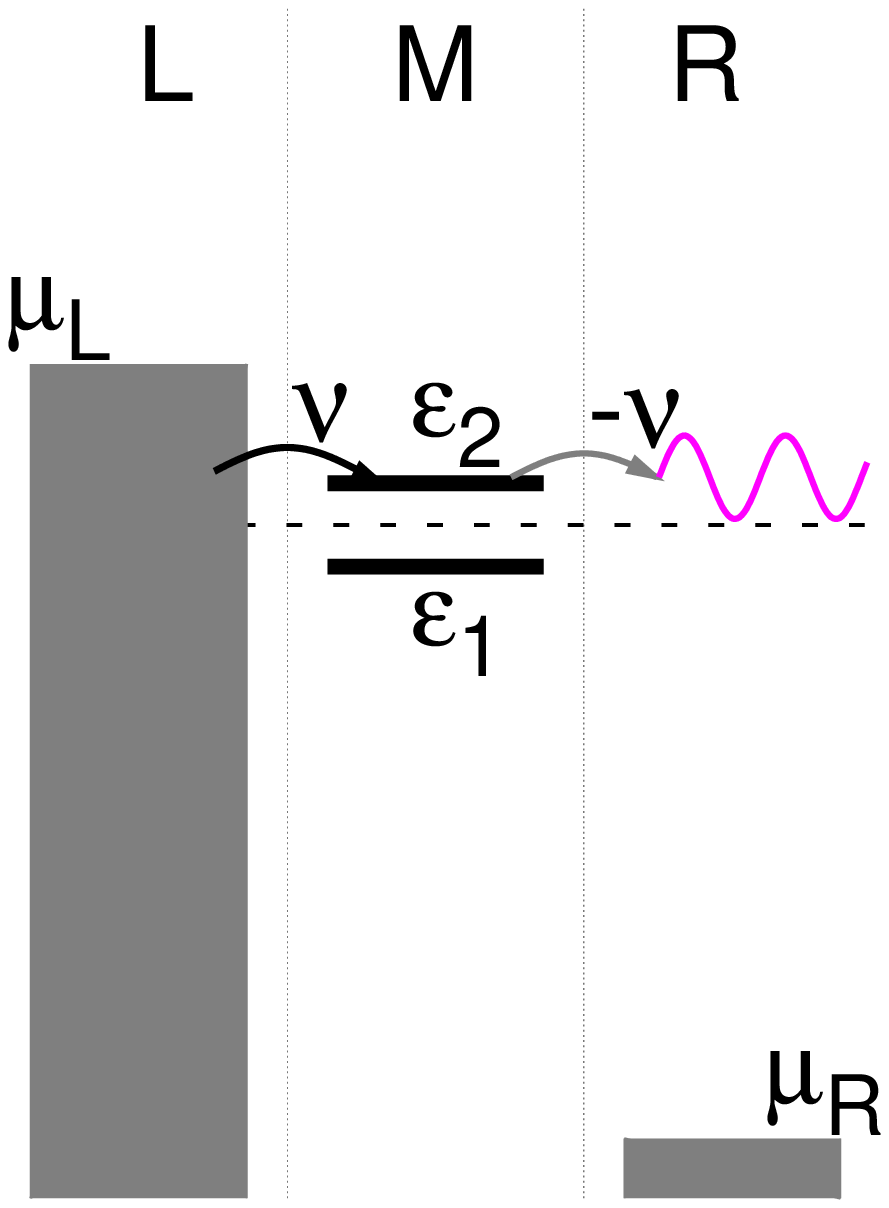}
\\ 
\end{tabular}
\end{center}
\caption{\label{TransportProcesses} Panel (a) and (b): Resonant and non-resonant 
electron transport processes through junction DES. 
The purple wiggly lines depict the coherent sum of the outgoing wavefunctions that are 
associated with electron tunneling through each of the two electronic states. 
Panel (c) and (d): Resonant electron transport through state $1$ and $2$ 
if the other state (\emph{i.e.}\ state $2$ and $1$, respectively) would not be present. 
Due to destructive interference, the corresponding tunneling amplitudes 
(purple wiggly lines) are much larger than 
their coherent sum (depicted in Panel (a)). 
}
\end{figure}

Quantum interference effects are destructive in this system, 
because the outgoing wavefunction, which is associated with 
an electron tunneling event through state $2$, differs by a phase of $\pi$ from the 
the one, which is associated with tunneling through state $1$. This is a result of 
the different spatial structure of the two states and is reflected in the different signs of the 
respective molecule-lead coupling strengths to the right lead: 
$\nu_{\text{R},1}=-\nu_{\text{R},2}$. 
The two outgoing wave functions associated with electron tunneling through state $1$ and $2$, 
therefore, destructively interfere with each other, leading to a strong 
suppression of the respective tunnel current. This is illustrated by an example 
for a resonant and a non-resonant transport process in Figs.\ \ref{TransportProcesses}a 
and \ref{TransportProcesses}b, respectively, where the coherent sum of the outgoing wavefunctions 
is depicted. Figs.\ \ref{TransportProcesses}c and \ref{TransportProcesses}d show the two outgoing 
wave functions associated with the resonant transport process in Fig.\ \ref{TransportProcesses}a. 
Note that these interference effects are more pronounced the closer the two electronic states 
are in energy.

Employing a different representation, 
the suppression of the current level in model DES can also be understood 
on different grounds. Thereby, 
the two eigenstates of this system are unitarily 
transformed to two localized states (see Figs.\ \ref{LinConduct}a and \ref{LinConduct}b for a schematic 
representation of the two equivalent representations). These states have the same energy $\epsilon_{0}$ 
and are coupled with each other by the coupling strength $\Delta=(\epsilon_{2}-\epsilon_{1})/2$. 
They represent, for example, the left and the right part of a molecular conductor.  
Using this local representation, electron transport through this junction involves a sequence of 
processes: electron transfer from the left electrode to left part of the molecule, followed by an 
intramolecular electron transfer process from the left part of the molecule to the right part, and finally 
a transfer process from the right part of the molecule to the right electrode. 
For $|\nu_{\text{L/R},1/2}|\gtrsim|\Delta|$, the bottleneck for
electron transport is the intramolecular electron transfer process. 
Accordingly, the suppression of the electrical current flowing through 
junction DES can also be understood in the local representation 
by the small effective coupling $\Delta$ 
rather than in terms of quantum interference effects. 
At high bias voltages, for example, the current flowing through the junction is given by 
\begin{eqnarray}
 I &=& 4e\frac{\Gamma}{2}  \frac{\left(\frac{\Delta}{\Gamma}\right)^{2}}{\left(\frac{\Delta}{\Gamma}\right)^{2}+1},  
\end{eqnarray}
according to the exact results derived by Gurvitz and Prager \cite{Gurvitz96,Gurvitz98}, where  
the wide-band approximation $\Gamma=\Gamma_{\text{L},11}(\mu_{\text{L}})$ is employed. 
In this expression, $4e\frac{\Gamma}{2}$ represents the current in the limit $\Delta\rightarrow\infty$.  
The current suppression for finite $\Delta$ is given by the factor 
$\frac{\left(\frac{\Delta}{\Gamma}\right)^{2}}{\left(\frac{\Delta}{\Gamma}\right)^{2}+1}$. It is 
solely determined by the ratio $\Delta/\Gamma$. 
Other aspects of the problem are, however, more straightforwardly understood in
the eigenstate representation. For example, if $\Gamma_{\text{L},11}\Gamma_{\text{R},11}\gtrsim\Delta^{2}$,   
a reduction in the coupling to the leads results in an increase of the current 
that is flowing through junction DES. This is illustrated by the solid
turquoise line in 
Fig.\ \ref{BasicInterferenceLinearFig}, 
which shows the current-voltage characteristic of model DES with a reduced
coupling 
to the leads, $\nu_{\text{L},1/2}=\nu_{\text{R},1}=-\nu_{\text{R},2}=0.07$\,eV. 
While this is easily understood in terms of a suppression 
of destructive interference effects due to a less pronounced broadening of the electronic states, 
it may hardly be anticipated using the localized state picture, where transport is understood 
rather as a sequence of intramolecular electron transfer processes.

\begin{figure}
\begin{tabular}{l}
\hspace{-0.5cm}
(a) \\[-1cm]
\resizebox{\newwidth}{\newheight}{
\includegraphics{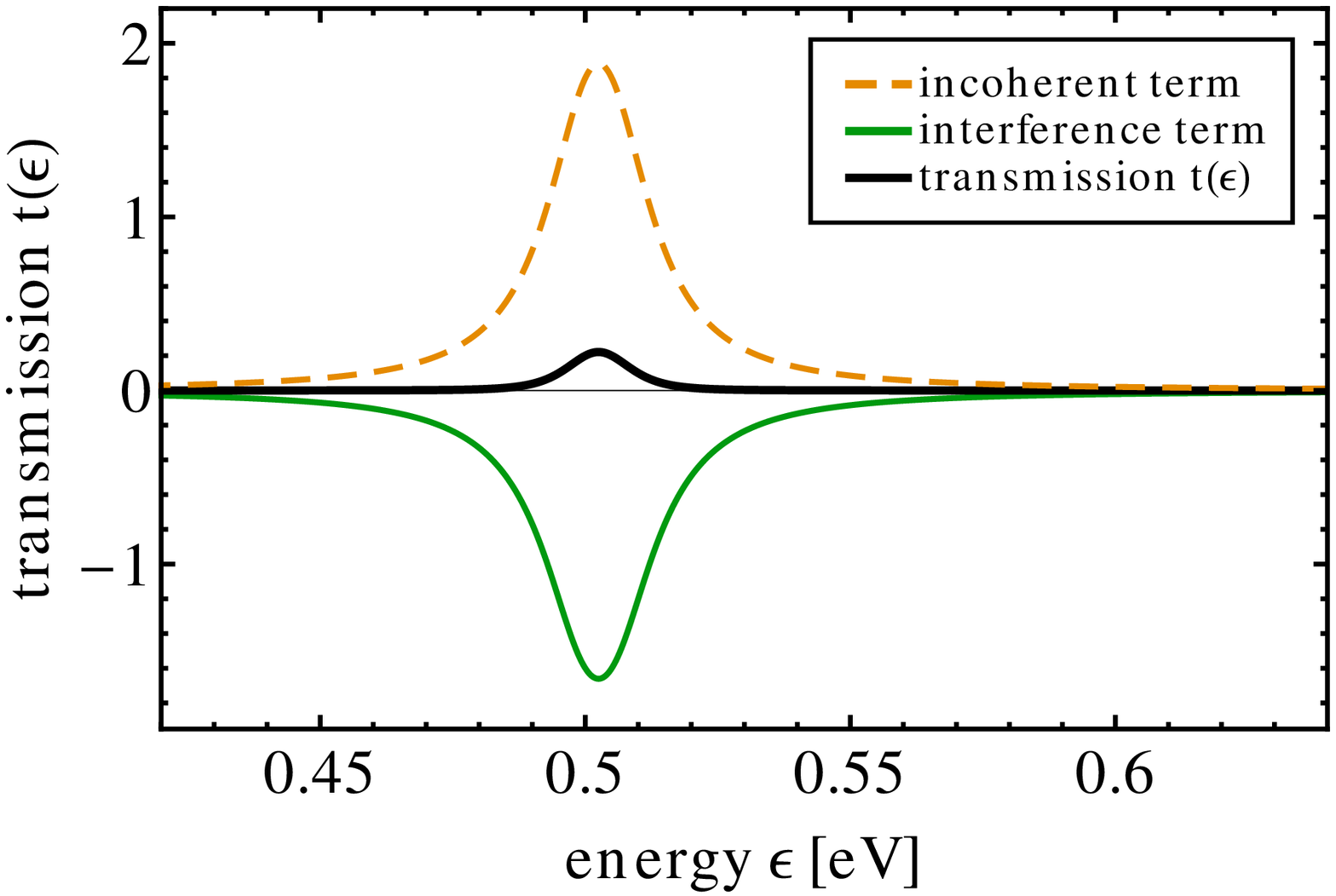}
}\\
\hspace{-0.5cm}
(b) \\[-1cm]
\resizebox{\newwidth}{\newheight}{
\includegraphics{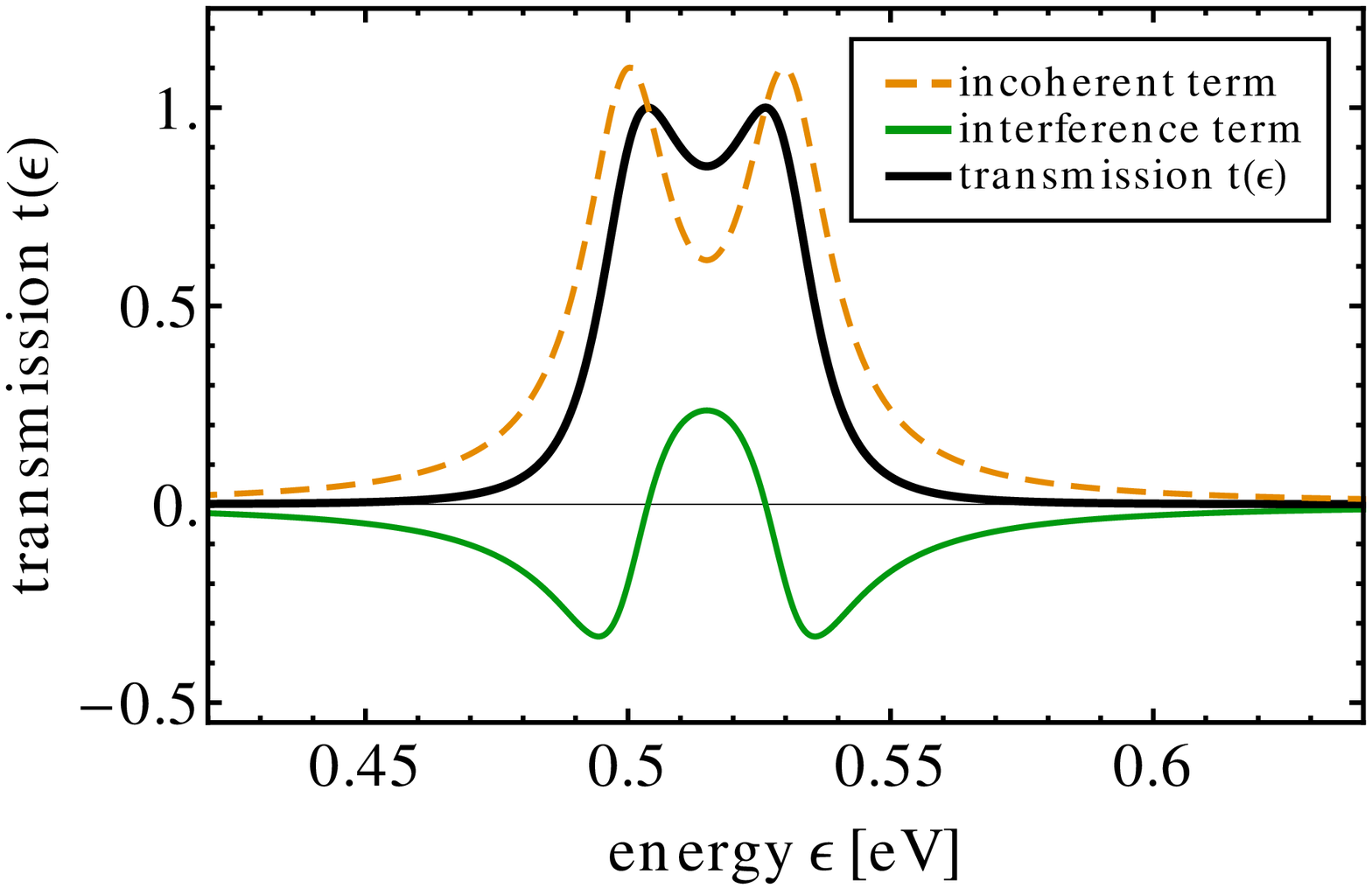}
}
\end{tabular}
\caption{(Color online)\label{BasicInterferenceLinear2Fig} 
Panel (a): The solid black line shows the transmission function of the 
linear molecular conductor DES (graphically illustrated in Figs.\ \ref{LinConduct}a and \ref{LinConduct}b). 
In addition, the corresponding incoherent (dashed orange line) and interference term (solid green line) are  
depicted. 
Panel (b): The same as Panel (a), where, however, the energy of the second electronic state is shifted 
from $\epsilon_{2}=0.505$\,eV to $\epsilon_{2}=0.53$\,eV.  
}
\end{figure}

To further analyze quantum interference effects in model DES and to facilitate the
discussion of more complex models, 
we employ the transmission function $t(\epsilon)$ of model DES, 
which, in general, is defined by \cite{Meir92,Nitzan01} 
\begin{eqnarray}
\label{transdes}
 t(\epsilon)  = 4 \pi^{2} \sum_{k\in\text{R},k'\in\text{L}} \vert  t_{kk'}(\epsilon) \vert^{2} \delta(\epsilon-\epsilon_{k}) \delta(\epsilon-\epsilon_{k'}) 
= \sum_{mnop\in\text{M}} \Gamma_{\text{L},mn}(\epsilon) G_{no}^{\text{r}}(\epsilon) \Gamma_{\text{R},op}(\epsilon) G_{pm}^{\text{a}}(\epsilon) .
\end{eqnarray}
Thereby, $t_{kk'}(\epsilon) = \sum_{mn} V_{mk}^{*} G^{\text{r}}_{mn} V_{nk'} $ 
denote the transition matrix elements
that represent the transmission amplitudes of the conduction process.
Accordingly, the transmission function can be split into 
an incoherent term  
\begin{eqnarray}
\label{incterm}
 t_{\text{inc}}(\epsilon) &=& 4 \pi^{2} \sum_{k\in\text{R},k'\in\text{L}} \sum_{mn} \vert  V_{mk}^{*} G^{\text{r}}_{mn} V_{nk'} \vert^{2} \delta(\epsilon-\epsilon_{k}) \delta(\epsilon-\epsilon_{k'}), \\
&=& 
\label{incterm-aux}
\frac{\Gamma^{2}}{\vert \epsilon - \epsilon_{1} + i \Gamma \vert^{2}} + \frac{\Gamma^{2}}{\vert \epsilon - \epsilon_{2} + i \Gamma \vert^{2}}, 
\end{eqnarray}
which represents the incoherent sum of the transmission amplitudes $V_{mk}^{*} G^{\text{r}}_{mn} V_{nk'}$, 
and an interference term 
\begin{eqnarray}
\label{intterm}
 t_{\text{int}}(\epsilon) &=& t(\epsilon) - t_{\text{inc}}(\epsilon) \,=\, 
- 2\text{Re}\left[\frac{\Gamma^{2}}{( \epsilon - \epsilon_{1} + i \Gamma )( \epsilon - \epsilon_{2} - i \Gamma )}\right] , 
\end{eqnarray}
which encodes the respective interference effects. 
The structure of these expressions suggests 
an equivalent decomposition of the transmission function $t(\epsilon)$ 
in terms of transmission amplitudes $\Lambda_{1/2}$ 
\begin{eqnarray}
 t(\epsilon) &=& \vert\Lambda_{1} - \Lambda_{2}\vert^{2}, \\
 t_{\text{inc}}(\epsilon) &=& \vert\Lambda_{1} \vert^{2} + \vert \Lambda_{2}\vert^{2}, \\
 t_{\text{int}}(\epsilon) &=& \Lambda_{1}^{*} \Lambda_{2} + \Lambda_{2}^{*} \Lambda_{1}, 
\end{eqnarray}
with 
\begin{subequations}
\label{conductanceampl}
\begin{eqnarray}
 \Lambda_{1} &=& \frac{\Gamma}{\epsilon-\epsilon_{1}+i\Gamma}, \\
 \Lambda_{2} &=& \frac{\Gamma}{\epsilon-\epsilon_{2}+i\Gamma}.  
\end{eqnarray}
\end{subequations} 
These amplitudes describe electron transport through state $1/2$ 
if the other state $2/1$ is not present, \emph{i.e.}\ in the limit 
$\epsilon_{2/1}\rightarrow\infty$. 
It should be noted that the definition of the incoherent term (\ref{incterm}) 
and the interference term (\ref{intterm}) is not unique, and depends, for example, on the basis 
that is used to represent the electronic degrees of freedom of the molecular bridge. 
Other decomposition schemes, which employ, \emph{e.g.}, molecular conductance orbitals, 
are possible and have already been used to study interference effects \cite{Solomon2008b}. 
In this work, however, we seek for an understanding of the 
transport properties of a molecular conductor in terms of its eigenstates, 
which, in contrast to molecular conductance orbitals, represent an energy-independent basis. 
For later reference, it is noted that 
the incoherent term (\ref{incterm}) 
can also be obtained 
by disregarding the off-diagonal elements of the self-energy matrix 
$\Sigma^{<,>}_{\text{L}}$ in the current formula (\ref{currentformula}) (cf.\ App.\ \ref{transDESVIB}).

Fig.\ \ref{BasicInterferenceLinear2Fig}a shows the transmission function of junction DES  
(solid black line) as well as the corresponding incoherent (dashed orange line) 
and interference term (solid green line). Due to the
quasi-degeneracy of the two states, the interference term is almost as 
large as the incoherent term. It shows only negative values, 
which indicate destructive interference effects. 
Thus, the incoherent term, which describes electron transport through this system without 
quantum interference effects, and the interference term almost cancel each other, leaving 
only a small peak at $\epsilon\approx(\epsilon_{2}-\epsilon_{1})/2$ 
in the respective transmission function. This peak is associated with 
the low current levels that are obtained for this model system (cf.\ Fig.\ \ref{BasicInterferenceLinearFig}).

Increasing the level spacing in this model by shifting the energy of the second 
state to higher values, $\epsilon_{2}=0.53$\,eV, the tranmission function 
becomes more complex, as shown in Fig.\ \ref{BasicInterferenceLinear2Fig}b. The
interference term is no longer restricted to negative values, but turns from negative 
to positive values if the energy of the electron is located between 
the energy of the first and the second state. Hence, for a larger level spacing, junction DES 
exhibits both destructive and constructive interference effects. 
This result is more straightforwardly understood based on the interference picture in the
eigenstate representation than in terms of the intramolecular 
coupling $\Delta$ in the local representation, 
especially if these constructive interference effects 
are dominant, such as, for example, if $\epsilon_{1}<\epsilon_{\text{F}}<\epsilon_{2}$.

\begin{figure}
\begin{center}
\begin{tabular}{l}
\resizebox{\newwidth}{\newheight}{
\includegraphics{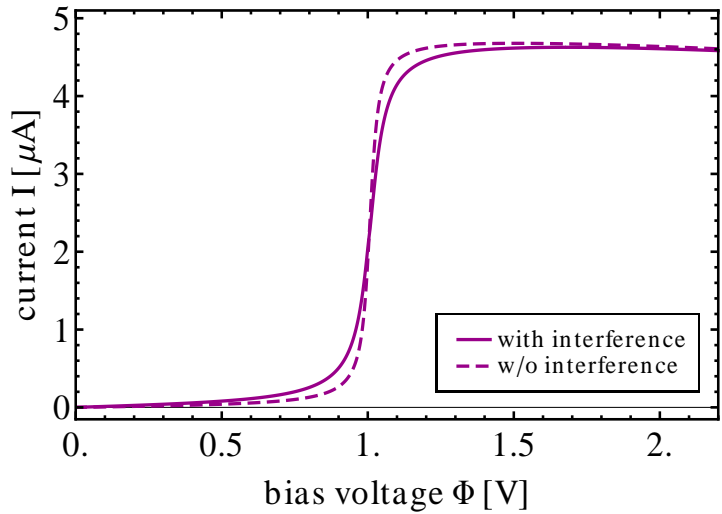}
} \\
\end{tabular}
\end{center}
\caption{(Color online)\label{BasicInterferenceBranchedFig} 
Current-voltage characteristics of the branched molecular conductor described by model CON  
(see Figs.\ \ref{LinConduct}c and \ref{LinConduct}d). 
The solid purple line depicts the current-voltage characteristic of this junction, including 
quantum interference effects. The dashed purple line is obtained by discarding them.  
}
\end{figure}

Next, we consider model CON. It 
differs from the previous system only by the sign of the coupling 
strength $\nu_{\text{R},2}$ (see Tab.\ \ref{parameters}). The 
two electronic states of model CON are thus coupled to the right lead by the same coupling strengths  
($\nu_{\text{R},1}=\nu_{\text{R},2}$). Due to the different sign, model CON and model DES 
describe very different types of molecular conductors. While model DES
corresponds in the local representation to a linear 
molecular conductor, model CON describes a branched molecular conductor 
(cf.\ Figs.\ \ref{LinConduct}c and \ref{LinConduct}d). 
Similar to model DES, it can also be unitarily transformed and represented by 
two localized molecular states that are mutually 
coupled with each other by the coupling strength $\Delta=(\epsilon_{2}-\epsilon_{1})/2$. 
While in model DES, however, each of these states is coupled to a different lead, 
model CON can be mapped onto a delocalized state, which corresponds to the backbone of a  
molecular conductor that is connected to both electrodes, and a localized state, 
which corresponds to a side branch of the molecular conductor that  
is not directly connected to the electrodes of the junction.

The current-voltage characteristic of this junction is represented in 
Fig.\ \ref{BasicInterferenceBranchedFig} by the solid purple line. 
In contrast to model DES, it does not show pronounced quantum 
interference effects. The current flowing 
through this junction is almost the same as the incoherent sum of the currents 
that are flowing through each of the two states. This can be inferred by comparison with the 
dashed purple line, which depicts the respective incoherent sum current. 
The only difference between these two characteristics is an enhanced  
broadening of the step that separates the non-resonant and the resonant transport regime 
at $e\Phi\approx2\epsilon_{1/2}$.

\begin{figure}
\begin{center}
\begin{tabular}{l}
\resizebox{\newwidth}{\newheight}{
\includegraphics{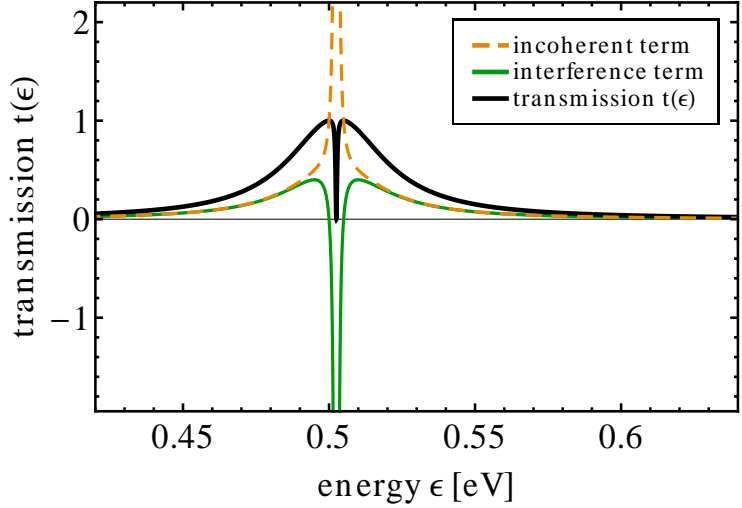}
} \\
\end{tabular}
\end{center}
\caption{(Color online)\label{BasicInterferenceBranchedTransFig} 
The solid black line shows the transmission function of the branched molecular conductor CON  
(graphically illustrated in Figs.\ \ref{LinConduct}c and \ref{LinConduct}d). In addition, 
the corresponding incoherent (dashed orange line) and interference term (solid green line) are depicted. 
}
\end{figure}

To understand this result, we analyze
the corresponding transmission function. 
As for model DES (cf.\ Eqs.\ \ref{transdes}), it can also be decomposed into  
an incoherent and an interference term (cf.\ App.\ \ref{transCONVIB}):  
\begin{eqnarray}
 t_{\text{inc}}(\epsilon) 
&=& \vert \Lambda_{1}(\epsilon) \vert^{2} +  \vert \Lambda_{2}(\epsilon) \vert^{2}, \\
 t_{\text{int}}(\epsilon) 
&=& \Lambda_{1}^{*}(\epsilon) \Lambda_{2}(\epsilon)+\Lambda_{2}^{*}(\epsilon) \Lambda_{1}(\epsilon). \nonumber
\end{eqnarray} 
with the transmission amplitudes 
\begin{eqnarray}
  \Lambda_{1} &=& \frac{\Gamma}{\epsilon-\epsilon_{1}+i\Gamma\frac{2\epsilon-\epsilon_{1}-\epsilon_{2}}{\epsilon-\epsilon_{2}}},\\
 \Lambda_{2} &=& \frac{\Gamma}{\epsilon-\epsilon_{2}+i\Gamma\frac{2\epsilon-\epsilon_{1}-\epsilon_{2}}{\epsilon-\epsilon_{1}}},   
\end{eqnarray}
which, similar to the amplitudes (\ref{conductanceampl}), 
correctly describe electron transport through junction CON 
in the two limits $\epsilon_{2/1}\rightarrow\infty$. 
Fig.\ \ref{BasicInterferenceBranchedTransFig} depicts the transmission function of this system 
and the corresponding incoherent and interference term. 
For low energies $\epsilon<\epsilon_{1}$ as well as high energies $\epsilon>\epsilon_{2}$, 
constructive interference effects, which are associated with a positive 
interference term, are indeed active and double the 
transmittance of the junction. This leads to significantly larger current levels 
in the non-resonant transport regime, \emph{i.e.}\ for $e\Phi<2\epsilon_{1}$ 
(cf.\ Fig.\ \ref{BasicInterferenceBranchedFig}). 
For intermediate energies $\epsilon_{1}<\epsilon<\epsilon_{2}$, however, 
the interference term indicates a transition 
from constructive to destructive interference effects, similar as 
for an increased level spacing in model DES 
(see Fig.\ \ref{BasicInterferenceLinear2Fig}b). 
This is accompanied by a steep increase and decrease of the incoherent and 
the interference term, respectively. At $\epsilon=(\epsilon_{2}+\epsilon_{1})/2$, 
the two terms are exactly the same and cancel each other. 
The transmission function of model CON thus drops to zero at this energy, exhibiting 
an antiresonance. These strong destructive interference 
effects reduce the increase of the current level at the onset of the resonant 
transport regime ($e\Phi\approx(\epsilon_{2}+\epsilon_{1})$), leading, in conjunction 
with the increased current level in the non-resonant transport regime, to a much broader 
step in the current-voltage characteristic. At higher bias voltages, 
destructive and constructive interference effects cancel each other such that the 
same current level is obtained as if interference effects would play no role in this system.

\subsection{Vibrationally Induced Decoherence in the Resonant Transport Regime}
\label{SequTunReg}

In this section, we study the influence of electronic-vibrational coupling on the 
quantum interference effects that we found in the resonant transport regime of 
model DES and model CON (\emph{i.e.}\ for bias voltages $e\Phi\gtrsim2\epsilon_{1/2}$, 
cf.\ Sec.\ \ref{BasIntSec}). The influence of this coupling on the corresponding effects 
in the non-resonant transport regime ($e\Phi<2\epsilon_{1/2}$) are addressed 
in Sec.\ \ref{CoTunEffects}. 

\begin{figure}
\begin{center}
\begin{tabular}{l}
\hspace{-0.5cm}(a) \\
\resizebox{\newwidth}{\newheight}{
\includegraphics{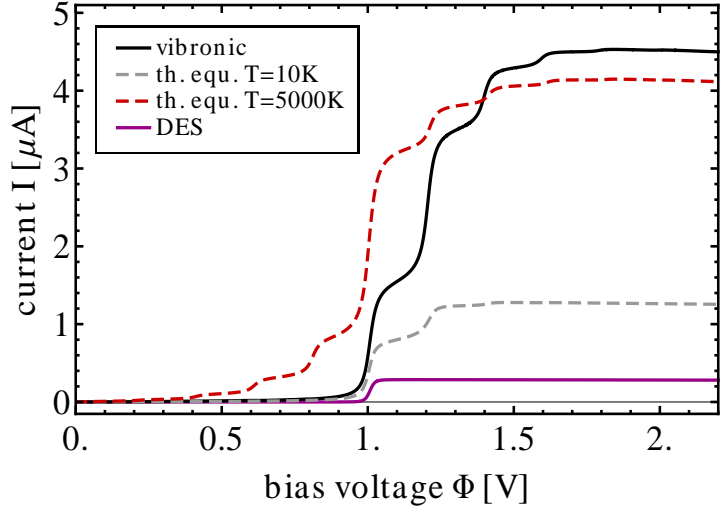}
}\\
\hspace{-0.5cm}(b) \\
\resizebox{\newwidth}{\newheight}{
\includegraphics{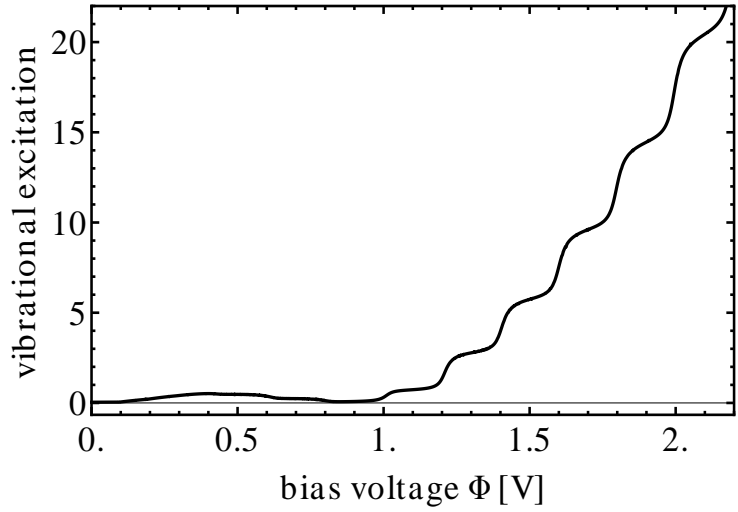}
}
\end{tabular}
\end{center}
\caption{(Color online)\label{DecoherenceSeqTunLinearFig} 
Panel (a): Current-voltage characteristics of the linear molecular conductor DESVIB (solid black line). 
The dashed gray and red lines depict the 
current-voltage characteristic of junction DESVIB that is obtained with the vibrational 
degree of freedom kept in thermal equilibrium at $10$ and $5000$\,K, respectively. 
The solid purple line shows the current-voltage characteristic of model DES. Comparison 
with this line shows the increase of the current level due to vibrationally induced decoherence in 
junction DESVIB.  
Panel (b): Vibrational excitation characteristic corresponding to the vibronic 
current-voltage characteristic shown in Panel (a). 
}
\end{figure}

To study vibrational effects in model DES and model CON, we extend these model systems   
by coupling to a single vibrational degree of freedom. Thereby, we consider first scenarios, 
where the vibrational mode is coupled to one of the states 
only, $\lambda_{11}=0$\,eV and $\lambda_{21}=0.05$\,eV. The resulting model systems 
are referred to as model DESVIB and model CONVIB 
(see Tab.\ \ref{parameters} for the complete set of parameters).
The specific choice of the electronic-vibrational coupling strengths in these systems 
is used to simplify the discussion of the corresponding decoherence mechanisms, where, 
in addition, the energy of the electronic states in models DESVIB and CONVIB have been 
chosen such that the polaron-shifted levels $\overline{\epsilon}_{m}$ agree 
with those of models DES and CON. This allows to separate static effects of the 
vibrations, due to a polaron shift of the energies, from dynamical decoherence 
effects (vide infra). 
In addition, we discuss results, where $\lambda_{11}\neq0$. This may represent 
more realistic coupling scenarios but, as will be shown, the corresponding decoherence mechanism 
can be understood on the same grounds as for the simplified coupling scenario.  
Note that in Ref.\ \cite{Hartle2011b} we have already studied and analyzed the 
transport characteristics of a single-molecule junction, which is very similar 
to junction DESVIB.

\begin{figure}
\begin{center}
\begin{tabular}{llll}
(a)&(b)&(c)&(d)\\
\resizebox{\newwidthprime}{\newheightprime}{
\includegraphics{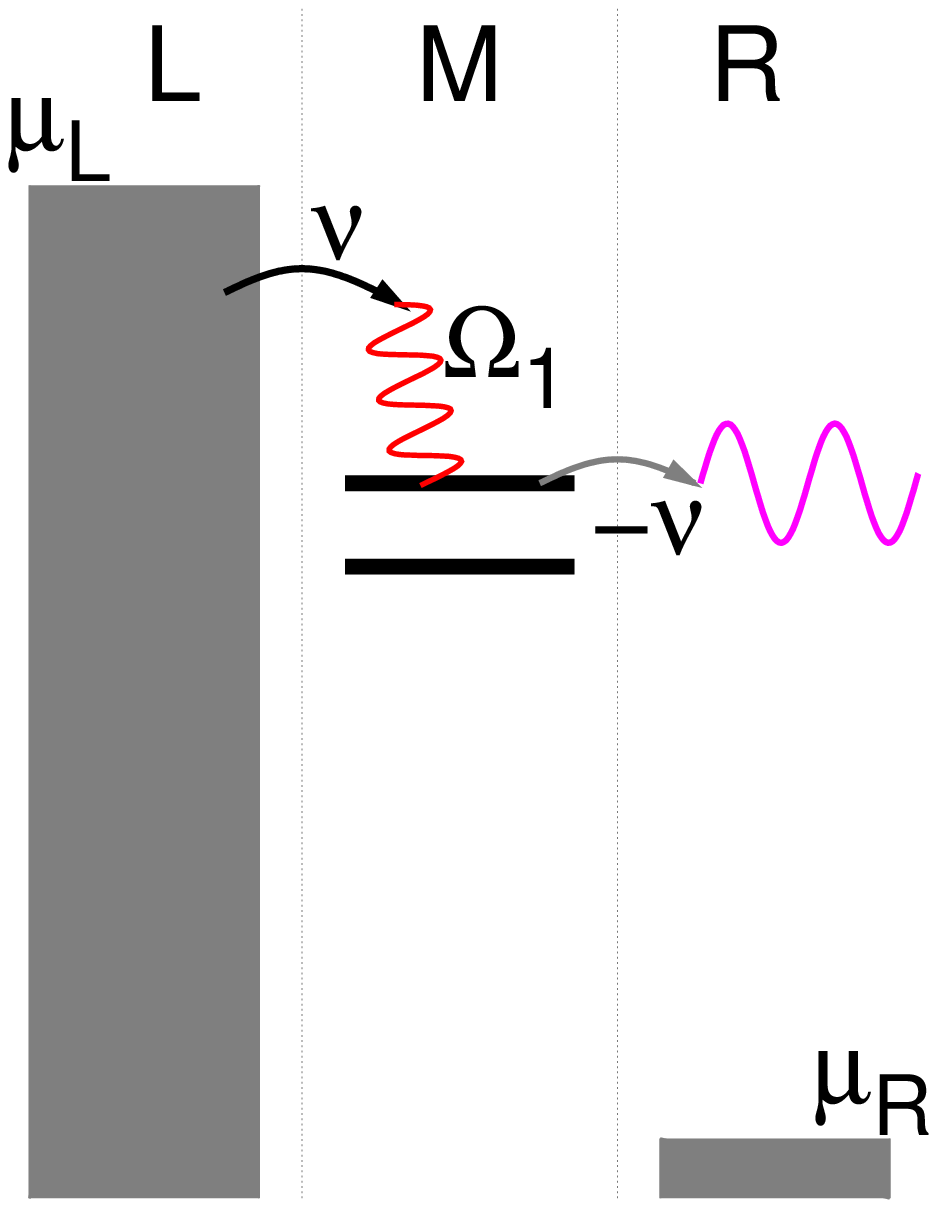}
}&
\resizebox{\newwidthprime}{\newheightprime}{
\includegraphics{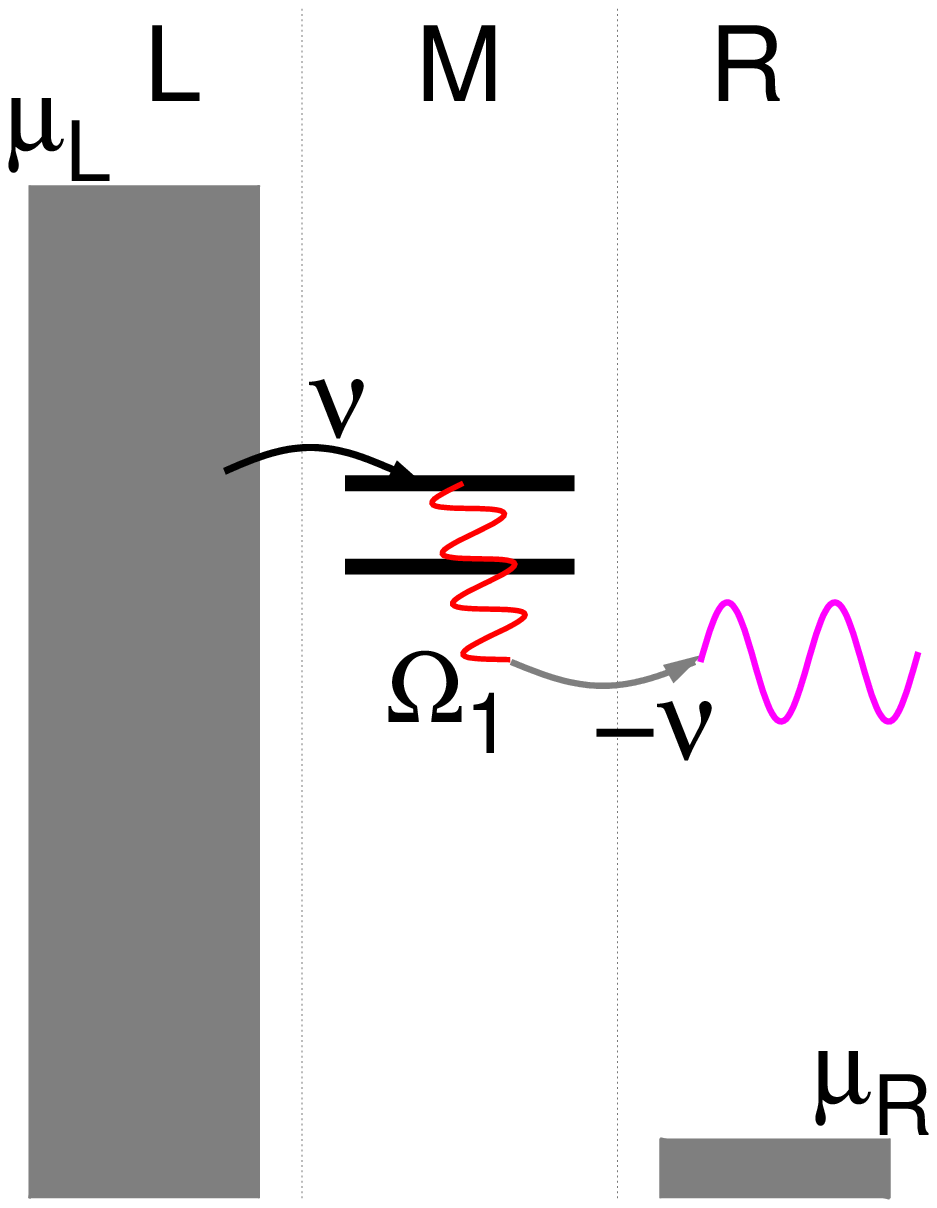}
}&
\resizebox{\newwidthprime}{\newheightprime}{
\includegraphics{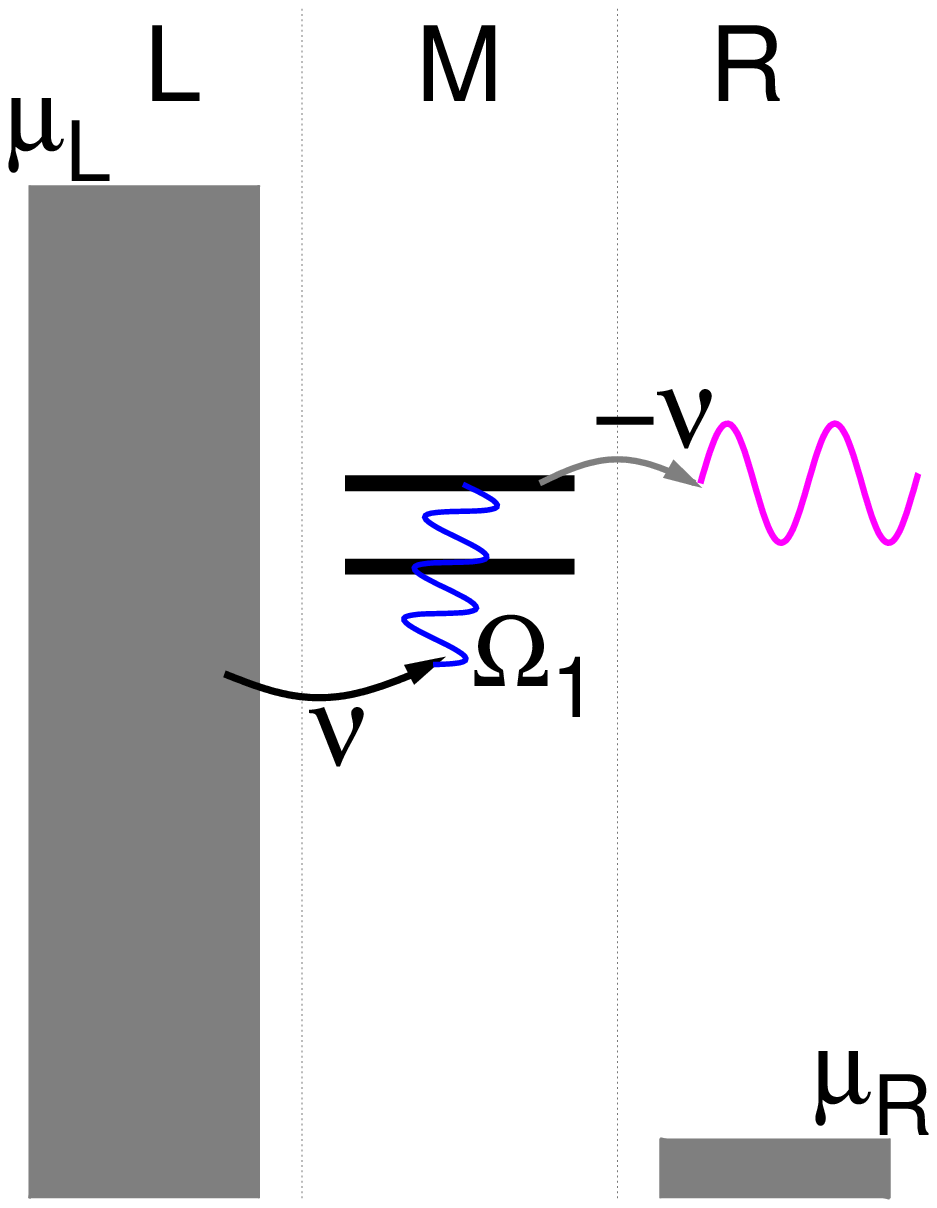}
}&
\resizebox{\newwidthprime}{\newheightprime}{
\includegraphics{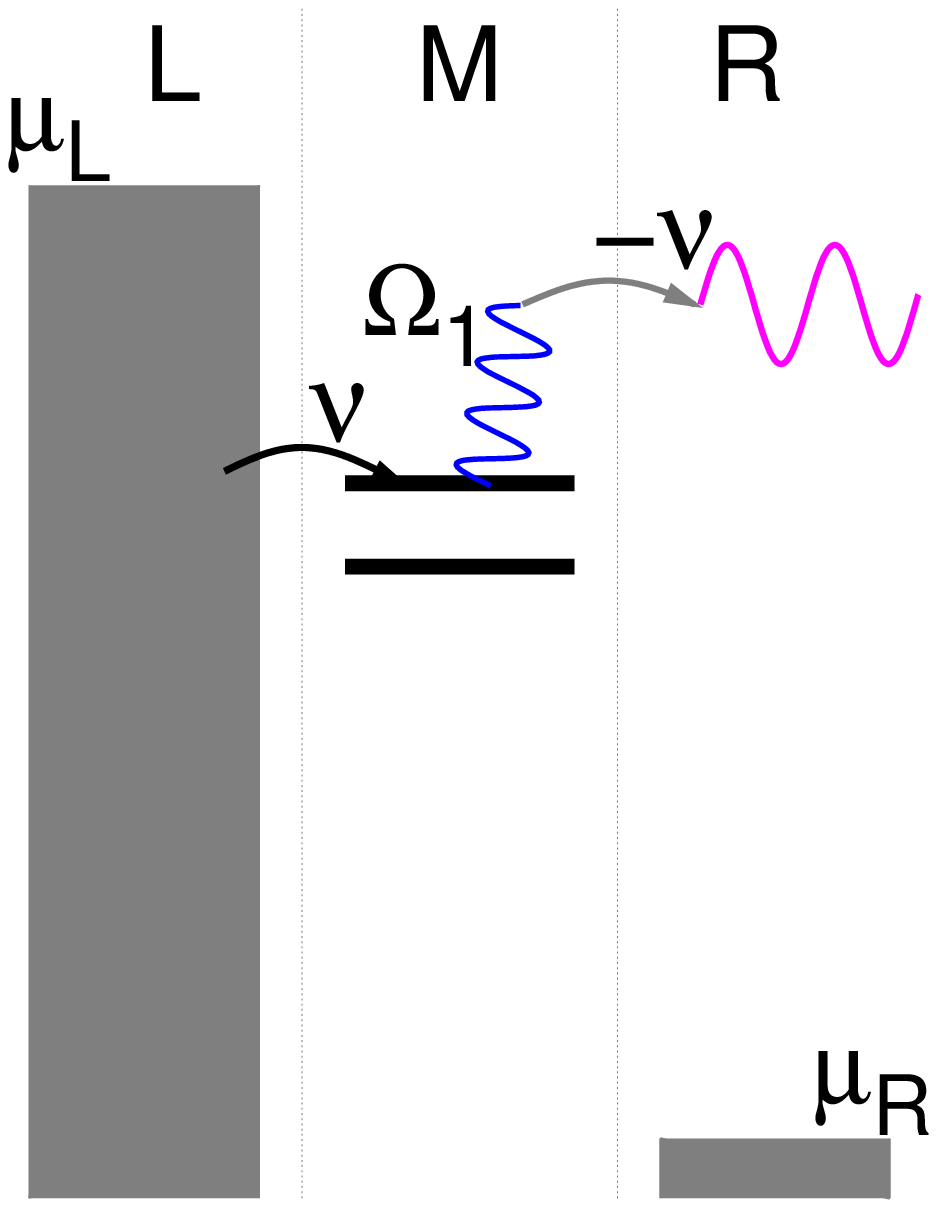}
}
\\ 
\end{tabular}
\end{center}
\caption{\label{basmech} 
Examples for resonant inelastic electron transport processes, where the 
vibrational degree of freedom of the junction is singly excited (Panel (a) and (b)) and 
deexcited (Panel (c) and (d)) upon an electron transfer process from the left electrode onto the molecular bridge 
(Panel (a) and (c)) or from the molecular bridge to the right lead (Panel (b) and (d)). 
}
\end{figure}

The current-voltage characteristic of model DESVIB is represented in 
Fig.\ \ref{DecoherenceSeqTunLinearFig}a by the solid black line. It is referred to as the vibronic 
current-voltage characteristic in the following. 
Similar to the current-voltage characteristic of model DES (solid purple line), 
it shows a step at $e\Phi=2\overline{\epsilon}_{1/2}$, 
which indicates the onset of resonant transport processes (cf.\ Sec.\ \ref{BasIntSec}). 
Besides this step, a number of additional steps can be seen at higher bias 
voltages, $e\Phi=2(\overline{\epsilon}_{1/2}+n\Omega_{1})$ 
($n\in\mathbb{N}$). They correspond to the onset of resonant excitation processes \cite{Hartle,Hartle2010b}, 
where electrons tunnel resonantly onto the molecular bridge, exciting the vibrational 
mode by $n$ vibrational quanta. An example of such an excitation process 
is graphically depicted in Fig.\ \ref{basmech}a.

It is also observed that the current level of model DESVIB is much larger than the one of model DES. 
This is caused by vibrationally induced decoherence that originates from the different 
vibronic coupling strengths $\lambda_{11}$ and $\lambda_{21}$. Since, in particular, the electronic-vibrational coupling 
of state $1$ is $\lambda_{11}=0$\,eV, 
electrons traversing the junction can only interact with the vibrational 
degree of freedom if they pass through state $2$. The interaction with 
the vibrational mode thus provides 'which-path-information' that 
destroys the coherence of the electrons \cite{Hartle2011b}. 
Without these decoherence effects, 
the current levels of model DESVIB would be an order of magnitude smaller and, 
actually, be the same as for model DES. 
This is not straightforwardly evident from the corresponding model parameters, 
since the level spacing in model DESVIB ($\epsilon_{2}-\epsilon_{1}=30$\,meV) 
is larger than in model DES ($\epsilon_{2}-\epsilon_{1}=5$\,meV). 
The polaron-shifted energy levels $\overline{\epsilon}_{1}$ and $\overline{\epsilon}_{2}$ of 
both model systems, however, are the same. As these are the relevant energies for electron tunneling, 
destructive quantum interference effects should be as pronounced in model DESVIB as in model DES. 
As this is not the case and the current level of model DESVIB even reaches the same 
values as the incoherent sum current of model DES (solid purple line in Fig.\ \ref{BasicInterferenceLinearFig}) 
at higher bias voltages ($\Phi\gtrsim1.5$\,V), 
we conclude that in model DESVIB electronic-vibrational coupling strongly quenches destructive 
quantum interference effects and that this quenching becomes stronger with increasing bias voltages.

The significantly enhanced electrical current of junction DESVIB is 
not the only manifestation of vibrationally induced decoherence in this system. 
Also, the relative step heights in the corresponding current-voltage characteristic 
are strongly influenced by quantum interference 
effects and vibrationally induced decoherence. Typically, the relative step heights are 
correlated with the Franck-Condon matrix elements 
$F_{0n}=\text{exp}(-\lambda_{21}^{2}/\Omega_{1}^{2}) (\lambda_{21}^{2n}/\Omega_{1}^{2n})/n!$. 
These matrix elements determine the probability for a transition from the 
vibrational ground state to the $n$th excited state, which, for example, may be part of 
an inelastic excitation process (Figs.\ \ref{basmech}a). Indeed, there is 
no strict one to one correspondence between the step heights and the Franck-Condon factors $F_{0n}$, 
because other processes, like the ones depicted in Figs.\ \ref{basmech}b -- Fig.\ \ref{basmech}d,  
and the respective nonequilibrium excitation of the vibrational mode lead, in general, 
to a suppression of the current level (see Ref.\ \cite{Hartle2010b}). 
But the first steps in the vibronic current-voltage characteristic of model DESVIB are, 
nevertheless, far too low. Instead of a ratio of $1:1.2$ between the first and the second step, 
as it is expected for transport through a single level \cite{Hartle2010b}, one observes a ratio of $1:2.2$. 
The successive steps show a similar behavior. 
This indicates that at the onset of the resonant transport regime 
destructive quantum interference effects decrease the current level of this system and 
that, at higher bias voltages, these effects become gradually suppressed as 
more inelastic processes become active. This is due to an increase of both the 
number of active excitation and deexcitation processes (Figs.\ \ref{basmech}a -- \ref{basmech}d), 
where the latter is particularly enhanced by the corresponding level of vibrational excitation 
(depicted in Fig.\ \ref{DecoherenceSeqTunLinearFig}b), which increases rapidly with the applied 
bias voltage in the resonant transport regime.

\begin{figure}
\begin{center}
\begin{tabular}{l}
\hspace{-0.5cm}(a) \\
\resizebox{\newwidth}{\newheight}{
\includegraphics{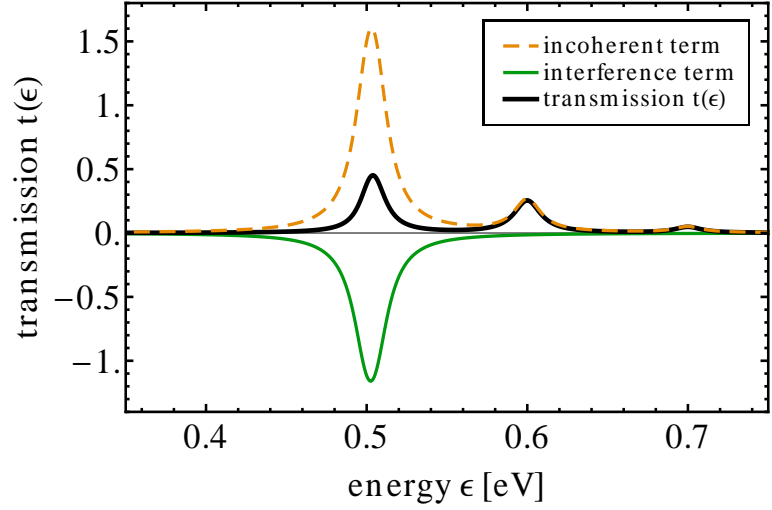}
}\\
\hspace{-0.5cm}(b) \\
\resizebox{\newwidth}{\newheight}{
\includegraphics{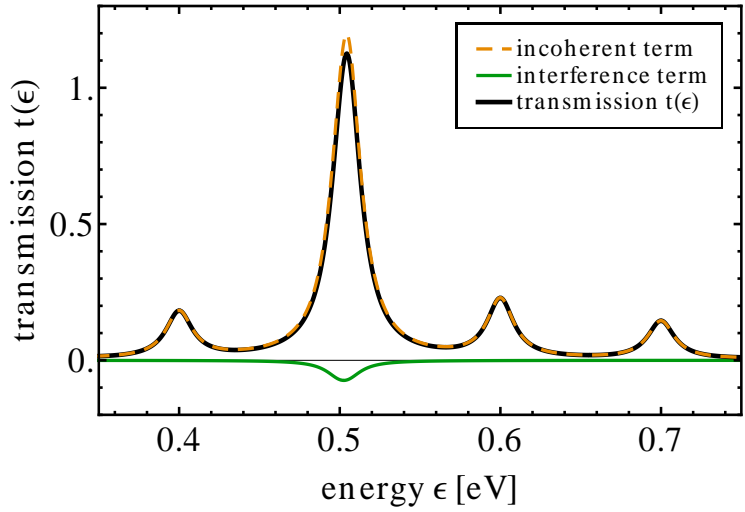}
}
\end{tabular}
\end{center}
\caption{(Color online)\label{DecoherenceSeqTunLinearFig2} 
Panel (a): The solid black line shows the transmission function of the linear molecular 
conductor DESVIB for an effective temperature of the vibrational degree of freedom 
of $10$\,K. 
In addition, the corresponding incoherent (dashed orange lines) and interference terms 
(solid green lines) are depicted. 
Panel (b): The same as for Panel (a) but for an effective temperature of the vibrational 
degree of freedom of $5000$\,K. 
}
\end{figure}

As before in Sec.\ \ref{BasIntSec}, these findings can be analyzed in more detail, considering  
the transmission function of this system. To define a transmission function in the presence of 
electronic-vibrational coupling, we restrict the vibrational degree of freedom 
to its thermal equilibrium state at temperature $T$, employ the wide-band approximation 
and consider high bias voltages 
($e\Phi\gg\epsilon_{1/2}$). 
With these assumptions, the current through junction DESVIB can be expressed 
in a form similar to Landauer theory 
\begin{eqnarray}
I\approx2e\int_{-\infty}^{\infty}(\text{d}\epsilon/2\pi)\,t(\epsilon) , 
\end{eqnarray}
with the transmission function 
\begin{eqnarray}
t(\epsilon)\equiv i\,\text{tr}\lbrace \mathbf{\Gamma}_{\text{L}}  
\textbf{G}^{>} \rbrace.
\end{eqnarray}
A decomposition of this transmission function in terms of transmission amplitudes 
is not obvious. Instead, we use the observation (cf.\ Sec.\ \ref{BasIntSec}) 
that the incoherent term of the transmission function
can also be obtained by neglecting the off-diagonal elements 
of the self-energy matrix in the current formula (\ref{currentformula}). 
The interference term is then 
given by $t_{\text{int}}(\epsilon)=t(\epsilon)-t_{\text{inc}}(\epsilon)$.  
For our specific model system, we thus obtain (see App.\ \ref{transDESVIB} for details)
\begin{subequations}
\label{condesparts}
\begin{eqnarray}
 t_{\text{inc}}(\epsilon) &=& \frac{\Gamma^{2}}{\left\vert 
\epsilon-\overline{\epsilon}_{1} + i \Gamma\right\vert^{2}}  +  A^{2}  \sum_{l=-\infty}^{\infty} I_{l}(x) 
\text{e}^{\beta l\Omega_{1}/2} \frac{\Gamma^{2}}{\left\vert \epsilon-\overline{\epsilon}_{2} - 
l \Omega_{1} + i \Gamma \right\vert^{2}}, \\
\label{condesparts2}
 t_{\text{int}}(\epsilon) &=& -2 A^{2}  \Gamma^{2} \text{Re}\left[ \frac{1}{(\epsilon-
\overline{\epsilon}_{2} + i \Gamma )( \epsilon-\overline{\epsilon}_{1} - 
i \Gamma )} \right]. 
\end{eqnarray}
\end{subequations}
Thereby, the prefactor $A=\text{e}^{-(\lambda_{21}^{2}/(2\Omega_{1}^{2}))(2N_{\text{vib}}+1)}$ 
is determined by the average vibrational excitation 
$N_{\text{vib}}=(\text{e}^{\beta\Omega_{1}}-1)^{-1}$ and the inverse temperature 
$\beta=(k_{\text{B}}T)^{-1}$, while  
$I_{l}(x)=I_{l}(2(\lambda_{21}^{2}/\Omega_{1}^{2})\sqrt{N_{\text{vib}}(N_{\text{vib}}+1)})$ 
denotes the $l$th modified Bessel function of the first kind.

The transmission function of model DESVIB, which is obtained according to this scheme, 
is depicted in Fig.\ \ref{DecoherenceSeqTunLinearFig2} by the solid black line. 
The corresponding incoherent and interference terms 
are given by the dashed orange and solid green line, respectively. 
Thereby, two scenarios are distinguished: (i) the effective temperature of the 
vibrational degree of freedom is assumed to be $10$\,K, \emph{i.e.}\ it is
effectively not excited, and (ii) 
the temperature is assumed to be $5000$\,K, which results in a level of
vibrational excitation that is comparable to the one obtained in the resonant transport 
regime (cf.\ Fig.\ \ref{DecoherenceSeqTunLinearFig}b). In both cases, the interference term 
is significantly weaker with respect to the incoherent term 
than in the respective electronic transport scenario, which, 
due to the polaron shift of the electronic energy levels, is the one of model DES 
(cf.\ Fig.\ \ref{BasicInterferenceLinear2Fig}a). Accordingly, the
transmittance of the junction is also larger and the corresponding current levels are higher. 
This can bee seen in Fig.\ \ref{DecoherenceSeqTunLinearFig}a by comparison of the dashed gray and red line, 
which depict the current-voltage characteristic of model DESVIB with the
vibrational mode kept
in thermal equilibrium at $10$ and $5000$\,K, 
respectively, with the solid purple line, which depicts the 
current-voltage characteristic of model DES. 

The suppression of the interference term indicates a strong quenching of quantum 
interference effects in this system. This quenching originates from both a static 
mechanism that constitutes an effective renormalization of the level-width 
functions $\Gamma_{K,mm}$ and vibrationally induced decoherence. 
The factor $A^{2}$, which precedes the interference term (\ref{condesparts2}), 
includes both of these effects. It describes, on one hand, 
the suppression of electronic tunneling events 
(as, \emph{e.g.}, depicted in Fig.\ \ref{TransportProcesses}) due to 
electronic-vibrational coupling or, equivalently, the decreased overlap 
between the vibrational states of different charge states of the molecular bridge. 
At $10$\,K, where the vibrational mode is effectively restricted to its ground state, 
this suppression corresponds to the Franck-Condon matrix element $F_{00}$.  
On the other hand, the prefactor $A^{2}$ also includes the effect of vibrationally induced decoherence, 
which originates from inelastic tunneling processes, 
where electrons tunnel resonantly through the junction at energies
$\approx\overline{\epsilon}_{1/2}$. 
Examples for such processes are depicted in 
Figs.\ \ref{basmech}b and \ref{basmech}d.
Inelastic processes, where electrons tunnel 
resonantly through the junction at energies 
$\epsilon\approx\overline{\epsilon}_{2}+n\Omega_{1}$ with
$n\in\mathbb{Z}/{\{0\}}$ 
(see Figs.\ \ref{basmech}a and \ref{basmech}c), 
complement the effect of vibrationally induced decoherence and result in
additional side peaks in the transmission function and the incoherent term 
but not in the interference term. 
Both the suppression of the interference term by the prefactor $A^{2}$ and the vibrational 
side peaks become stronger at higher levels of vibrational excitation, which increases 
the probability for inelastic processes, in particular deexcitation processes.  
As a result, the main peak in the interference term, which appears at $\epsilon\approx\overline{\epsilon}_{1/2}$  
and is clearly visible in the electronic case, is somewhat reduced in the vibronic case with $10$\,K and  
vanishes almost completely in the vibronic case with $5000$\,K. 
Thus, vibrationally induced decoherence leads to a complete 
suppression of interference effects in this system as the effective temperature of the vibrational mode 
(or, equivalently, $N_{\text{vib}}$) increases.

\begin{figure}
\begin{center}
\begin{tabular}{l}
\resizebox{\newwidth}{\newheight}{
\includegraphics{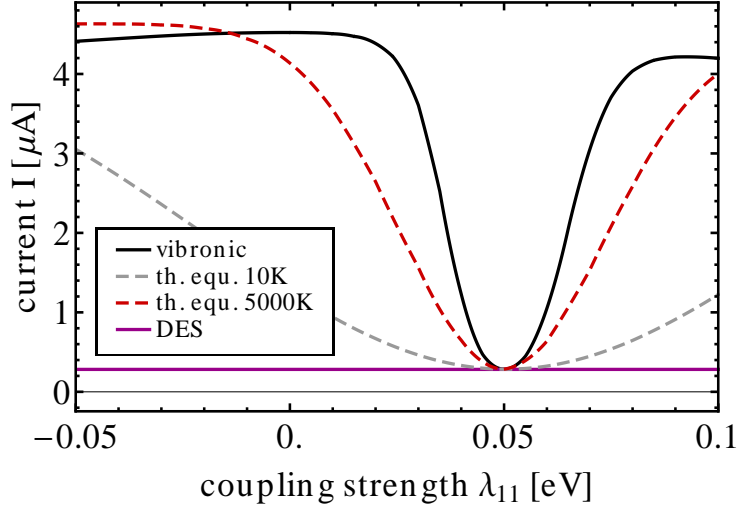}
}
\end{tabular}
\end{center}
\caption{(Color online)\label{DecoherenceSeqTunLinearFig3} 
Current flowing through junction DESVIB at a bias voltage of $\Phi=2$\,V 
as a function of the electronic-vibrational coupling strength $\lambda_{11}$
(solid black line). All the other parameters of model DESVIB are fixed, 
in particular, $\lambda_{21}=0.05$\,eV. 
Comparison to the current of model DES (solid purple line) shows that 
at $\lambda_{11}=\lambda_{21}=0.05$\,eV interaction of the tunneling electrons with the vibrations 
gives no 'which-path' information and, thus, quantum interference effects suppress the current level of 
both systems in the same way. 
}
\end{figure}

So far, we have restricted our discussion of decoherence effects in model DESVIB to the 
special scenario, where only one of the electronic states is coupled to the vibrational 
mode ($\lambda_{11}=0$ while $\lambda_{21}=0.05$\,eV$\neq0$). 
In most realistic situations both states will be coupled to the vibrational mode.   
If these couplings are similar, interaction of the tunneling electrons with the vibrational degree of freedom 
provides less accurate 'which-path' information and, therefore, the quenching of destructive interference is 
also less pronounced. 
This is demonstrated in Fig.\ \ref{DecoherenceSeqTunLinearFig3}, where the current level of model DESVIB 
(at $\Phi=2$\,V) is shown as a function of the electronic-vibrational coupling strength $\lambda_{11}$. 
While vibrationally induced decoherence results in an almost complete suppression of destructive 
quantum interference effects, if the electronic-vibrational coupling strengths of both states 
are very different, $\vert\lambda_{11}-\lambda_{21}\vert > 0.02$\,eV, destructive interference effects 
are active and lead to a strong suppression of the current level, if the coupling strengths of both states 
are similar, that is for $\vert\lambda_{11}-\lambda_{21}\vert < 0.01$\,eV. 
The magnitude of the current suppression is thereby determined by the ratio of the electronic parameters 
$\Delta=\epsilon_{1}-\epsilon_{2}/2$ and $\Gamma$ (as already discussed in Sec.\ \ref{BasIntSec}). 
The corresponding width is, to a large extent, determined by nonequilibrium effects. 
This can be seen by comparison of the solid black, the dashed gray and the dashed red lines, 
which correspond to a full nonequilibrium calculation and calculations where the vibrational 
degree of freedom is restricted to its thermal equilibrium state at $10$ and $5000$\,K, respectively. 
For larger levels of vibrational excitation, the valley of current suppression around $\lambda_{11}\approx\lambda_{21}$ 
appears to be significantly narrower. 
This behavior can already be deduced from our analytical result, Eqs.\ (\ref{condesparts}). There, the  
interference term (\ref{condesparts2}) becomes strongly suppressed, as the prefactor $A$ becomes smaller 
for larger levels of vibrational excitation $N_{\text{vib}}$. Eqs.\ (\ref{condesparts}) also demonstrate that, 
besides the effective temperature of the vibration, vibrationally induced decoherence is more effective 
for larger differences of the dimensionless electronic-vibrational coupling strengths 
$\vert\lambda_{11}/\Omega_{1}-\lambda_{21}/\Omega_{1}\vert$. This can be seen by the dashed gray line, 
where the width of the associated current suppression is given by the frequency of the 
vibrational mode $\Omega_{1}$. 


\begin{figure}
\begin{center}
\begin{tabular}{l}
\resizebox{\newwidth}{\newheight}{
\includegraphics{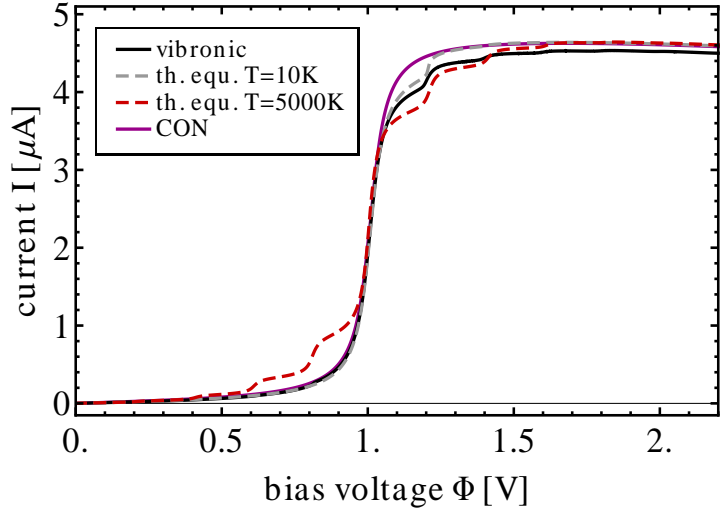}
}
\end{tabular}
\end{center}
\caption{(Color online)\label{DecoherenceSeqTunBranchedFig} 
Current-voltage characteristics of the branched molecular conductor $\text{CONVIB}$ (solid black line). 
The dashed gray and red lines depict the current-voltage characteristics that are obtained by 
restricting the vibrational degree of freedom to its thermal equilibrium state at $10$ and $5000$\,K, respectively. 
The solid purple line 
shows the current-voltage characteristic of model CON. 
}
\end{figure}

Next, we consider model CONVIB. 
The corresponding current-voltage characteristic is shown in 
Fig.\ \ref{DecoherenceSeqTunBranchedFig}. In contrast to model DESVIB, this characteristic 
does not exhibit pronounced decoherence effects, because in model CON interference effects 
play, a priori, a less significant role in the resonant transport regime (cf.\ Sec.\ \ref{BasIntSec}). 
The respective level of vibrational excitation (data not shown)  
is almost identical to the one of model DESVIB, since inelastic processes, due to the simplified 
electronic-vibrational coupling scenario ($\lambda_{11}=0$), are not influenced by 
interference effects in both systems. 
Analysis of the corresponding transmission function, however, reveals that 
electronic-vibrational coupling affects quantum interference effects that are active in 
model CONVIB in a very similar way as in model DESVIB.

\begin{figure}
\begin{center}
\begin{tabular}{l}
\hspace{-0.5cm}(a) \\
\resizebox{\newwidth}{\newheight}{
\includegraphics{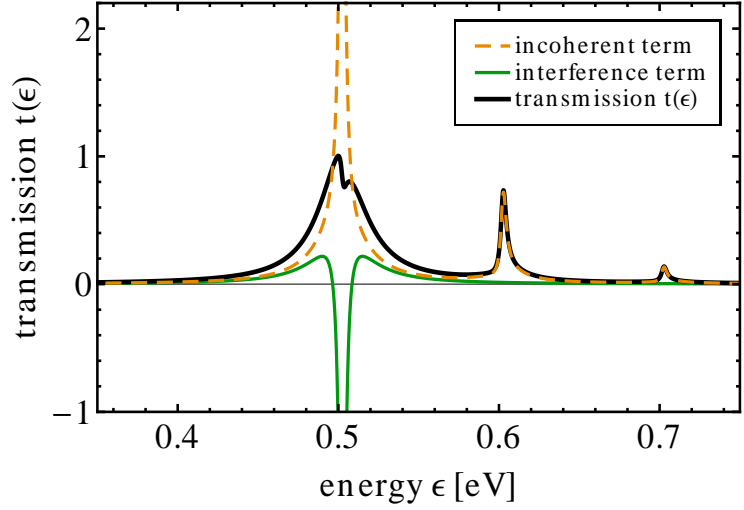}
}\\
\hspace{-0.5cm}(b) \\
\resizebox{\newwidth}{\newheight}{
\includegraphics{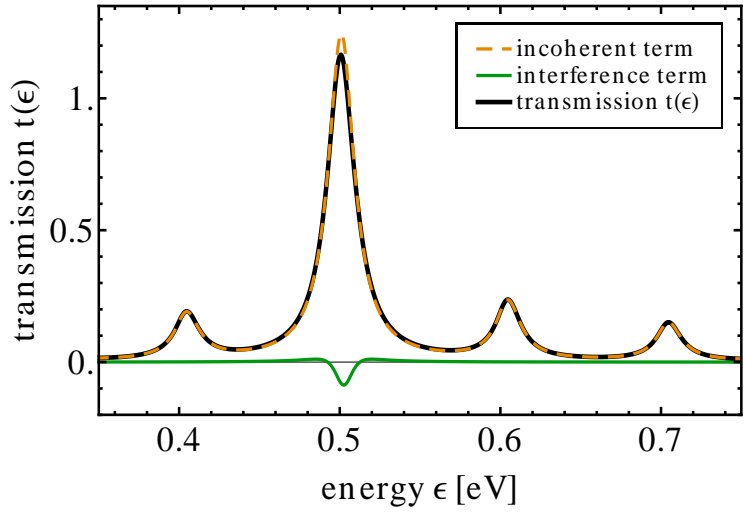}
}
\end{tabular}
\end{center}
\caption{(Color online)\label{DecoherenceSeqTunBranchedFig2} 
Panel (a): The solid black line shows the transmission function of the branched molecular 
conductor CONVIB for an effective temperature of the vibrational degree of freedom 
of $10$\,K. In addition, the corresponding incoherent (dashed orange lines) and 
interference terms (solid green lines) are depicted. 
Panel (b): The same as for Panel (a) but for an effective temperature of the vibrational degree of freedom of $5000$\,K. 
}
\end{figure}

Using the same scheme as before, the transmission function 
and the interference term of model CONVIB are given by (see App.\ \ref{transCONVIB}) 
\begin{subequations}
\label{condespartsII}
\begin{eqnarray}
 t(\epsilon) &=& 
c(\epsilon) \left[ \vert \epsilon-\overline{\epsilon}_{2} + i \Gamma \vert^{2} + 
2 A^{2} (\epsilon-\overline{\epsilon}_{1})(\epsilon-\overline{\epsilon}_{2}) - 2 A^{2} \Gamma^{2} (2-A^{2})  \right] \\
&& +  A^{2}  \sum_{l=-\infty}^{\infty} I_{l}(x) 
\text{e}^{\beta l\Omega_{1}/2} c(\epsilon-l\Omega_{1}) \left(\left\vert \epsilon-\overline{\epsilon}_{1} - 
l \Omega_{1} + i \Gamma \right\vert^{2} - \Gamma^{2} A^{2} \right) , \nonumber\\
 t_{\text{int}}(\epsilon) &=& 2 A^{2} c(\epsilon) 
\left[  (\epsilon-\overline{\epsilon}_{1})(\epsilon-\overline{\epsilon}_{2}) - \Gamma^{2} (1-A^{2})  \right], 
\end{eqnarray}
\end{subequations}
with 
\begin{eqnarray}
c(\epsilon) &=& \left\vert \frac{\Gamma}{(\epsilon-\overline{\epsilon}_{1}+i\Gamma)(\epsilon-\overline{\epsilon}_{2}+i\Gamma)+A^{2}\Gamma^{2}} \right\vert^{2}.
\end{eqnarray}
They are represented in Fig.\ \ref{DecoherenceSeqTunBranchedFig2}, 
including the respective incoherent term. Vibrationally induced decoherence causes the same 
effects as outlined before, that is, a suppression of the interference term and 
the appearance of side peaks that are not counterbalanced by the interference term. 
As a result, the antiresonance at $\epsilon=(\overline{\epsilon}_{1}+\overline{\epsilon}_{2})/2$ is quenched 
and even vanishes as the level of vibrational excitation increases.

\subsection{Vibrationally Induced Decoherence in the Non-Resonant Transport Regime}
\label{CoTunEffects}

While in Sec.\ \ref{SequTunReg} we have focused on the resonant transport regime 
of a molecular junction, here we investigate the effect of electronic-vibrational coupling 
on quantum interference effects in the non-resonant transport regime. To this end, 
we discuss the transport characteristics of the same model systems as in Sec.\ \ref{SequTunReg} 
but for lower bias voltages: $e\Phi<0.25$\,eV$\ll2\epsilon_{1/2}$.

\begin{figure}
\begin{tabular}{l}
\resizebox{\newwidth}{\newheight}{
\includegraphics{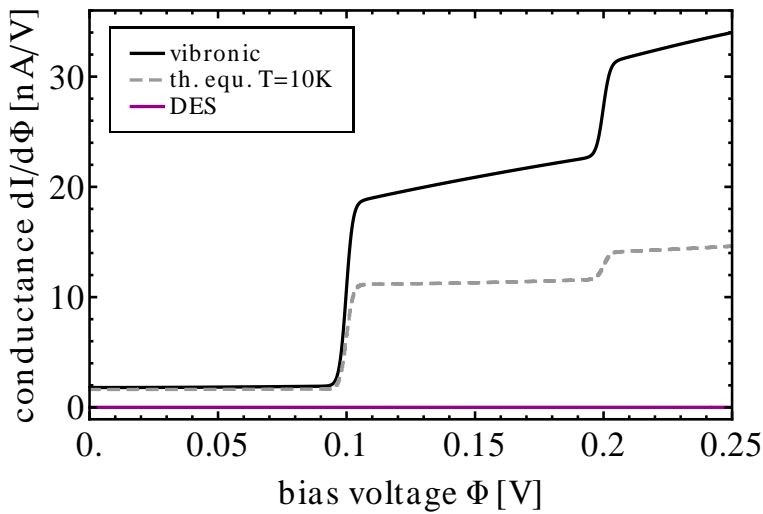}
}\\
\end{tabular}
\caption{(Color online)\label{DecoherenceCoTunLinearFig} 
Conductance-voltage characteristics of the linear molecular conductor 
DESVIB (solid black line). The solid purple line shows the conductance-voltage characteristic of junction DES. 
The dashed gray line depicts the conductance-voltage characteristic of junction DESVIB 
with the vibrational degree of freedom kept in thermal equilibrium at $10$\,K. 
}
\end{figure}

\begin{figure}
\begin{center}
\begin{tabular}{lll}
(a)&(b)&(c)\\
\resizebox{\newwidthprime}{\newheightprime}{
\includegraphics{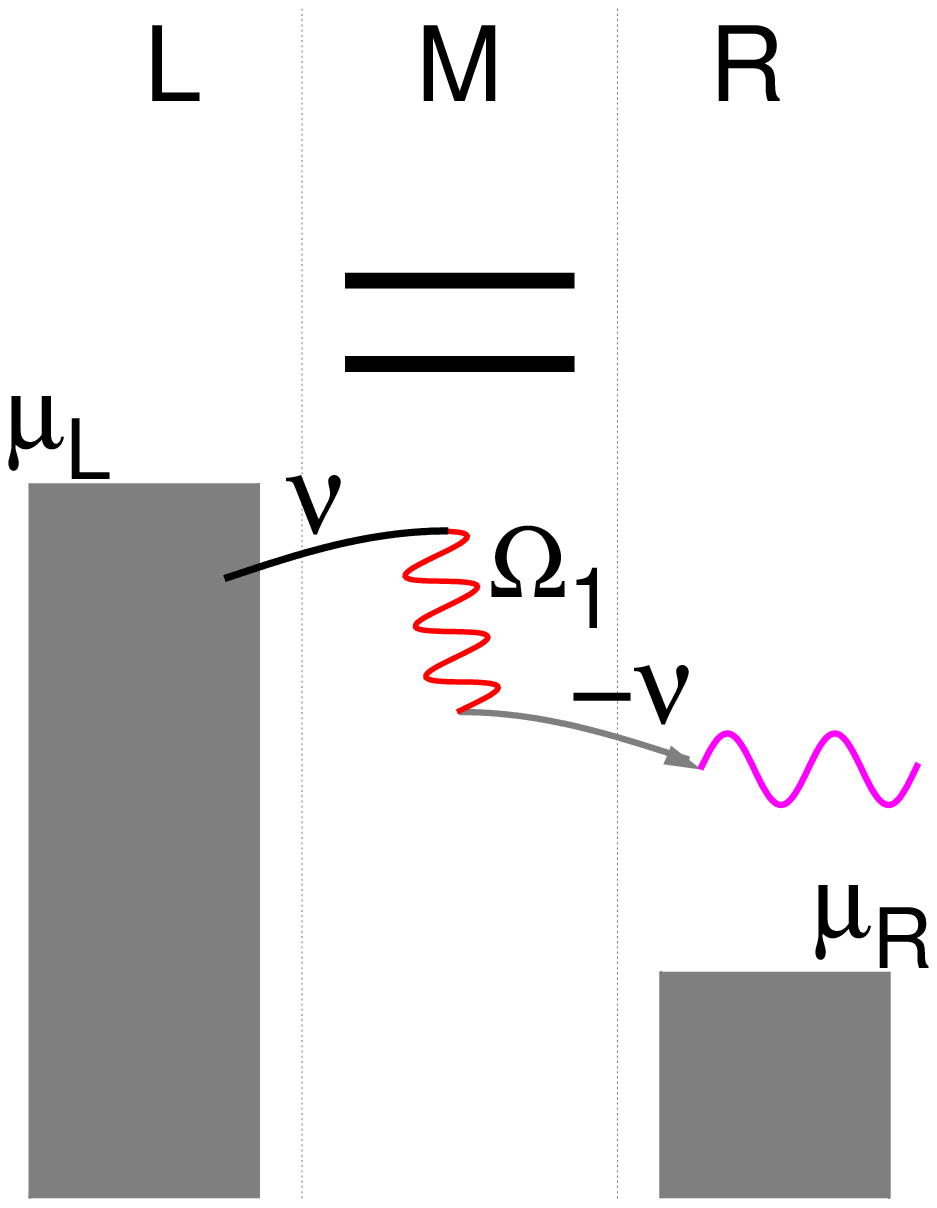}
}&
\resizebox{\newwidthprime}{\newheightprime}{
\includegraphics{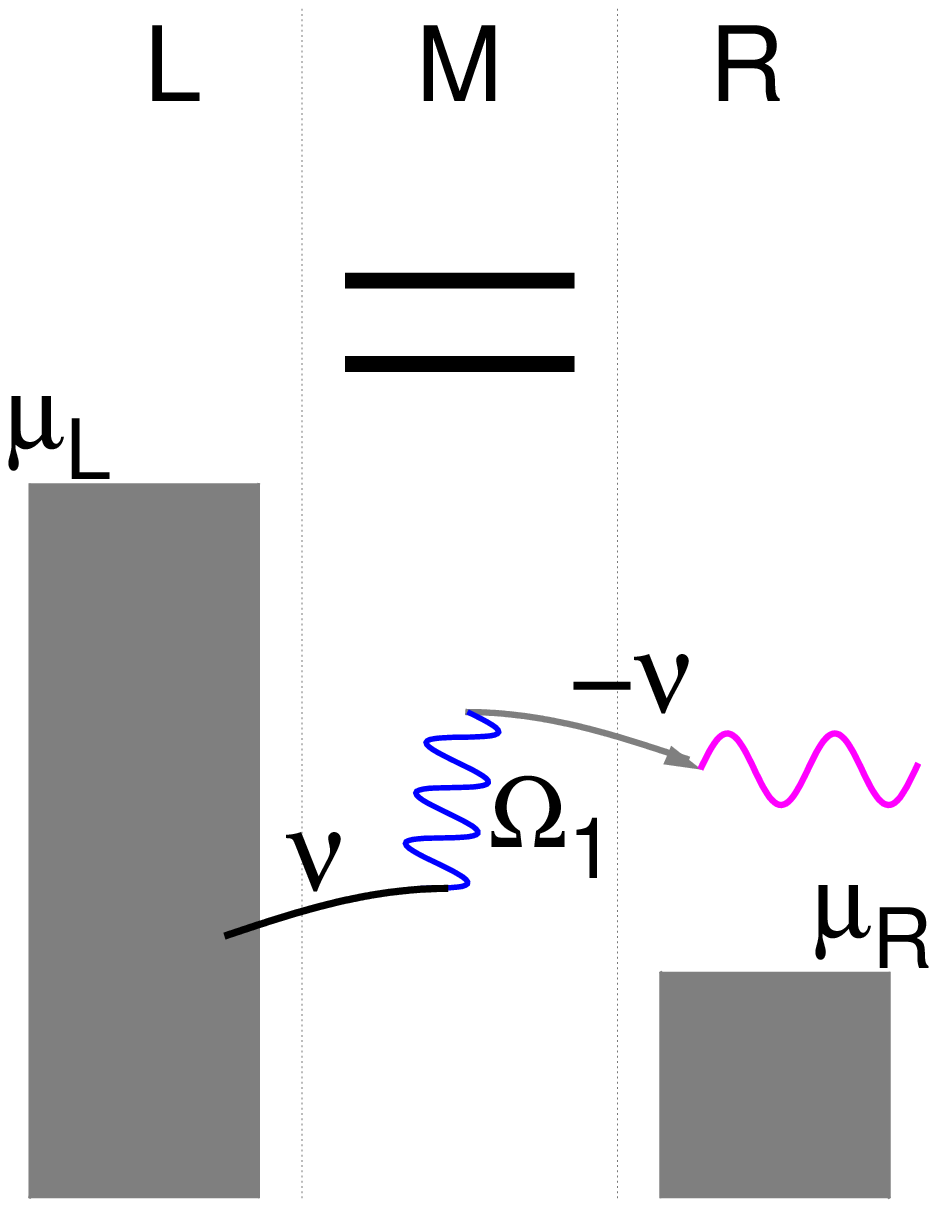}
}&
\resizebox{\newwidthprime}{\newheightprime}{
\includegraphics{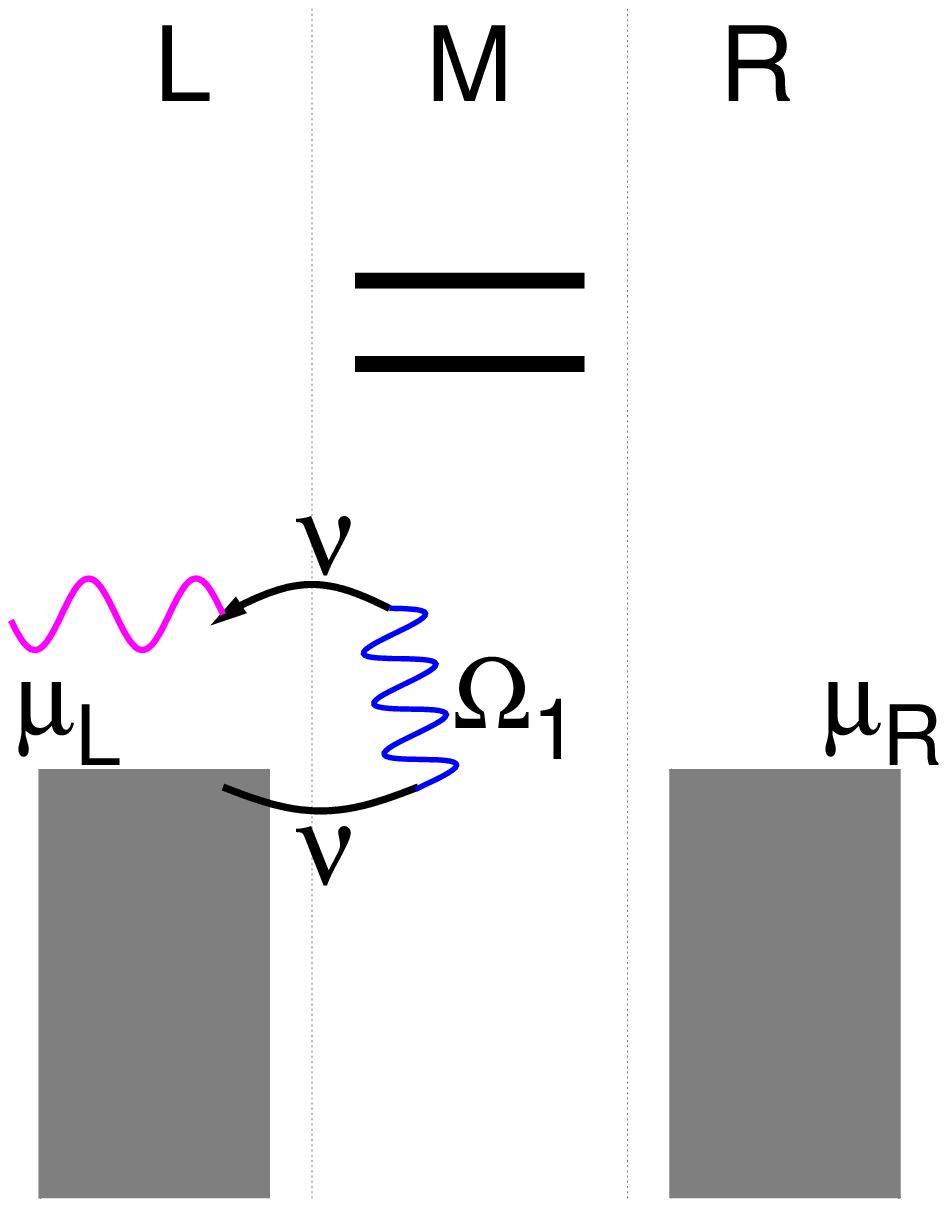}
}
\\ 
\end{tabular}
\end{center}
\caption{\label{basmechcotun} 
Examples of non-resonant inelastic processes, including transport (Panel (a) and (b)) and 
electron-hole pair creation processes (Panel (c)). Note that transport processes can occur 
at low temperatures (\emph{i.e.}\ $k_{\text{B}}T_{\text{L/R}}\gg\Omega_{1}$) via the excitation (Panel (a), $e\Phi>\Omega_{1}$) 
or deexcitation of the vibrational degree of freedom (Panel (b)) while, in contrast, 
electron-hole pair creation processes involve predominantly deexcitation processes,  
because the respective excitation processes are suppressed by Pauli blocking. 
}
\end{figure}

First, we consider model DESVIB. The corresponding conductance-voltage characteristic 
(\emph{i.e.}\ the differential conductance $\text{d}I/\text{d}\Phi$) is represented by the solid black line 
in Fig.\ \ref{DecoherenceCoTunLinearFig}. 
In the non-resonant transport regime, the conductance of junction DESVIB is 
almost constant, because the probability of non-resonant transport processes, 
like the one depicted in Fig.\ \ref{TransportProcesses}b), does not significantly vary with the energy of the 
tunneling electrons and because the energy window, where these processes can occur, increases 
linearly with the applied bias voltage. Deviations of this constant behavior arise, whenever the bias voltage exceeds 
a multiple of the vibrational frequency, $e\Phi\approx n\Omega_{1}$ ($n\in\mathbb{N}$). At these bias voltages 
steps appear in the conductance-voltage characteristics that are related to the onset of 
inelastic channels, where electrons are tunneling through the junction non-resonantly by exciting the 
vibrational degree of freedom (an example process is depicted in Fig.\ \ref{basmechcotun}a). 
This behavior is well known and is used, for example, to determine the frequency of vibrational modes 
in inelastic electron tunneling spectroscopy of molecular junctions 
\cite{Ho98,Lorente2000,Troisi2006,GalpScience08,Godby2012}.

Vibrationally induced decoherence modifies this picture in a very similar way as in 
the resonant transport regime. 
First of all, the current/conductance of junction DESVIB is significantly increased, as is evident 
by comparison with the conductance-voltage characteristics of junction DES (solid purple line). 
While, for $e\Phi<\Omega_{1}$, this increase can be solely attributed to the static effect of a 
decreased overlap between vibrational states of different charge states of the molecular bridge 
(see Sec.\ \ref{SequTunReg}), the opening of inelastic channels at $e\Phi\geq n\Omega_{1}$ leads 
to vibrationally induced decoherence effects that increase the conductance of junction DESVIB 
much more. 
Secondly, the relative step heights in the conductance-voltage characteristics, 
which should be correlated, in a similar way as the steps in the current-voltage characteristics, 
with the Franck-Condon matrix elements $F_{0n}$, appear far too large. This is again due 
to destructive quantum interference effects that suppress the current level at low bias voltages 
but become less effective the more inelastic processes become active. Thus, for example, 
the conductance ratio before and after the first step is roughly $1:10$ and not $1:1.3$, 
as for a comparable scenario with a single electronic state \cite{Hartle}. 
Moreover, the conductance of junction DESVIB is not a constant after the onset of 
the first inelastic channel at $e\Phi=\Omega_{1}$. This is shown by the solid black and 
the dashed gray line, where nonequilibrium effects or the heating of the vibrational 
degree of freedom is taken into account and discarded, respectively. As non-resonant 
excitation processes (an example of which is depicted in Fig.\ \ref{basmechcotun}a) become 
availabe for $e\Phi>\Omega_{1}$ and lead to a step-wise increase of the conductance, 
which is seen in both the black and the gray line, they also 
result in a finite level of vibrational excitation. Consequently, non-resonant deexcitation 
processes (Fig.\ \ref{basmechcotun}b) occur and increase the conductance of the junction even further 
(which is only visible in the black line). As the number of excitation processes 
increases continuously for $e\Phi>\Omega_{1}$, 
the level of vibrational excitation also increases continuously and so does the 
current contribution of these non-resonant deexcitation processes.

\begin{figure}
\begin{center}
\resizebox{\newwidth}{\newheight}{
\includegraphics{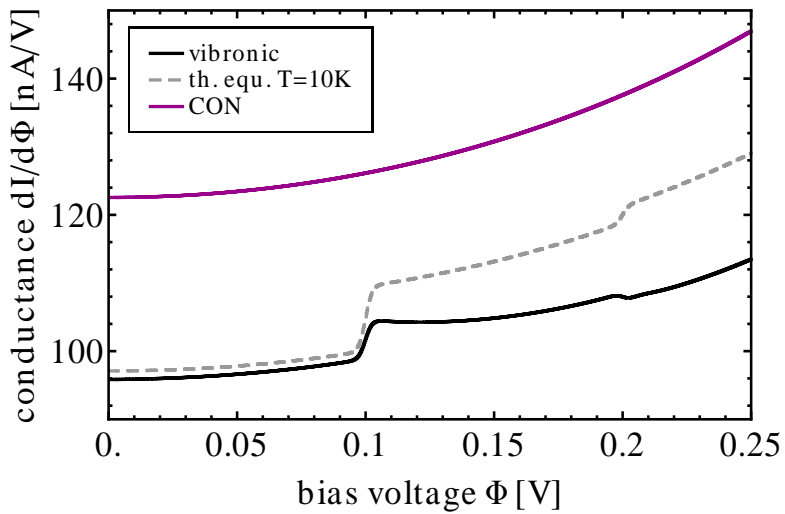}
}
\end{center}
\caption{(Color online)\label{DecoherenceCoTunBranchedFig} 
Conductance-voltage characteristics of the branched molecular conductor CONVIB. 
The solid purple line shows the conductance-voltage characteristic of junction CON. 
The dashed gray line depicts the 
conductance-voltage characteristic of junction CONVIB, which is obtained by 
evaluating the vibrational 
degree of freedom by the thermal equilibrium state that it would acquire at $10$\,K. 
}
\end{figure}

Next, we investigate 
non-resonant transport through model CONVIB.
The corresponding conductance-voltage 
characteristics is depicted in Fig.\ \ref{DecoherenceCoTunBranchedFig}. 
In contrast to junction DESVIB, the vibronic conductance-voltage characteristic exhibits  
overall smaller conductance values. This is due to both a reduction of the probability for 
electronic transfer events (Fig.\ \ref{TransportProcesses}b) as well as 
vibrationally induced decoherence, which quenches the constructive interference effects 
that are active in this system. 
Thereby, a higher level of vibrational excitation, due to a more important role of deexcitation processes, 
leads to a further reduction of conductance (compare the solid black and the dashed gray line). 
The steps, which appear very pronounced in the vibronic conductance-voltage 
characteristic of model DESVIB at $e\Phi\backsimeq n\Omega_{1}$ ($n\in\mathbb{N}$), 
are, therefore, substantially reduced. This reflects the competition 
between a reduction of conductance due to vibrationally induced decoherence and 
the increase of conductance due to the onset of non-resonant excitation processes.

Another effect of vibrationally induced decoherence is that 
the slope of the vibronic conductance-voltage characteristic can even be negative 
after the onset of an inelastic channel at $e\Phi\approx n\Omega_{1}$ ($n\in\mathbb{N}$). 
This can be seen, for example, after the first step in the solid black line of 
Fig.\ \ref{DecoherenceCoTunBranchedFig} and is related to the number of inelastic processes, 
which increases with the applied bias voltage $\Phi$. 
Recall that, for the same reasons, the slope of the conductance-voltage characteristic 
of junction DESVIB is positive at these bias voltages. 
In addition, 
at the opening of the second inelastic channel, $e\Phi\approx2\Omega_{1}$, 
the reduction of conductance due to vibrationally induced decoherence may also prevail 
over the increase of conductance due to the onset of non-resonant excitation processes. This 
leads to a drop in the conductance of junction CONVIB at these bias voltages. 
It should be noted that such a decrease of conductance (due to the opening of inelastic transport channels)  
has been reported before \cite{Lorente2000,Haupt2009,Avriller2009,Schmidt2009} 
but only for highly conducting molecular junctions ($\text{d}I/\text{d}\Phi\gtrsim 20$\,$\mu$A/V). 
We thus conclude that quantum interference effects and vibrationally induced decoherence have 
a strong impact on the inelastic tunneling spectra of a molecular junction.

Note that antiresonances such as they appear in model CON  
have been studied extensively before and have already been considered in 
the context of molecular spintronics \cite{Herrmann2010,Herrmann2011} 
or thermoelectric devices \cite{Bergfield2009,Bergfield2010b}. Thereby, it is desirable that  
the antiresonance is located near the Fermi energy of the junction. 
Our results suggest that vibrationally induced decoherence puts 
a strong constraint on these applications. As dynamical excitation and deexcitation 
processes are weakening such antiresonances in both the resonant and the non-resonant transport 
regime, they should be located in a rather narrow range of energies around the Fermi 
energy of the junction, $[\epsilon_{\text{F}}-\Omega_{\text{low}},\epsilon_{\text{F}}+\Omega_{\text{low}}]$, 
which is determined by the frequency $\Omega_{\text{low}}$ of the low-frequency modes of the 
respective system.

\subsection{Suppression of Local Heating by Destructive Quantum Interference Effects} 
\label{localcool}

In Secs.\ \ref{BasIntSec}--\ref{CoTunEffects}, we have shown that destructive interference effects 
suppress the electrical current flowing through a molecular conductor. In this section, 
we demonstrate that the respective level of vibrational excitation may also be strongly suppressed, 
in particular in the resonant transport regime, where the level of current-induced vibrational excitation 
can be very high 
(cf.\ Figs.\ \ref{DecoherenceSeqTunLinearFig}b or, \emph{e.g.}, 
Refs.\ \onlinecite{Semmelhack,Hartle,Hartle09,Romano10,Hartle2010,Hartle2010,Volkovich2011b,Ankerhold2011}).

\begin{figure}
\begin{center}
\begin{tabular}{lll}
(a)&(b)\\
\resizebox{\newwidthprime}{\newheightprime}{
\includegraphics{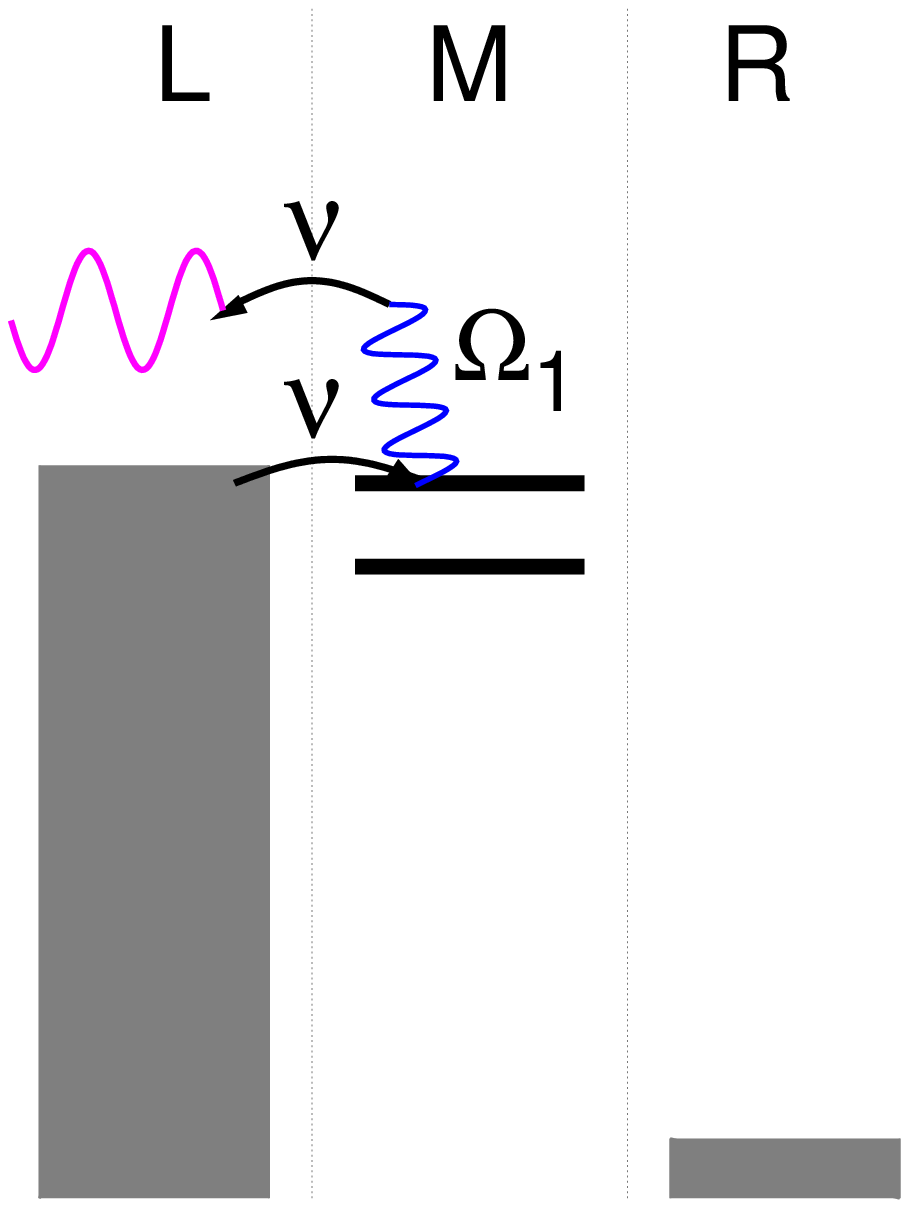}
}&
\resizebox{\newwidthprime}{\newheightprime}{
\includegraphics{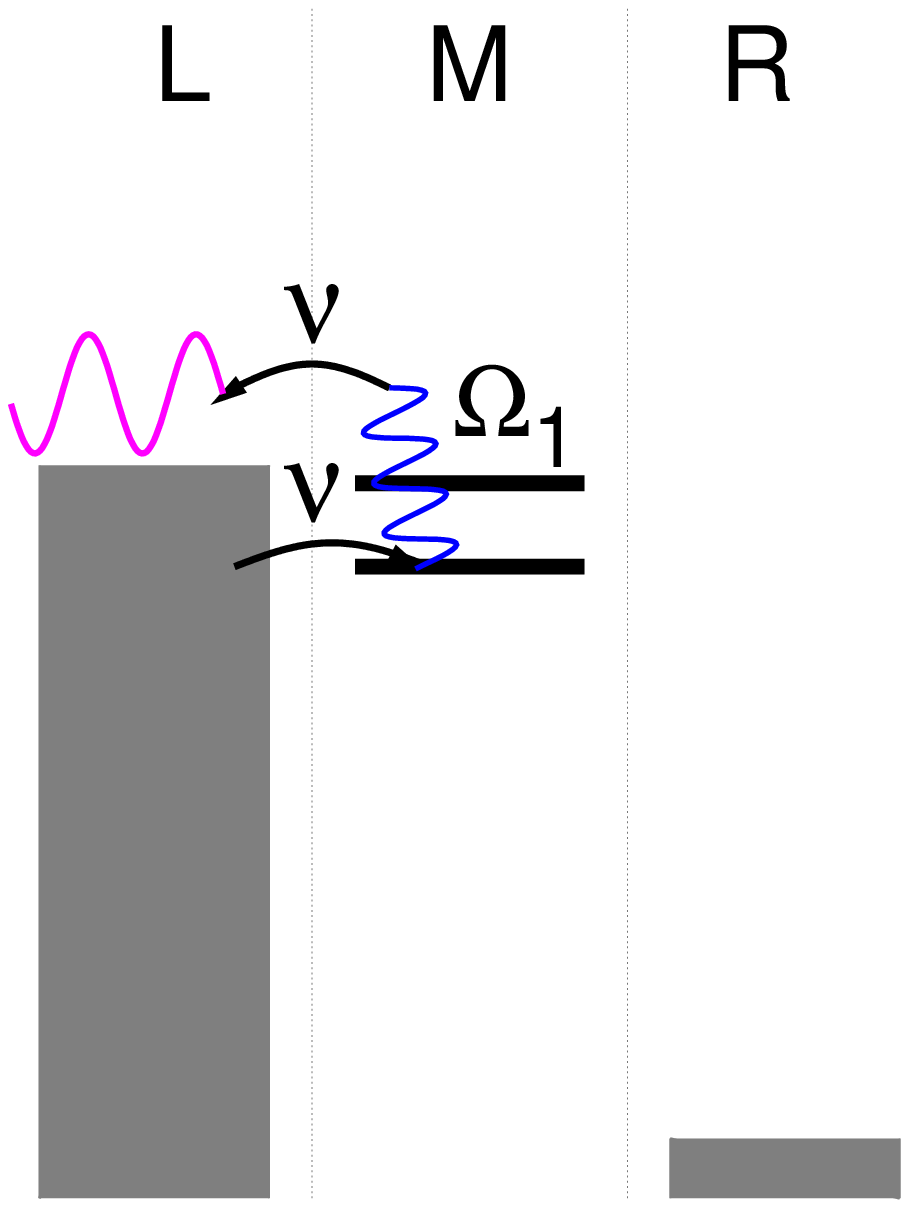}
}
\\ 
\end{tabular}
\end{center}
\caption{\label{el-hole-pair}  
Examples of resonant electron-hole pair creation processes. In contrast to transport 
processes, which involve both electrodes, these processes involve just one of the leads. 
They are thus not subject to the destructive quantum interference effects 
that are active in model DESVIB and, consequently, if destructive interference effects 
are very pronounced, lead to a strong suppression of the current-induced level of 
vibrational excitation in such a system (which originates, \emph{e.g.}, from the resonant excitation 
processes depicted in Figs.\ \ref{basmech}a and \ref{basmech}b). 
}
\end{figure}

To understand the vibrational excitation characteristics of a molecular conductor, in general, 
one has to take into account excitation and deexcitation 
processes due to electron transport (cf.\ Figs.\ \ref{basmech}a -- \ref{basmech}d for 
resonant transport processes and Figs.\ \ref{basmechcotun}a -- \ref{basmechcotun}b for 
non-resonant transport processes) but also deexcitation processes, 
which are associated with electron-hole pair creation processes in the 
leads \footnote{At higher temperatures of the electrodes, electron-hole pair 
creation processes can also occur via vibrational excitation processes. At lower temperatures, 
however, these processes are typically Pauli-blocked.}. 
Examples for such electron-hole pair creation processes are depicted 
in Fig.\ \ref{basmechcotun}c and Fig.\ \ref{el-hole-pair}, where 
non-resonant and resonant pair creation processes are shown, respectively. 
In the course of a pair creation process, an electron is not transferred from one electrode to the other, 
as in corresponding transport processes, 
but transfers onto the molecular bridge and, from there, back to the electrode where it came from. 
Although these processes do not directly contribute to the electrical current that is flowing 
through the junction, they represent a very important cooling mechanism for the vibrational degrees 
of freedom \cite{Hartle2010,Hartle2010b,Hartle2011,Volkovich2011b}. Therefore, as the efficiency of 
transport processes is, inter alia, determined by the level of vibrational excitation, 
they also have an indirect 
influence on the electrical transport properties of a molecular conductor. As we have already outlined 
in Ref.\ \cite{Hartle2010b}, this influence can, nonetheless, be substantial.

In the present context, it is important to note that in model DESVIB 
inelastic transport processes are not suppressed by destructive quantum 
interference effects due to the specific electronic-vibrational coupling scenario 
in this junction (\emph{i.e.}\ because $\lambda_{11}=0$). 
Therefore, we consider another model system in this section, model DESVIB2. 
It is very similar to model DESVIB but, 
in contrast, includes electronic-vibrational coupling also in state $1$:  
$\lambda_{11}\approx\lambda_{21}$ but $\lambda_{11}\neq\lambda_{21}$ 
(see Tab.\ \ref{parameters} for a complete list of parameters). 
As a consequence of the coupling to both states, inelastic transport processes are, similar 
to electronic transport processes, also suppressed by destructive quantum interference effects. 
In contrast, however, electron-hole pair creation processes are not suppressed, 
because they involve the coupling of an electronic state to just one of the electrodes and not 
to both, and, even if the respective coupling strength is negative, its square is not.

\begin{figure}
\begin{center}
\begin{tabular}{l}
\resizebox{\newwidth}{\newheight}{
\includegraphics{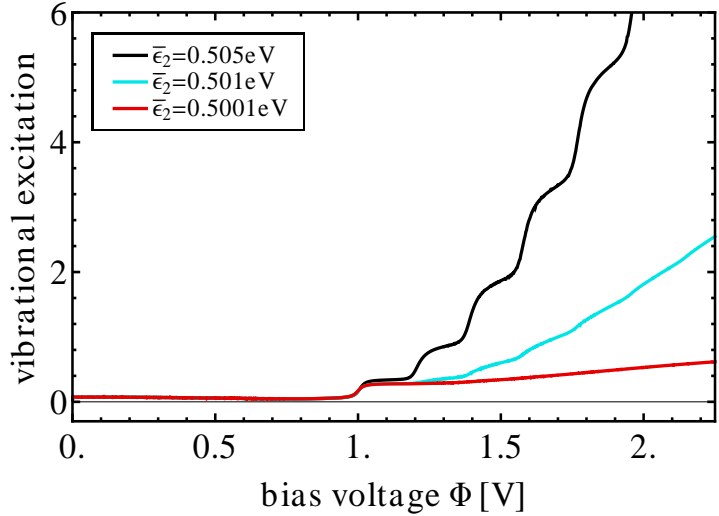}
}
\end{tabular}
\end{center}
\caption{(Color online)\label{FigCoolingbyInt} 
Vibrational excitation characteristics of the linear molecular conductor DESVIB2 
for different values of the polaron-shifted energy level of state $2$, $\overline{\epsilon}_{2}$. }
\end{figure}

This has a strong impact on the excitation levels of junction DESVIB2. The corresponding 
vibrational excitation characteristic is shown in Fig.\ \ref{FigCoolingbyInt} by the solid black line. 
It exhibits very similar features as the one shown by model DESVIB but approximately 
only half of the corresponding vibrational excitation levels (cf.\ Fig.\ \ref{DecoherenceSeqTunLinearFig}b). 
This is related to the fact that inelastic transport processes are suppressed by destructive quantum 
interference effects in this system while electron-hole pair creation processes are not. 
Thus, the ratio between heating and cooling processes, which is typically between 1:2 or 1:1 
for resonant electron transport through a single electronic state \cite{Hartle2011}, is shifted towards 
much lower values, resulting in much lower levels of vibrational excitation.

This cooling effect is even more pronounced, if the level spacing 
between the two electronic states is smaller and, consequently, the suppression of inelastic transport 
processes is more pronounced. The solid blue and red line in Fig.\ \ref{FigCoolingbyInt}, 
for example, show the excitation characteristics of model DESVIB2 for smaller
splittings of the (polaron-shifted) energy levels. 
The corresponding excitation levels are substantially smaller. The only major contribution 
that remains in the limit $\overline{\epsilon}_{2}\rightarrow\overline{\epsilon}_{1}$ is the one due 
to the formation of a polaronic state, $\sum_{mn}(\lambda_{m1}\lambda_{n1}/\Omega_{1})\langle 
c_{m}^{\dagger}c_{m} c_{n}^{\dagger}c_{n} \rangle_{\overline{H}}$ (cf.\ Eq.\ (\ref{formulavibex})). 
This, however, is only a static contribution that originates from charging processes and  
cannot be used in deexcitation processes. In other words, 
the current-induced excitation of the vibrational mode is almost completely 
suppressed in this limit.

Having in a mind that states $1$ and $2$ of model DESVIB2 
may represent just symmetric and antisymmetric combinations of localized molecular orbitals 
(see Figs.\ \ref{LinConduct}a and \ref{LinConduct}b), their electronic structure 
may be very similar and, accordingly, also the respective electronic-vibrational coupling 
strengths. Indeed, such pairs of electronic states are often found in realistic models of 
single-molecule junctions \cite{Benesch06,Benesch08,Hartle2011b,Ballmann2012}. 
The cooling mechanism described in this section can, therefore, be expected to occur 
in a large variety of single-molecule junctions. 

\subsection{Decoherence-Induced Temperature Dependence of the Current} 
\label{anomal}

In Sec.\ \ref{localcool}, we have shown that
electron-hole pair creation processes facilitate 
an important cooling mechanism for the vibrational degrees of freedom of a single-molecule 
contact \cite{Hartle2010,Hartle2010b,Hartle2011,Volkovich2011b}, 
particularly in the presence of strong destructive quantum interference effects. 
Within the local representation of the linear 
molecular conductor by two localized molecular orbitals (as in Fig.\ \ref{LinConduct}b), this 
can be understood in an alternative way: If the intramolecular coupling $\Delta$ becomes very weak, 
$\Delta\rightarrow0$, the left and the right part of the molecular conductor decouple. In that sense, the 
two parts of the molecule behave as being part of the left or right electrode, that is they also  
acquire the same equilibrium state. 
This thermalization is facilitated by electron-hole pair creation processes and  
is comparable to the thermalization of a molecule adsorbed on a surface with the substrate. 
Thus, the temperature in the electrodes can be used to control the 
excitation levels of the vibrational modes of the left and the right part of such a molecular conductor. 
Moreover, as higher levels of vibrational excitation enhance the effect of vibrationally induced decoherence 
(see Sec.\ \ref{SequTunReg}), 
the temperature in the electrodes can, in turn, be used to control quantum interference effects in 
single-molecule junctions. This will be demonstrated in this section.

In general, a variation of the temperature of the electrodes is accompanied by a change in the contact geometry 
of a single-molecule contact, either due to thermal expansion of the electrodes or 
irreversible drifts of single gold atoms. In experiments, the stability of the contact
geometry can often only be maintained for low temperature variations \cite{Ballmann2012}. 
In order to demonstrate control of quantum interference effects via the temperature of the electrodes, 
we therefore consider a model of a single-molecule contact that includes low-frequency 
vibrational modes, which, in the given range of temperatures, exhibit significant changes in their 
excitation levels. To this end, we consider a mode with a frequency of $\Omega_{1}=5$\,meV, 
which represents a frequency value at the lower end of the vibrational spectrum 
of a larger molecule \cite{Benesch06,Benesch08,Hartle2011b}. Moreover,  
we use a range from $10$\,K to $100$\,K for the temperature of the electrodes in the following, where 
the corresponding excitation level of the vibrational mode varies between $N_{\text{vib}}=0.003$ 
and $N_{\text{vib}}=1.3$, respectively. 
This model system is referred to as model DESCOOL (see Tab.\ \ref{parameters} for a complete 
list of parameters). Besides the low-frequency mode, it includes two electronic states that are 
almost degenerate, that is $\overline{\epsilon}_{2}-\overline{\epsilon}_{1}\ll\Gamma$.

\begin{figure}
\begin{tabular}{l}
\hspace{-0.5cm}(a) \\
\resizebox{\newwidth}{\newheight}{
\includegraphics{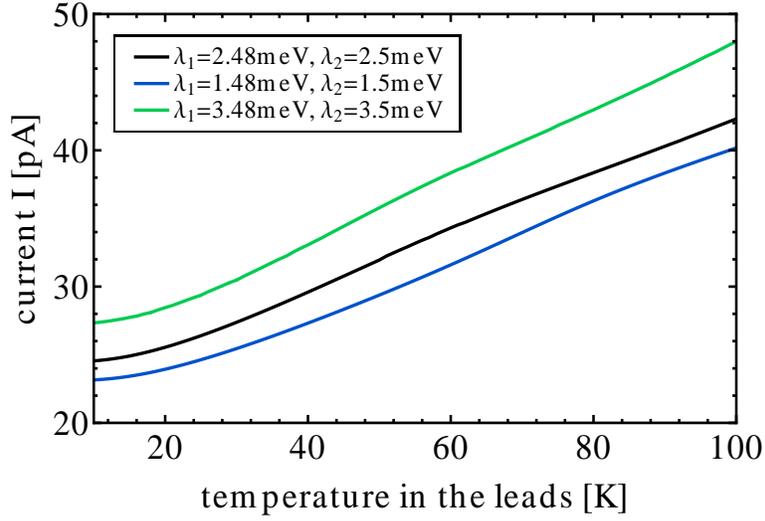}
}\\
\hspace{-0.5cm}(b) \\
\resizebox{\newwidth}{\newheight}{
\includegraphics{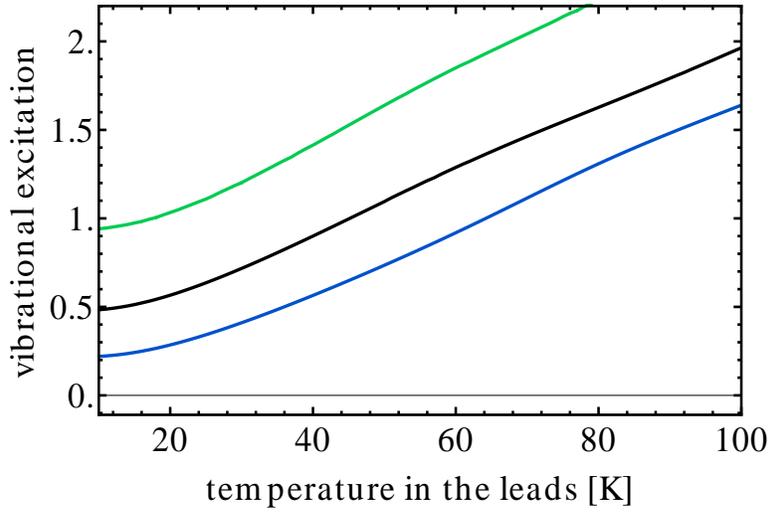}
}
\end{tabular}
\caption{(Color online)\label{FigTempDep} 
Panel (a): Current-temperature characteristics of the linear molecular conductor DESCOOL. 
Panel (b): Vibrational excitation characteristic corresponding to the 
current-temperature characteristic shown in Panel (a). 
}
\end{figure}

The current-temperature characteristic of this model system 
\footnote{It is noted that the current levels of model DESCOOl are rather low 
(tens of pA). This is because we have chosen
the level broadening $\Gamma=0.8$\,meV significantly smaller than the  
frequency of the vibrational mode $\Omega_{1}=5$\,meV in order to
ensure the validity of the effective factorization of the 
single-particle Green's function, Eq.\ (\ref{decoupling}). While this is a prerequisite for our 
nonequilibrium Green's function formalism, the underlying physical 
processes described in this section are effective also for larger molecule-lead 
coupling strengths.} 
and the corresponding level of vibrational 
excitation are shown by the solid black lines in Fig.\ \ref{FigTempDep}. Thereby, we have used a bias voltage 
of $e\Phi=0.2$\,eV$\gg2\overline{\epsilon}_{1/2}$, which corresponds to the resonant transport regime of this 
junction. Note that, at this bias voltage, 
the current level of junction DESCOOL is influenced neither by the thermal broadening of the Fermi 
distribution functions in the leads \cite{Selzer2004,Poot2006,Choi2008,Sedghi2011} nor vibrational 
sidepeaks that are associated with the onset of resonant excitation processes (Fig.\ \ref{basmech}a). 
As can be seen in Fig.\ \ref{FigTempDep}, however, the current and the excitation level 
of the vibrational mode increases substantially with the temperature in the electrodes. 
We attribute this behavior to the quenching of destructive quantum interference effects 
by inelastic transport processes, which is enhanced, via the thermalization mechanism 
described above, for higher 
temperatures of the electrodes. This shows that the temperature in the electrodes can be used to control 
destructive quantum interference effects in a single-molecule junction. Note that, very 
recently, this control mechanism has been demonstrated by Ballmann \emph{et al.}\ in mechanically 
controlled break junction experiments for a variety of different molecules \cite{Ballmann2012}.

It is interesting at this point to study the temperature dependence of model DESCOOL 
for different absolute values of the electronic-vibrational coupling strengths 
$\lambda_{11}$ and $\lambda_{21}$. Thereby, as we have seen in Sec.\ \ref{SequTunReg}, 
it is expedient to keep the difference in the dimensionless couplings constant, \emph{i.e.}\ 
$\vert \lambda_{11} - \lambda_{21} \vert/\Omega_{1}$. Accordingly, 
the solid blue/green lines in Fig.\ \ref{FigTempDep} 
represent the temperature characteristics of model DESCOOL that we obtain by 
reducing/increasing the two coupling strengths by $1$\,meV. 
These results corroborate that vibrationally induced decoherence 
is, to a large extent, driven by nonequilibrium effects 
(see, e.g., the discussion at the end of Sec.\ \ref{SequTunReg}), because 
for smaller electronic-vibrational coupling strengths we find lower levels of vibrational excitation 
and, due to the consequently less effective quenching of destructive interference effects, 
a more pronounced suppression of the current level. 
As a result, the relative temperature dependence 
$\Delta I= I(100\,\text{K})-I(10\,\text{K})$ is stronger for weaker electronic-vibrational coupling strengths 
in this system. 
It should be noted at this point that, in systems where destructive interference effects are not active, 
weaker electronic-vibrational coupling is typically associated with larger levels of vibrational excitation \cite{Hartle2010b}. 
This counterintuitive phenomenon is related to the suppression electron-hole pair creation processes, 
which, in the resonant transport regime, is more pronounced for higher bias voltages but also 
for weaker electronic-vibrational coupling. This suppression of pair creation processes is counterbalanced 
in transport through junction DESCOOL due to the presence of destructive interference effects.

\section{Conclusion}

We have investigated the role of vibrations in electron transport through 
single-molecule junctions, which are governed by strong quantum interference effects. 
To this end, we have employed both analytical and numerical results that are obtained from   
nonequilibrium Green's function theory \cite{Galperin06,Hartle,Hartle09,Hartle2010,Volkovich2011b}. 

Our results show that electronic-vibrational coupling has a strong and non-trivial influence 
on the transport properties of a molecular conductor, in particular, if quantum 
interference effects play a dominant role. This has been demonstrated by investigating 
models of molecular junctions, 
where pronounced quantum interference effects arise due to the existence of quasidegenerate states. 
On one hand, these states provide different pathways for a tunneling electron, 
which gives rise to strong interference effects in the respective transport properties. 
On the other hand, however, each of these states is, in general, also very specifically 
coupled to the vibrational degrees of freedom of the junction 
such that interaction of the tunneling electrons with the vibrations gives information about their 
tunneling pathway. This which-path information quenches the interference effects that are inherent to 
transport through quasidegenerate states. 


As we have shown, this decoherence mechanism is more pronounced for higher 
levels of vibrational excitation or, equivalently, higher effective temperatures of the vibrations, 
because the probability for an interaction of the tunneling electron with the vibrations 
increases with temperature. For example, at high bias voltages, where resonant transport 
processes may result in high levels of vibrational excitation, vibrationally induced decoherence 
is therefore very pronounced and quantum interference effects play only a minor role. 
At the onset of the resonant transport regime, however, 
interference effects may result in sizable effects, 
such as, for example, a reorganization of step heights in the current-voltage 
characteristics or a modification of the 
widths of these steps. At low bias voltages or in the non-resonant transport regime, 
where the excitation levels of the vibrational modes are typically rather low, 
vibrationally induced decoherence is less pronounced. 
In this regime, the interplay between interference and decoherence may, 
nevertheless, result in a strong modification 
of the signals that are associated with inelastic electron tunneling. 
For example, in the presence of destructive interference effects, 
the onset of inelastic processes is, in general,  
more pronounced, while in the presence of constructive interference effects the respective 
signals are suppressed or even reversed. As a result, 
negative jumps in the conductance-voltage characteristic can appear, 
although the conductance of the molecule is much smaller than the 
conductance quantum \cite{Lorente2000,Haupt2009,Avriller2009,Schmidt2009}. 
A correct and thorough analysis of vibrational signals in the 
transport characteristics of a single-molecule contact needs to take into account these 
effects, as they may lead to strong deviations from the commonly employed Franck-Condon picture.

We have also elucidated the role of electron-hole pair creation processes in the presence 
of strong destructive quantum interference effects. In contrast to transport processes, 
electron-hole pair creation processes
are not suppressed by interference effects and, therefore, 
constitute the dominant inelastic processes in such junctions.  
At low temperatures of the electrodes, they are solely cooling the vibrational 
degrees of freedom. In this case, 
the effective temperature of the vibrational modes is only very weakly influenced by 
inelastic electron transport processes and rather determined by the temperature 
of the electrodes. 
Because the level of vibrational excitation determines, inter alia, the 
efficiency of vibrationally induced 
decoherence in such a junction, electron-hole pair creation processes thus facilitate 
an effective mechanism to control quantum interference effects in single-molecule junctions 
by an external control parameter, that is the temperature of the electrodes. 
This has recently been demonstrated in experiments on 
a variety of different molecular junctions by Ballmann \emph{et al.}\ \cite{Ballmann2012}. 

While we have presented a comprehensive study of vibrationally induced decoherence 
in single-molecule junctions, further research is required that addresses 
not only the anti-adiabatic regime but also the adiabatic and the respective cross-over regime. 
As quantum interference effects appear in particular for quasi-degenerate 
electronic states, it is also interesting to investigate the effect of 
non-adiabatic electronic-vibrational coupling \cite{Domcke04,Reckermann2008,Repp2009} 
in these systems. Moreover, the analysis should be extended to regimes, where the renormalized 
electron-electron interactions $\overline{U}_{mn}$ (cf.\ Sec.\ \ref{hamiltonian})  are not negligible. These
extensions require the development of theoretical methods that allow a
nonperturbative description of electron-electron interactions 
as well as electronic-vibrational coupling, including higher order effects and correlations. 



\section*{Acknowledgement}

We thank S.\ Ballmann, S.\ Brisker-Klaiman, P.\ B.\ Coto, B.\ Kubala, A.\ Nitzan, U.\ Peskin and 
H.\ B.\ Weber for many fruitful and inspiring discussions. 
MT gratefully acknowledges the hospitality of the IAS at the Hebrew University Jerusalem 
within the workshop on molecular electronics. 
The generous allocation of computing time by the computing centers 
in Erlangen (RRZE), Munich (LRZ), and J\"ulich (JSC) is gratefully acknowledged. 
This work has been supported by the 
German-Israeli Foundation for Scientific Development (GIF) and the 
Deutsche Forschungsgemeinschaft (DFG) 
through the DFG-Cluster of Excellence 'Engineering of Advanced Materials' and SFB 953. 
RH acknowledges support by the Alexander von Humboldt Foundation.

\appendix

\section{Transmission function for junction DESVIB}
\label{transDESVIB}

In this appendix, we derive the 
transmission function of model DESVIB and the corresponding incoherent and 
interference term (cf.\ Eqs.\ (\ref{condesparts}), Sec.\ \ref{SequTunReg}). 
To this end, we employ the wide-band approximation, 
$\Gamma\approx\Gamma_{\text{L},11}(\mu_{\text{L}})$, 
and consider large bias voltages ($\Phi\rightarrow\infty$). Furthermore, we 
evaluate the vibrational degree of freedom in thermal equilibrium, that is we 
restrict it to an effective (finite) temperature $T$. Given these assumptions, 
the real-time projections of the corresponding electronic self-energy matrices 
(\ref{elselfen}) can be written as 
\begin{eqnarray}
 \Sigma^{<}_{\text{L}}(\epsilon) &=& \left( \begin{array}{cc}
i \Gamma & i \Gamma A \\ 
i A \Gamma & i \Gamma \\ 
\end{array}
\right), \\
 \Sigma^{>}_{\text{L}}(\epsilon) &=& 0, \\
 \Sigma^{<}_{\text{R}}(\epsilon) &=& 0, \\ 
 \Sigma^{>}_{\text{R}}(\epsilon) &=& \left( \begin{array}{cc}
-i \Gamma & i \Gamma A \\ 
i A \Gamma & -i \Gamma \\ 
\end{array}
\right), \\
 \Sigma^{\text{r}}_{\text{L}}(\epsilon) &=& \left( \begin{array}{cc}
- i \Gamma & -i \Gamma A \\ 
-i \Gamma A & - i \Gamma \\ 
\end{array}
\right) = \left(\Sigma^{\text{a}}_{\text{L}}(\epsilon) \right)^{\dagger}, \\
 \Sigma^{\text{r}}_{\text{R}}(\epsilon) &=& \left( \begin{array}{cc}
- i \Gamma & i \Gamma A \\ 
i \Gamma A & - i \Gamma \\ 
\end{array}
\right) = \left(\Sigma^{\text{a}}_{\text{R}}(\epsilon) \right)^{\dagger}, 
\end{eqnarray}
with 
\begin{eqnarray}
 A &=& \text{e}^{-\frac{\lambda_{21}^{2}}{2\Omega_{1}}(2N_{\text{vib}}+1)}, \\
N_{\text{vib}} &=& \frac{1}{\text{e}^{\frac{\Omega_{1}}{k_{\text{B}}T}}-1}. 
\end{eqnarray} 
The retarded/advanced projection of the electronic part of the single-particle Green's function 
(cf.\ Eq.\ (\ref{elDy})) is therefore given by 
\begin{eqnarray}
 \bar{G}^{\text{r/a}}(\epsilon) &=& \left( \begin{array}{cc}
g_{1}^{\text{r/a}}(\epsilon)  & 0 \\ 
0 & g_{2}^{\text{r/a}}(\epsilon)  \\ 
\end{array}
\right) = \left( \begin{array}{cc}
\frac{1}{\epsilon-\overline{\epsilon}_{1}\pm i\Gamma}  & 0 \\ 
0 & \frac{1}{\epsilon-\overline{\epsilon}_{2}\pm i\Gamma}  \\ 
\end{array}
\right) 
\end{eqnarray}
and, according to the Keldysh equations (\ref{elKe}), we write the corresponding 
greater Green's function as 
\begin{eqnarray}
 \bar{G}^{>}(\epsilon) &=& \left( \begin{array}{cc}
-i\Gamma \vert g_{1}^{\text{r}}(\epsilon) \vert^{2}  & i\Gamma A g_{1}^{\text{r}}(\epsilon) g_{2}^{\text{a}}(\epsilon) \\ 
i\Gamma A g_{2}^{\text{r}}(\epsilon) g_{1}^{\text{a}}(\epsilon) & -i\Gamma \vert g_{2}^{\text{r}}(\epsilon) \vert^{2}  \\ 
\end{array}
\right). \nonumber
\end{eqnarray}
The greater projection of the single-particle Green's function matrix 
(see Eq.\ (\ref{decoupling})) thus reads 
\begin{eqnarray}
 G^{>}(\epsilon) &=& \left( \begin{array}{cc}
-i\Gamma \vert g_{1}^{\text{r}}(\epsilon) \vert^{2}  & i\Gamma A^{2} g_{1}^{\text{r}}(\epsilon) g_{2}^{\text{a}}(\epsilon) \\ 
i\Gamma A^{2} g_{2}^{\text{r}}(\epsilon) g_{1}^{\text{a}}(\epsilon) & -i\Gamma A^{2} \sum_{l=-\infty}^{\infty} 
B_{l}  \vert g_{2}^{\text{r}}(\epsilon-l\Omega_{1}) \vert^{2}  \\ 
\end{array}
\right), 
\end{eqnarray}
with 
\begin{eqnarray}
 B_{l} &=& I_{l}\left(2\frac{\lambda_{21}^{2}}{\Omega_{1}^{2}}\sqrt{N_{\text{vib}}(N_{\text{vib}}+1)}\right) \text{e}^{\frac{l\Omega_{1}}{2k_{\text{B}}T}}
\end{eqnarray}
and $I_{l}$ the $l$th Bessel function of the first kind. 
Thereby, we have used the lesser projection of the shift operator correlation function \cite{Mahan81}
\begin{eqnarray}
\label{shiftopfunc}
 \langle X_m(t)X_{n}^\dagger(t') \rangle_{\overline{H}}^{>} &=& \delta_{m2} \delta_{n2} 
\text{exp}\left[ \frac{\lambda_{21}^{2}}{\Omega_{1}^{2}} \left( N_{\text{vib}} (\text{e}^{i\Omega_{1}(t-t')}-1)  + (1+ N_{\text{vib}}) (\text{e}^{-i\Omega_{1}(t-t')}-1) \right)   \right]  \nonumber\\
&=& \delta_{m2} \delta_{n2}  A^{2} \sum_{l=-\infty}^{\infty} B_{l} \text{e}^{-i l \Omega_{1} (t-t')},
\end{eqnarray}
and convoluted this function with the lesser projection of the electronic part of 
the single-particle Green's function according to Eq.\ (\ref{decoupling}). 
Finally, the transmission function is obtained using the Meir-Wingreen-like formula (\ref{currentformula}), 
which can be rewritten as 
\begin{eqnarray}
I &=& 2e\int\frac{\text{d}\epsilon}{2\pi}\, \sum_{mn} 
\left( \Sigma_{\text{L},mn}^{(0),<}(\epsilon) G^{>}_{nm}(\epsilon)-\Sigma_{\text{L},mn}^{(0),>}(\epsilon) G^{<}_{nm}(\epsilon) \right)  
\end{eqnarray}
with  
\begin{eqnarray}
  \Sigma^{(0),<}_{\text{L}}(\epsilon) &=& \left( \begin{array}{cc}
i \Gamma & i \Gamma  \\ 
i \Gamma & i \Gamma \\ 
\end{array}
\right), \\
 \Sigma^{(0),>}_{\text{L}}(\epsilon) &=& 0. 
\end{eqnarray}
For large bias voltages, this expression can be used to define a transmission function 
by 
\begin{eqnarray} 
\label{deftrans}
 I &\stackrel{\Phi\rightarrow\infty}{=}&2e\int_{\infty}^{\infty}\frac{\text{d}\epsilon}{2\pi}\,  
\text{Tr}\left[ \Sigma_{\text{L}}^{(0),<}(\epsilon) G^{>}(\epsilon) \right] \equiv 2e\int_{\infty}^{\infty}\frac{\text{d}\epsilon}{2\pi}\, t(\epsilon).  
\end{eqnarray}
This transmission function is given by 
\begin{eqnarray}
\label{auxua}
 t(\epsilon) &=& \text{Tr}\left[ \Sigma_{\text{L}}^{(0),<}(\epsilon) G^{>}(\epsilon) \right] \\ 
&=& \text{Tr}\left[ 
\left( \begin{array}{cc}
i\Gamma  & i\Gamma  \\ 
i\Gamma  & i\Gamma  \\ 
\end{array}
\right)
\left( \begin{array}{cc}
-i\Gamma \vert g_{1}^{\text{r}}(\epsilon) \vert^{2}  & i\Gamma A^{2} g_{1}^{\text{r}}(\epsilon) g_{2}^{\text{a}}(\epsilon) \\ 
i\Gamma A^{2} g_{2}^{\text{r}}(\epsilon) g_{1}^{\text{a}}(\epsilon) & -i\Gamma A^{2} \sum_{l=-\infty}^{\infty} 
B_{l}  \vert g_{2}^{\text{r}}(\epsilon-l\Omega_{1}) \vert^{2}  \\ 
\end{array}
\right)\right] \nonumber\\
&=& \Gamma^{2} \vert g_{1}^{\text{r}}(\epsilon) \vert^{2} - \Gamma^{2} A^{2} \left( 
g_{2}^{\text{r}}(\epsilon) g_{1}^{\text{a}}(\epsilon) + g_{1}^{\text{r}}(\epsilon) g_{2}^{\text{a}}(\epsilon)  
\right) + \Gamma^{2} A^{2} \sum_{l=-\infty}^{\infty} 
B_{l}  \vert g_{2}^{\text{r}}(\epsilon-l\Omega_{1}) \vert^{2}. \nonumber
\end{eqnarray}
The respective incoherent term is obtained by neglecting the off-diagonal 
elements of $\Sigma^{<,(0)}_{\text{L}}$ in Eq.\ (\ref{auxua}) (cf.\ the discussion given in Sec.\ \ref{BasIntSec}) 
\begin{eqnarray}
 t_{\text{inc}}(\epsilon) &=& \Gamma^{2} \vert g_{1}^{\text{r}}(\epsilon) \vert^{2} + \Gamma^{2} A^{2} \sum_{l=-\infty}^{\infty} 
B_{l}  \vert g_{2}^{\text{r}}(\epsilon-l\Omega_{1}) \vert^{2}, 
\end{eqnarray}
which also allows to identify the corresponding interference term as 
\begin{eqnarray}
 t_{\text{int}}(\epsilon) &=&  t(\epsilon) - t_{\text{inc}}(\epsilon) \\
&=& - \Gamma^{2} A^{2} \left( 
g_{2}^{\text{r}}(\epsilon) g_{1}^{\text{a}}(\epsilon) + g_{1}^{\text{r}}(\epsilon) g_{2}^{\text{a}}(\epsilon)  
\right).  \nonumber
\end{eqnarray}

\section{Transmission function for junction CONVIB}
\label{transCONVIB}

In this appendix, we derive the transmission function of model CONVIB and the 
corresponding incoherent and interference term (cf.\ Eqs.\ (\ref{condespartsII}), Sec.\ \ref{SequTunReg}). 
To this end, we employ the same approximations as in appendix \ref{transDESVIB}. 
The corresponding real-time projections of the 
electronic self-energy matrices (\ref{elselfen}) are given by 
\begin{eqnarray}
 \Sigma^{<}_{\text{L}}(\epsilon) &=& \left( \begin{array}{cc}
i \Gamma & i \Gamma A \\ 
i \Gamma A & i \Gamma \\ 
\end{array}
\right), \\
 \Sigma^{>}_{\text{L}}(\epsilon) &=& 0, \\
 \Sigma^{<}_{\text{R}}(\epsilon) &=& 0, \\ 
 \Sigma^{>}_{\text{R}}(\epsilon) &=& \left( \begin{array}{cc}
-i \Gamma & - i \Gamma A \\ 
-i \Gamma A & -i \Gamma \\ 
\end{array}
\right), \\
 \Sigma^{\text{r}}_{\text{L}}(\epsilon) &=& \left( \begin{array}{cc}
- i \Gamma & -i \Gamma A \\ 
-i \Gamma A & - i \Gamma \\ 
\end{array}
\right) = \left(\Sigma^{\text{a}}_{\text{L}}(\epsilon) \right)^{\dagger}, \\
 \Sigma^{\text{r}}_{\text{R}}(\epsilon) &=& \left( \begin{array}{cc}
- i \Gamma & -i \Gamma A \\ 
-i \Gamma A & - i \Gamma \\ 
\end{array}
\right) = \left(\Sigma^{\text{a}}_{\text{R}}(\epsilon) \right)^{\dagger}, 
\end{eqnarray}
leading to the following retarded/advanced projection of the electronic part of the single-particle Green's function 
(cf.\ Eq.\ (\ref{elDy})):  
\begin{eqnarray}
 \bar{G}^{\text{r}}(\epsilon) &=& \left( \begin{array}{cc}
g_{11}^{\text{r}}(\epsilon)  & g_{12}^{\text{r}}(\epsilon) \\ 
g_{21}^{\text{r}}(\epsilon) & g_{22}^{\text{r}}(\epsilon)  \\ 
\end{array}
\right) \\
&=& \frac{1}{(\epsilon-\overline{\epsilon}_{1}+i\Gamma)(\epsilon-\overline{\epsilon}_{2}+i\Gamma)+\Gamma^{2}A^{2}} \left( \begin{array}{cc}
\epsilon-\overline{\epsilon}_{2}+i\Gamma  & -i\Gamma A \\ 
-i\Gamma A & \epsilon-\overline{\epsilon}_{1}+i\Gamma   \\ 
\end{array}
\right). \nonumber
\end{eqnarray}
According to the Keldysh equations (\ref{elKe}), we can write the greater projection of the (electronic) 
Green's function as 
\begin{eqnarray}
 \bar{G}^{>}(\epsilon) 
&=& -i\Gamma \left\vert \frac{1}{(\epsilon-\overline{\epsilon}_{1}+i\Gamma)(\epsilon-\overline{\epsilon}_{2}+i\Gamma)+\Gamma^{2}A^{2}} 
\right\vert^{2} \cdot\nonumber\\
&& \left( \begin{array}{cc}
\epsilon-\overline{\epsilon}_{2}+i\Gamma  & -i\Gamma A \\ 
-i\Gamma A & \epsilon-\overline{\epsilon}_{1}+i\Gamma   \\ 
\end{array}
\right)
\left( \begin{array}{cc}
\epsilon-\overline{\epsilon}_{2}-i\Gamma(1-A^{2})  & A(\epsilon-\overline{\epsilon}_{1}) \\ 
A(\epsilon-\overline{\epsilon}_{2}) & \epsilon-\overline{\epsilon}_{1}-i\Gamma(1-A^{2})  \\ 
\end{array}
\right) \nonumber\\
&=& -i\Gamma c(\epsilon)   
\left( \begin{array}{cc}
Z_{11}(\epsilon)  & Z_{12}(\epsilon) \\ 
Z_{21}(\epsilon) & Z_{22}(\epsilon)  \\ 
\end{array}
\right),   \nonumber
\end{eqnarray}
with 
\begin{eqnarray}
c(\epsilon) &=& \left\vert \frac{1}{(\epsilon-\overline{\epsilon}_{1}+i\Gamma)(\epsilon-\overline{\epsilon}_{2}+i\Gamma)+\Gamma^{2}A^{2}} 
\right\vert^{2}, \\
Z_{11}(\epsilon) &=& \vert \epsilon-\overline{\epsilon}_{2}+i\Gamma \vert^{2} - \Gamma^{2}A^{2}, \\
Z_{12}(\epsilon) &=& A (\epsilon-\overline{\epsilon}_{1}) (\epsilon-\overline{\epsilon}_{2}) -\Gamma^{2} A (1-A^{2}),  \\
Z_{21}(\epsilon) &=& A (\epsilon-\overline{\epsilon}_{1}) (\epsilon-\overline{\epsilon}_{2}) -\Gamma^{2} A (1-A^{2}),  \\
Z_{22}(\epsilon) &=& \vert \epsilon-\overline{\epsilon}_{1}+i\Gamma \vert^{2} - \Gamma^{2}A^{2}. 
\end{eqnarray}
Convolution of this expression with the greater projection of the shift operator correlation functions 
(\ref{shiftopfunc}) leads, according to Eq.\ (\ref{decoupling}), to the greater projection of the 
single-particle Green's function matrix 
\begin{eqnarray}
 G^{>}(\epsilon) &=& -i\Gamma \left( \begin{array}{cc}
c(\epsilon) Z_{11}(\epsilon)  & A c(\epsilon) Z_{12}(\epsilon) \\ 
A c(\epsilon) Z_{21}(\epsilon) & A^{2} \sum_{l=-\infty}^{\infty} 
B_{l}  c(\epsilon-l\Omega_{1}) Z_{22}(\epsilon-l\Omega_{1})  \\ 
\end{array}
\right). 
\end{eqnarray}
The corresponding transmission function, as defined by Eq.\ (\ref{deftrans}), thus reads
\begin{eqnarray}
\label{auxua2}
 t(\epsilon) &=& \text{Tr}\left[ \Sigma_{\text{L}}^{(0),<}(\epsilon) G^{>}(\epsilon) \right] \\ 
&=& \Gamma^{2} \text{Tr}\left[ 
\left( \begin{array}{cc}
1  & 1  \\ 
1  & 1  \\ 
\end{array}
\right)
\left( \begin{array}{cc}
c(\epsilon) Z_{11}(\epsilon)  & A c(\epsilon) Z_{12}(\epsilon) \\ 
A c(\epsilon) Z_{21}(\epsilon) & A^{2} \sum_{l=-\infty}^{\infty} 
B_{l}  c(\epsilon-l\Omega_{1}) Z_{22}(\epsilon-l\Omega_{1})  \\ 
\end{array}
\right)\right] \nonumber\\
&=& \Gamma^{2} \left( c(\epsilon) Z_{11}(\epsilon) +  A c(\epsilon) Z_{21}(\epsilon) + A c(\epsilon) Z_{12}(\epsilon) + 
A^{2} \sum_{l=-\infty}^{\infty} 
B_{l}  c(\epsilon-l\Omega_{1}) Z_{22}(\epsilon-l\Omega_{1}) \right). \nonumber
\end{eqnarray}
Considering the contributions of the off-diagonal elements of $\Sigma^{<,(0)}_{\text{L}}$ 
in Eq.\ (\ref{auxua2}) (cf.\ the discussion given in Sec.\ \ref{BasIntSec}), 
it can be decomposed into the interference term 
\begin{eqnarray}
 t_{\text{int}}(\epsilon) &=& A \Gamma^{2} c(\epsilon) \left( Z_{21}(\epsilon) +  Z_{12}(\epsilon)  \right),
\end{eqnarray}
and the incoherent term 
\begin{eqnarray}
 t_{\text{inc}}(\epsilon) &=& \Gamma^{2} \left( c(\epsilon) Z_{11}(\epsilon)  + A^{2} \sum_{l=-\infty}^{\infty}  
B_{l}  c(\epsilon-l\Omega_{1}) Z_{22}(\epsilon-l\Omega_{1}) \right). 
\end{eqnarray}

\end{document}